%% file: Main.tex
\documentclass[manuscript,screen]{acmart}

\usepackage{makecell}
\usepackage[most]{tcolorbox}
\usepackage{fontawesome}
\usepackage{pifont}
\usepackage{multirow}
\usepackage{tikz}
\usepackage{pgfplots}
\usepackage{xcolor}
\usepackage{subcaption}
\usepackage{tabularx}
\usepackage[table]{xcolor}
\usepackage{placeins}
\pgfplotsset{compat=1.18}
\usetikzlibrary{shapes.geometric, arrows.meta, positioning, calc, patterns}

\definecolor{symptom1}{RGB}{66, 133, 244}   
\definecolor{symptom2}{RGB}{234, 67, 53}    
\definecolor{symptom3}{RGB}{251, 188, 4}    
\definecolor{symptom4}{RGB}{52, 168, 83}    
\definecolor{symptom5}{RGB}{153, 102, 204}  
\definecolor{symptom6}{RGB}{255, 152, 0}    

\definecolor{root1}{RGB}{66, 133, 244}      
\definecolor{root2}{RGB}{234, 67, 53}       
\definecolor{root3}{RGB}{251, 188, 4}       
\definecolor{root4}{RGB}{52, 168, 83}       
\definecolor{root5}{RGB}{153, 102, 204}     
\definecolor{root6}{RGB}{255, 152, 0}       

\definecolor{fastmcp}{RGB}{66, 133, 244}
\definecolor{tssdk}{RGB}{234, 67, 53}
\definecolor{pysdk}{RGB}{52, 168, 83}
\definecolor{javasdk}{RGB}{251, 188, 4}

\definecolor{lightgray}{gray}{0.9}
\definecolor{lightpink}{rgb}{1.0, 0.9, 0.9}
\definecolor{lightyellow}{rgb}{1.0, 1.0, 0.9}
\definecolor{lightgreen}{rgb}{0.9, 1.0, 0.9}
\definecolor{lightblue}{rgb}{0.9, 0.95, 1.0}
\definecolor{lightbeige}{rgb}{0.97, 0.94, 0.88}
\definecolor{lightpurple}{rgb}{0.88, 0.97, 0.97}

\newtcolorbox{mybox}[2][]
  {colback = white, colframe = black,
    colbacktitle = gray, enhanced,
    attach boxed title to top left={yshift=-2mm,xshift = 4mm},
    title=#2,#1}

\AtBeginDocument{%
  }

\setcopyright{none}
\copyrightyear{2026}
\acmYear{2026}
\acmDOI{}

\acmJournal{TOSEM}
\acmVolume{1}
\acmNumber{1}
\acmArticle{1}
\acmMonth{6}

\begin{document}


\title{What Was Once Learned May Need to Be Unlearned: Machine Unlearning for Deprecated API Knowledge in Large Language Models}

\author{Jin Liu}
\email{jinliu@whu.edu.cn}
\affiliation{%
  \institution{School of Computer Science, Wuhan University}
  \city{Wuhan}
  \country{China}
}

\author{Yanzhong He}
\email{yanzhonghe@whu.edu.cn}
\affiliation{%
  \institution{School of Computer Science, Wuhan University}
  \city{Wuhan}
  \country{China}
}

\author{Guancheng Lin}
\email{guanchlin4-c@my.cityu.edu.hk}
\affiliation{%
  \institution{Department of Computer Science, City University of Hong Kong}
  \city{Hong Kong}
  \country{China}
}

\author{Xiao Liu}
\email{xiao.liu@deakin.edu.au}
\affiliation{%
  \institution{School of Information Technology, Deakin University}
  \city{Burwood}
  \country{Australia}
}

\author{Jacky Wai Keung}
\email{Jacky.Keung@cityu.edu.hk}
\affiliation{%
  \institution{Department of Computer Science, City University of Hong Kong}
  \city{Hong Kong}
  \country{China}
}

\author{Xiao Yu}
\email{xiao.yu@zju.edu.cn}
\affiliation{
  \institution{The State Key Laboratory of Blockchain and Data Security, Zhejiang University}
  \city{Hangzhou}
  \country{China}
}

\author{Xiaoxue Ma}
\email{kxma@hkmu.edu.hk}
\affiliation{%
  \institution{School of Science and Technology, Hong Kong Metropolitan University}
  \city{Hong Kong}
  \country{China}
}

 \renewcommand{\shortauthors}{Liu et al.}
\begin{abstract}

Large language models (LLMs) for code completion may generate deprecated APIs because their pre-training corpora contain code written against historical library versions. Existing approaches address this problem through inference-time intervention, model editing, or machine unlearning. However, code completion often admits multiple plausible solutions, and suppressing a deprecated API should not require the model to always generate a predefined up-to-date replacement. Moreover, existing deprecated-API unlearning studies primarily construct forget data from predefined samples without consistently verifying whether the target model actually exhibits the corresponding deprecated behavior, and provide limited evaluation of unintended changes to other and unrelated API behaviors. 
To address these limitations, we conduct a systematic empirical study of machine unlearning for deprecated API knowledge. We construct \texttt{MUDAPIBench}, a behavior-grounded benchmark containing more than 7,000 model-specific unlearning instances derived from 145 deprecated-to-up-to-date API mappings across eight Python libraries. Using a largely automated construction pipeline, we retain a candidate instance only when the original model actually generates the corresponding deprecated API. We evaluate eight representative machine unlearning methods across three code LLMs in terms of deprecated API forgetting, up-to-date API generation, preservation of other API behaviors, preservation of unrelated API behavior, general code-generation capability, and computational efficiency. 
Our results show that Gradient Difference (GD) achieves the most favorable overall trade-off among the evaluated methods, effectively suppressing deprecated API generation while better maintaining up-to-date API generation, other and unrelated API behaviors, and general code-generation capability. GD also maintains moderate training time and relatively low memory overhead. Further analyses reveal substantial variation across software libraries and show that APIs deprecated after a model's training-data cutoff are considerably more difficult to forget than those deprecated before the cutoff. Layer-wise representation and parameter analyses further show that GD achieves strong forgetting with comparatively controlled internal changes. Based on these findings, we derive implications for practitioners and researchers on applying, evaluating, and advancing machine unlearning for evolving API knowledge.

\end{abstract}




\maketitle

\section{Introduction}

Modern software development relies extensively on third-party libraries to reuse existing functionality through Application Programming Interfaces (APIs)~\cite{wang2020empirical,zhan2021research}. These libraries continuously evolve through refactoring~\cite{kula2018empirical}, bug fixes~\cite{hu2023empirical}, security updates~\cite{wang2020exploring}, and functional extensions, resulting in frequent changes to their APIs. During this evolution, some APIs are marked as \textit{deprecated APIs} and replaced by newer recommended alternatives, which we refer to as \textit{up-to-date APIs}. Developers are therefore expected to migrate from deprecated APIs to their corresponding up-to-date APIs, as deprecated APIs may eventually be removed or become incompatible with evolving library functionality.

Such API evolution poses a challenge for Large Language Models (LLMs) used for code completion. Although LLMs have demonstrated strong code-generation capabilities and are increasingly integrated into coding assistants \cite{yu2024fight,yu2025realisticcodebench,sun2026fly,jiang2026survey}, their knowledge is largely determined by the code available during training. Large-scale pre-training corpora may contain code written against historical library versions as well as mixed usages of deprecated and up-to-date APIs. Consequently, LLMs may retain outdated API usage patterns even after the corresponding libraries have evolved and may still recommend deprecated APIs during code completion. Wang et al.~\cite{wang2024llms}, for example, reported that 37.4\% of API calls generated by GPT-3.5 correspond to deprecated APIs. Such outdated recommendations require additional developer verification and migration effort and may eventually become invalid as the corresponding libraries continue to evolve.

Existing approaches address deprecated API generation primarily through inference-time intervention or model-level knowledge updating. At inference time, \texttt{REPLACEAPI}~\cite{wang2024llms} maintains mappings between deprecated and up-to-date APIs and uses them to revise LLM-generated completions. When a deprecated API is detected, the subsequent generated tokens are discarded, the corresponding up-to-date API is inserted to construct a new prefix, and the LLM is invoked again to continue completion. Although this strategy can correct detected deprecated API usages, it leaves the outdated knowledge encoded in the model unchanged. Moreover, inspecting generated completions and re-invoking the LLM for detected cases introduce additional latency and token consumption, which is undesirable for interactive code completion. 
To update deprecated API knowledge at the model level, \textsc{AdaLoRA-L}~\cite{guancheng2026don} formulates API evolution as a targeted model-editing problem that promotes the corresponding up-to-date API as a predefined editing target. To construct its editing instances, three completions are generated for each candidate prompt under deterministic decoding (temperature $=0$), and only prompts for which the deprecated API appears in all three completions are retained. As illustrated in the upper part of Figure~\ref{fig:motivation}, all three completions for the same prompt contain the deprecated API \verb|torch.svd|, while \verb|torch.linalg.svd| is specified as its corresponding up-to-date API and editing target. This construction focuses on contexts in which deprecated API behavior is consistently observed and naturally supports a directed deprecated-to-up-to-date API replacement objective.

\subsection{Motivation}

\begin{figure*}[t]
    \centering
    \includegraphics[width=\textwidth]{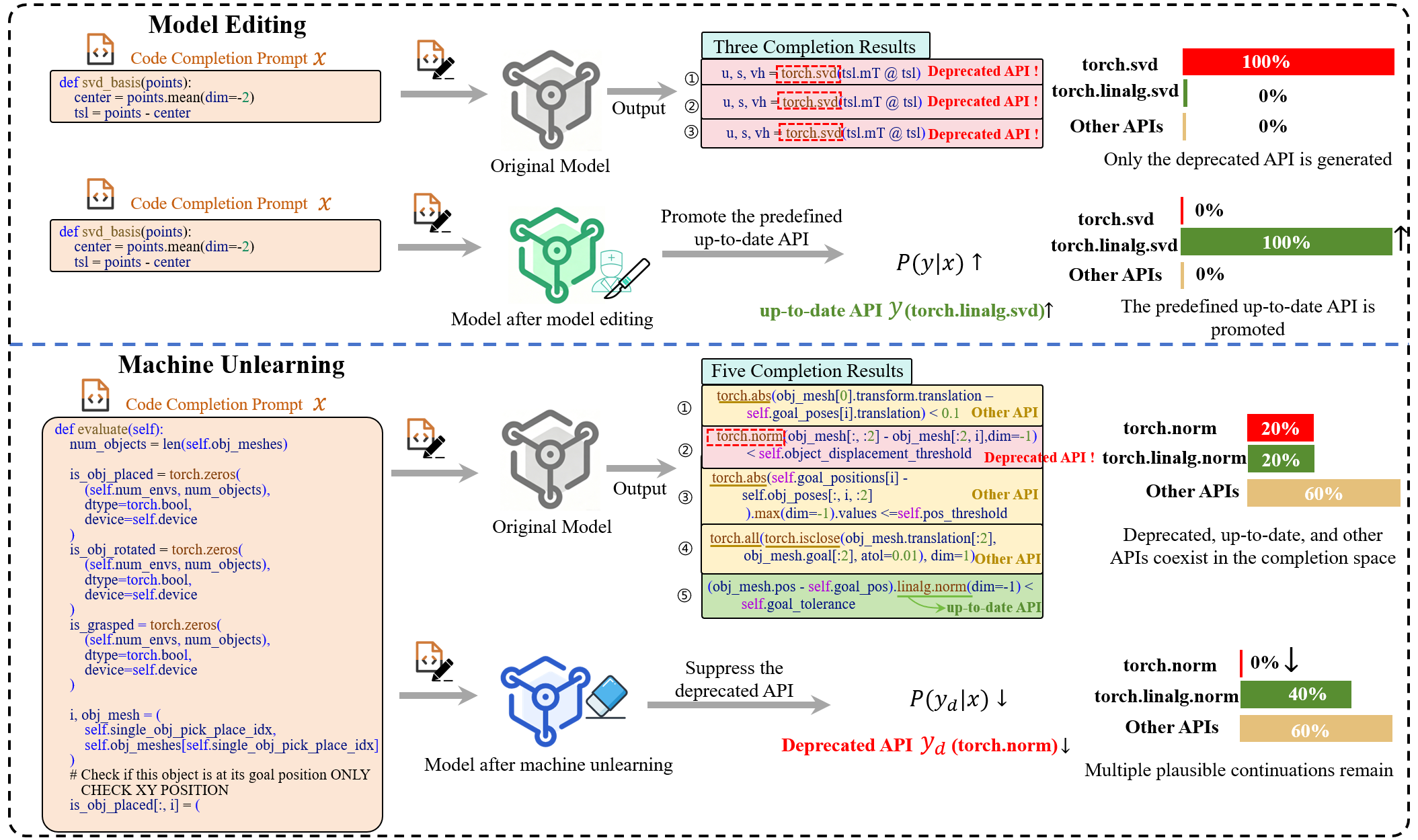}
    \caption{The motivation for machine unlearning of deprecated API knowledge in multi-solution code completion.}
    \label{fig:motivation}
\end{figure*}

However, deprecated API behavior in code completion does not necessarily follow such a one-to-one deprecated-to-up-to-date pattern. As illustrated in the lower part of Figure~\ref{fig:motivation}, among five completions generated for the same context, one contains the deprecated API \verb|torch.norm|, one contains its corresponding up-to-date API \verb|torch.linalg.norm|, and the remaining three involve APIs such as \verb|torch.abs|, \verb|torch.all|, and \verb|torch.isclose|. We refer to APIs other than the deprecated API and its corresponding up-to-date API as \textit{other APIs}. Here, ``other'' only distinguishes these APIs from the deprecated and up-to-date APIs and does not imply correctness or functional equivalence.

In this example, the five completions address the same high-level programming intent---determining whether an object is sufficiently close to its goal position---but implement different notions of positional closeness. While \verb|torch.norm| and \verb|torch.linalg.norm| use norm-based distance, the other completions employ alternative criteria, such as coordinate-wise absolute differences with \verb|torch.abs|, maximum coordinate deviation, or coordinate-wise closeness with \verb|torch.isclose| and \verb|torch.all|. These completions are therefore not necessarily functionally equivalent, but represent plausible ways of completing the high-level position-checking logic given the preceding code context. This example illustrates the \textit{implementation diversity} of code completion, where the same high-level programming intent can be realized through different APIs and code structures. Beyond this example, code completion may also exhibit \textit{intent diversity}, because a given code context may admit multiple plausible subsequent programming intents. Together, implementation and intent diversity make the completion space substantially broader than completions involving only the deprecated API or its corresponding up-to-date API.

This multi-solution characteristic makes a single-target replacement objective potentially restrictive. Let $x$ denote the code-completion context, $y_d$ a completion containing the deprecated API, and $y$ the corresponding completion using its up-to-date API. Directed model editing explicitly promotes $y$, conceptually increasing $P(y\mid x)$. While this encourages generation of the predefined up-to-date completion, it does not explicitly preserve other plausible continuations in the original completion space. This motivates a different model-level objective: rather than specifying what the model should generate after adaptation, we focus on removing the deprecated behavior itself. We therefore formulate deprecated API adaptation as a machine unlearning problem. Instead of specifying which completion the model should generate, the objective is to suppress the undesirable deprecated behavior by reducing $P(y_d\mid x)$, ideally toward zero, without prescribing a unique target for the remaining completion space. The model may still generate $y$ or other non-deprecated continuations. In other words, directed model editing promotes a predefined replacement, whereas machine unlearning focuses on removing what should no longer be generated. This formulation is better aligned with the multi-solution nature of code completion.

Recent studies have begun to apply machine unlearning to deprecated API adaptation. Jiang et al.~\cite{jiang2026large} introduced deprecated API unlearning as a code-unlearning task and proposed PROD, which suppresses deprecated code at the token level while redistributing probability mass over the remaining vocabulary. Tran et al.~\cite{tran2026towards} further combines SimNPO or PROD with a contrastive objective to suppress deprecated API completions while promoting their corresponding up-to-date replacements. Although these studies demonstrate the potential of machine unlearning for evolving API knowledge, several important issues remain underexplored. 
First, existing forget data are primarily constructed from predefined deprecated-API samples without requiring the corresponding deprecated-API behavior to be observable in every target model under the associated code context before unlearning. As a result, a candidate may be treated as a forget instance without first verifying that the targeted behavior is observable in the original model. Second, existing evaluations mainly assess capability preservation through general code-generation benchmarks such as HumanEval~\cite{chen2021evaluating}, providing limited insight into the localized impact of unlearning. In particular, they do not explicitly evaluate whether unlearning a deprecated API causes unintended changes to unrelated API-generation behavior, i.e., API behavior that does not involve either the deprecated API being forgotten or its corresponding up-to-date API. We refer to the preservation of such unrelated API behavior as \textit{Specificity}, which reflects whether the effect of unlearning remains localized to the targeted API knowledge.
Third, existing studies do not explicitly examine whether other API behaviors originally exhibited by the model remain preserved after unlearning. As discussed above, these other API behaviors may constitute plausible continuations in multi-solution code completion and should therefore not be unnecessarily disrupted when suppressing the deprecated API.

\subsection{Our Work and Contributions}

To address the above limitations, we conduct a systematic empirical study of machine unlearning for deprecated API knowledge in code LLMs. Our study is motivated by the multi-solution nature of code completion: successful deprecated API unlearning should not only suppress deprecated API generation, but should also avoid unnecessarily disrupting other desirable generation behaviors.

To support this study, we construct \texttt{MUDAPIBench}, a dedicated \textbf{M}achine \textbf{U}nlearning for \textbf{D}eprecated \textbf{API} \textbf{Bench}mark containing more than 7,000 model-specific unlearning instances. Starting from 145 verified deprecated-to-up-to-date API mappings across eight popular Python libraries~\cite{wang2024llms}, we collect real-world code-completion contexts from open-source repositories. Rather than directly treating the collected samples as forget instances, we perform model-specific filtering: for each candidate context, we sample five completions from the original model and retain the instance only when the corresponding deprecated API is actually generated. This ensures that every retained instance targets deprecated-API behavior observable in the model before unlearning. Based on these instances, we further construct Forget Data, Retain Data, Generalization Data, and Specificity Data for unlearning and evaluation.

Using \texttt{MUDAPIBench}, we systematically evaluate eight representative machine unlearning methods across three code LLMs: Gradient Ascent (GA), Gradient Difference (GD), Kullback--Leibler Divergence-based Unlearning (KL), Random Label Fine-tuning (RLFT), Direct Preference Optimization (DPO), Negative Preference Optimization (NPO), Simple Negative Preference Optimization (SimNPO), and Probabilistic Redistribution for Output Distribution (PROD), on Qwen2.5-Coder (3B), StarCoder2 (3B), and DeepSeek-Coder (1.3B). Our evaluation considers deprecated API forgetting, up-to-date API generation, preservation of other API behaviors within the same completion contexts, preservation of unrelated API behavior on separate inputs, and general code-generation capability. In particular, we introduce Other API Match Recall (OAMR) to explicitly measure the preservation of other API behaviors originally exhibited by the model. 
Our empirical results reveal a clear trade-off between deprecated API forgetting and capability preservation. GA, RLFT, and SimNPO strongly suppress deprecated API generation but cause greater disruption to desirable generation behaviors, whereas DPO and NPO provide stronger preservation at the cost of weaker forgetting. GD achieves the most favorable overall balance, combining strong deprecated API suppression with better preservation of up-to-date API generation, other API behaviors, unrelated API behavior, and general code-generation capability. GD also maintains moderate training time and relatively low memory overhead.

We further investigate factors associated with unlearning performance and the internal changes induced by different methods. Unlearning difficulty varies substantially across software libraries, with SciPy generally being more difficult to unlearn than TensorFlow. APIs deprecated after a model's training-data cutoff are also substantially more difficult to forget than those deprecated before the cutoff. Finally, our layer-wise analyses show that different forgetting--preservation trade-offs are associated with different degrees of internal perturbation: aggressive methods such as RLFT induce broader representation and parameter-importance changes, whereas GD combines strong forgetting with comparatively controlled internal changes.

Our main contributions are summarized as follows:
\begin{itemize}
    \item \textbf{Benchmark.}
    We construct \texttt{MUDAPIBench}, a behavior-grounded benchmark containing more than 7,000 model-specific unlearning instances derived from real API evolution across eight Python libraries. With a largely automated construction pipeline, \texttt{MUDAPIBench} provides a standardized and rigorous evaluation benchmark for deprecated API unlearning, supporting research in both software engineering and the broader machine unlearning community. 
    \item \textbf{Evaluation.}
    We conduct a comprehensive empirical evaluation of eight representative machine unlearning methods across three code LLMs, covering both deprecated API forgetting and preservation of desirable generation behaviors. 
    \item \textbf{Findings.}
    We identify GD as providing the most favorable overall forgetting--preservation trade-off and reveal substantial variations associated with software libraries, API deprecation timing, computational efficiency, and internal model changes. 
\end{itemize}

\subsection{Organization}

The remainder of this paper is organized as follows.
Section~\ref{sec: task definition} defines the deprecated API unlearning task.
Section~\ref{sec:study design} introduces the construction of \texttt{MUDAPIBench}, the evaluated machine unlearning methods, experimental settings, and evaluation metrics.
Section~\ref{sec:results} presents the empirical results for our research questions.
Section~\ref{sec:Discussion} provides a case study, discusses the implications of our findings, and presents threats to validity.
Section~\ref{sec:Related Work} reviews related work on machine unlearning and LLM adaptation to API evolution.
Finally, Section~\ref{sec:Conclusion} concludes the paper.

\section{Task Definition}
\label{sec: task definition}

Our study frames the code machine unlearning scenario as a code completion task, with the primary goal of removing deprecated API knowledge from LLMs while preserving their general code generation capabilities. In this context, we denote the pre-unlearning LLM as $f_0$ and the post-unlearning LLM as $f_u$. For each unlearning instance, we define a triplet $M=(x,y_d,y)$, where $x$ represents the unlearning input (a code snippet to be completed), $y_d$ denotes a deprecated API completion generated by the original LLM $f_0$ for $x$, and $y$ denotes the completion using its corresponding up-to-date replacement API. For a given input $x$, the original LLM $f_0$ can generate the deprecated API completion $y_d$, i.e., $f_0(x)=y_d$. The goal of code machine unlearning is to enable the unlearned LLM $f_u$ to forget the deprecated API knowledge and avoid generating $y_d$ for $x$, i.e., $f_u(x)\neq y_d$, while preserving its knowledge of unrelated APIs and general code generation capabilities. Notably, $f_u(x)$ is not required to equal $y$, as the same code completion context may admit other valid continuations. 
Following existing works on machine unlearning~\cite{maini2024tofu,tian2024forget}, we define an effective code machine unlearning process according to the following three key criteria.

\textbf{Effectiveness} ensures that the unlearned LLM $f_u$ successfully avoids generating the target deprecated API $y_d$ for the original unlearning input $x$:
\[
f_u(x)\neq y_d.
\]

\textbf{Generalization} requires that the forgetting effect extends beyond the original unlearning input $x$. Specifically, for an unseen input $x_g$ that is syntactically and semantically different from $x$, but for which the original LLM $f_0$ can generate the same deprecated API $y_d$, the unlearned LLM $f_u$ should also avoid generating $y_d$:
\[
f_0(x_g)=y_d
\Longrightarrow
f_u(x_g)\neq y_d.
\]

\textbf{Specificity} ensures that unlearning the target deprecated API does not adversely affect the LLM's knowledge of unrelated APIs and its general code generation capabilities. Let $\mathbb{U}=\{U_1,U_2,\ldots\}$ denote a set of unrelated code completion instances, where each $U=(x_s,y_s)$ consists of a non-target input $x_s$ and its corresponding completion output from the original LLM $f_0$, i.e., $y_s=f_0(x_s)$. Here, $x_s$ is unrelated to the target unlearning task and contains neither the target deprecated API $y_d$ nor its corresponding up-to-date API $y$. The unlearned LLM $f_u$ should preserve its original generation behavior on these unrelated inputs:
\[
f_u(x_s)=f_0(x_s), \quad \forall U\in\mathbb{U}.
\]

\section{Study Design}
\label{sec:study design}

\subsection{Machine Unlearning Subjects}
\label{sec:machine unlearning subjects}

We select three representative open-source code LLMs as the subjects for machine unlearning: Qwen2.5-Coder (3B), StarCoder2 (3B), and DeepSeek-Coder (1.3B). These models have been widely evaluated and adopted in software engineering research \cite{pan2024codev,sultana2024code,ji2025causality,sun2024ai}, while their different training corpora and data cutoffs provide diverse settings for studying deprecated API unlearning.

\textbf{Qwen2.5-Coder (3B)}~\cite{hui2024qwen2} is a code-oriented LLM developed by Alibaba. Its pre-training corpus contains approximately 5.5 trillion tokens, including source code, text–code grounding data, and synthetic data, with a training-data cutoff of February 2024.

\textbf{StarCoder2 (3B)}~\cite{lozhkov2024starcoder} is developed by BigCode and trained on The Stack v2. Its training corpus covers 17 programming languages and more than 3 trillion tokens, with data collected up to September 2023. The model is trained with a fill-in-the-middle objective to support code generation and completion.

\textbf{DeepSeek-Coder (1.3B)}~\cite{guo2024deepseek} is developed by DeepSeek and pre-trained from scratch on approximately 2 trillion tokens. Its training corpus consists of 87\% source code and 13\% natural-language data in English and Chinese, with a training-data cutoff of February 2023.

\subsection{MUDAPIBench Construction}

\input{table/benchmark_datanum}

We construct \texttt{MUDAPIBench}, which contains more than 7,000 deprecated-API unlearning instances. Following the problem definition in Section~\ref{sec: task definition}, each unlearning instance is represented as $M=(x,y_d,y)$, where $x$ denotes a code-completion prompt, $y_d$ denotes the target deprecated API that can be generated by the original LLM $f_0$ for $x$, and $y$ denotes its corresponding up-to-date API. Based on the filtered unlearning instances, we further construct Forget Data, Generalization Data, Retain Data, and Specificity Data for model unlearning and evaluation. Figure~\ref{fig:bench_construction} illustrates the overall construction process of \texttt{MUDAPIBench}.

\textbf{Step 1: Code Completion Prompt Collection.}
We first collect real-world code contexts associated with up-to-date API usages. We adopt the 145 verified deprecated-to-up-to-date API mappings from Wang et al.~\cite{wang2024llms}, covering eight popular Python libraries, as shown in Figure~\ref{fig:api_distribution}. For each up-to-date API $y$, we search open-source repositories using Sourcegraph~\cite{source-graph} and collect functions containing calls to the corresponding API. This process yields 65,596 candidate functions. For each function, the code preceding the target API invocation is extracted as the completion prompt $x$, while the observed API invocation identifies the corresponding up-to-date API $y$. Together with its mapped deprecated API $y_d$, each prompt forms a candidate unlearning instance $M=(x,y_d,y)$ for subsequent model-specific filtering.

\textbf{Step 2: Model-Specific Data Filtering.}
We next identify candidate instances for which the original LLM still exhibits the target deprecated-API behavior. For each candidate prompt $x$, we independently query each of the three code LLMs introduced in Section~\ref{sec:machine unlearning subjects} and sample five completions with the temperature set to $0.8$. A candidate instance is retained for a particular model if the corresponding deprecated API $y_d$ appears in at least one of the five completions. The retained instances constitute the model-specific unlearning instances in \texttt{MUDAPIBench}. This criterion ensures that the target deprecated-API behavior remains observable in the original model without requiring it to dominate all possible completions.
For example, as shown in Figure~\ref{fig:bench_construction}, the candidate prompt
\verb|def evaluate(self): ... is_obj_placed[:, i] = (| 
is retained because the original LLM generates a completion containing the deprecated API \verb|torch.norm()| in at least one of the five sampled completions.

\textbf{Step 3: Data Preparation.}
Based on the filtered unlearning instances, we prepare four types of data: Forget Data, Retain Data, Generalization Data, and Specificity Data. Forget Data and Retain Data are used during machine unlearning, whereas Forget Data, Generalization Data, and Specificity Data are used to evaluate Forgetting Effectiveness, Generalization, and Specificity, respectively.

\textbf{(1) Forget Data and Generalization Data.}
Forget Data and Generalization Data are derived from the same pool of filtered unlearning instances. For each deprecated-to-up-to-date API mapping $(y_d,y)$, we randomly split the corresponding instances into two disjoint subsets of equal size. One subset is used as Forget Data, while the other is held out as Generalization Data and is not involved in unlearning. 
For an unlearning instance $M=(x,y_d,y)$ assigned to Forget Data, the objective is to reduce the tendency of the unlearned model $f_u$ to generate a completion containing the deprecated API $y_d$ for context $x$. After unlearning, the same Forget Data are used to evaluate whether the target deprecated-API behavior has been effectively suppressed. 
Generalization Data contain the same deprecated-to-up-to-date API mappings as the corresponding Forget Data but use different code-completion contexts. Because these contexts are held out from unlearning, they allow us to assess whether the forgetting effect generalizes beyond the specific contexts seen during training. Specifically, after unlearning on the Forget Data, we examine whether $f_u$ also suppresses the same deprecated API $y_d$ on the held-out Generalization Data. This setting allows us to distinguish generalization of the forgetting effect across contexts from context-specific suppression. 
For example, as shown in Figure~\ref{fig:bench_construction}, for the mapping
\verb|torch.norm()| $\rightarrow$ \verb|torch.linalg.norm()|,
the instance
\verb|def evaluate(self): ...|
may be assigned to Forget Data, while another filtered instance,
\verb|def normalize_vectors(embeddings, center=False): ...|,
is held out as Generalization Data.

\textbf{(2) Retain Data.}
Retain Data provide a knowledge-preservation constraint during unlearning and are constructed only for instances included in Forget Data. Their purpose is to constrain unintended changes to non-target generation behavior during unlearning. 
For each Forget Data instance $M=(x,y_d,y)$, we reuse the five completions generated by the original model $f_0$ during the model-specific filtering in Step~2. We identify all non-deprecated APIs appearing in these completions and count their occurrence frequencies across the five generations. The completion $y_r$ containing the most frequently occurring non-deprecated API is selected as the corresponding Retain Data. In this way, Retain Data capture non-deprecated generation behavior frequently exhibited by the original model and provide a preservation constraint against unintended behavioral changes during unlearning 
For example, for the Forget Data input
\verb|def evaluate(self): ... is_obj_placed[:, i] = (|,
the completion
\verb|torch.abs(obj_mesh[0].transform.|
\verb|translation - self.goal_poses[i].translation) < 0.1|
is selected as $y_r$, because four of the five completions generated in Step~2 contain non-deprecated APIs, among which \verb|torch.abs| occurs most frequently.

\textbf{(3) Specificity Data}  are used to evaluate whether unlearning the target deprecated API causes unintended changes to unrelated code-generation behavior. Following prior studies~\cite{wang2024llms,guancheng2026don}, we consider a challenging setting in which the non-target inputs are semantically similar to the Forget Data inputs but do not involve the corresponding target API pair. 
We first randomly sample 100,000 Python files from The Stack v2~\cite{lozhkov2024starcoder} to form a candidate pool of non-target code. We then use CodeBERT~\cite{feng2020codebert} to obtain vector representations of the Forget Data inputs and candidate non-target inputs. For each Forget Data input $x$, we retrieve the five candidate inputs
$x_s^1,x_s^2,x_s^3,x_s^4,x_s^5$
with the smallest embedding distances, while ensuring that they contain neither the target deprecated API $y_d$ nor its corresponding up-to-date API $y$. 
Before unlearning, each selected non-target input is completed by the original model $f_0$, and the resulting completion is recorded as its pre-unlearning reference output. These non-target inputs and their reference completions jointly constitute the Specificity Data. After unlearning, we compare the behavior of $f_u$ on these inputs with that of $f_0$ to assess whether non-target API knowledge and general code-generation capabilities are preserved. For example, as shown in Figure \ref{fig:bench_construction}, the non-target API for one piece of Specificity Data is \verb|torch.matmul()|.

\textbf{Dataset Statistics.}  
Table~\ref{tab:bench_statistics} summarizes the numbers of unlearning instances and API mappings in \texttt{MUDAPIBench}. After model-specific filtering, 81 unique deprecated-to-up-to-date API pairs remain from the original 145 mappings provided by Wang et al.~\cite{wang2024llms}. The Generalization Data contain slightly fewer instances than the Forget Data because some API mappings contain only one filtered unlearning instance and therefore cannot provide a distinct held-out context for generalization evaluation. The Specificity Data are five times larger than the Forget Data because five non-target inputs are retrieved for each Forget Data instance. Figure~\ref{fig:api_distribution} illustrates the distribution of unlearning instances across the eight Python libraries. 

\subsection{Machine Unlearning Methods}

\begin{table}[!thbp]
\centering
\caption{The brief descriptions of the eight machine unlearning methods.}

\label{tab:methods_description}
\small
\begin{tabularx}{\textwidth}{c|X}
\hline
\multicolumn{2}{c}{\cellcolor{lightgray}\textbf{Gradient Optimization}} \\ \hline
\textbf{GA}~\cite{golatkar2020eternal} & It performs unlearning through gradient ascent on the loss of deprecated API code, driving the model away from its original fitting direction. By directly weakening the model's ability to fit deprecated API knowledge, GA reduces the probability of generating deprecated API completions and thereby suppresses the corresponding knowledge. As a direct gradient-ascent strategy, GA focuses on deprecated API knowledge without explicitly imposing constraints on non-target knowledge. \\
\hline
\textbf{GD}~\cite{liu2022continual} & It extends GA by introducing a gradient descent objective on retain data alongside the gradient ascent objective on deprecated API knowledge. While gradient ascent weakens the model's fitting capability for deprecated API knowledge, gradient descent on retain data reinforces its ability to fit non-target code. By jointly optimizing these two opposing gradient directions, GD balances effective deprecated API forgetting with the preservation of general code generation capability. \\
\hline
\multicolumn{2}{c}{\cellcolor{lightgray}\textbf{Distribution Regularization}} \\ \hline
\textbf{KL} \cite{kullback1951information} & It extends GA by introducing a Kullback--Leibler divergence constraint on retain data alongside the forgetting objective. Specifically, it encourages the unlearned model to maintain an output distribution close to that of the original model on non-target code contexts. This constraint helps preserve general and non-target API knowledge while reducing deprecated API knowledge. \\
\hline
\multicolumn{2}{c}{\cellcolor{lightgray}\textbf{Random Supervision}} \\ \hline
\textbf{RLFT }\cite{golatkar2020eternal} & It disrupts the original correspondence between code completion prompts and deprecated API code by randomly replacing the deprecated API code with unrelated labels. Fine-tuning on these randomly supervised examples prevents the model from maintaining the association between the code completion prompts and deprecated API completions. Through randomized supervision, RLFT weakens the learned association between code completion prompts and deprecated API knowledge. \\
\hline
\multicolumn{2}{c}{\cellcolor{lightgray}\textbf{Preference Optimization}} \\ \hline
\textbf{DPO}~\cite{maini2024tofu} & It formulates machine unlearning as a preference optimization problem by constructing preferred and non-preferred responses for each unlearning instance. Following prior work \cite{maini2024tofu,jiang2026large}, ``I don't know'' is treated as the preferred response, while the completion containing the deprecated API is treated as the non-preferred response. DPO then optimizes their relative preference with respect to the original model, encouraging the unlearned model to assign lower preference to deprecated API completions. \\
\hline
\textbf{NPO}~\cite{zhang2024negative} & It builds on DPO by reformulating machine unlearning as a negative preference optimization problem, directly suppressing the model's preference for deprecated API completions. Using the original model as a reference, NPO compares the relative likelihood of deprecated API completions under the unlearned and original models and optimizes the model to decrease this relative preference. This enables targeted suppression of deprecated API knowledge without requiring an explicitly constructed preferred response. \\
\hline
\textbf{SimNPO}~\cite{fan2026simplicity} & It simplifies NPO by removing the need for a reference model. Instead of optimizing the relative likelihood between the unlearned and original models, SimNPO constructs the forgetting objective solely from the normalized likelihood of deprecated API completions under the unlearned model. A forgetting margin is introduced to control the optimization boundary, simplifying the unlearning procedure and reducing the computational overhead associated with maintaining a reference model. \\
\hline
\multicolumn{2}{c}{\cellcolor{lightgray}\textbf{Target Distribution}} \\ \hline
\textbf{PROD}~\cite{jiang2026large} & It performs unlearning by constructing a modified target distribution from the original model’s output distribution. Specifically, PROD identifies and removes the highest-probability token corresponding to the deprecated API code completion and redistributes its probability mass among the remaining candidate tokens while preserving their relative probabilities. The resulting distribution serves as a soft supervision signal, guiding the model away from deprecated API code while retaining its original generation behavior on other code as much as possible.  \\
\hline
\end{tabularx}
\end{table}

We select eight representative machine unlearning methods that have been widely adopted as baselines in prior studies~\cite{maini2024tofu,yao2024machine,tian2024forget,xu2025unlearning}. Following the taxonomy of Le et al.~\cite{le2025survey}, we organize these methods into five categories according to their underlying optimization mechanisms: \textit{Gradient Optimization}, \textit{Distribution Regularization}, \textit{Random Supervision}, \textit{Preference Optimization}, and \textit{Target Distribution}. 
\textbf{Gradient Optimization} includes Gradient Ascent (GA)~\cite{golatkar2020eternal} and Gradient Difference (GD)~\cite{liu2022continual}. GA directly increases the loss on Forget Data to reduce the likelihood of deprecated-API completions, while GD additionally incorporates Retain Data to balance forgetting with the preservation of non-target knowledge. 
\textbf{Distribution Regularization}, represented by Kullback--Leibler Divergence-based Unlearning (KL)~\cite{kullback1951information}, constrains the unlearned model to remain close to the original model on Retain Data, thereby reducing unintended changes to non-target behavior. 
\textbf{Random Supervision}, represented by Random Label Fine-tuning (RLFT)~\cite{golatkar2020eternal}, replaces the original forgetting targets with random supervision, disrupting the association between forgetting inputs and deprecated-API completions. 
\textbf{Preference Optimization}, including Direct Preference Optimization (DPO)~\cite{maini2024tofu}, Negative Preference Optimization (NPO)~\cite{zhang2024negative}, and Simple Negative Preference Optimization (SimNPO)~\cite{fan2026simplicity}, reformulates forgetting as a preference optimization problem that reduces the model's preference for undesirable deprecated-API completions; SimNPO further removes the dependence on a reference model. 
Finally, \textbf{Target Distribution}, represented by Probabilistic Redistribution for Output Distribution (PROD)~\cite{jiang2026large}, constructs a modified target distribution by reducing the probability mass assigned to target forget tokens and redistributing it among alternative tokens, thereby steering the model away from deprecated-API completions while retaining non-target generation behavior. 
Together, these eight methods cover substantially different unlearning mechanisms and enable us to systematically compare their effectiveness for deprecated API unlearning. Brief descriptions of the individual methods are provided in Table~\ref{tab:methods_description}.

\subsection{Experiment Settings}

All experiments are conducted on an NVIDIA RTX A6000 GPU. Following existing code completion studies~\cite{yu2024fight,zhang2023repocoder}, we use incomplete code snippets directly as prompts for code completion without adding any additional instructions. We adopt the hyperparameter settings for the machine unlearning methods based on prior studies~\cite{jiang2026large,maini2024tofu,dorna2026openunlearning,yao2024machine} that reported strong performance in their respective experiments. Specifically, we set $\beta$ to 1 for both DPO and NPO, while setting $\beta$ to 1 and $\delta$ to 0 for SimNPO. To reduce the computational and memory overhead of machine unlearning, we adopt Low-Rank Adaptation (LoRA) \cite{hu2021lora} based parameter-efficient fine-tuning for all machine unlearning methods. Specifically, we set the LoRA rank $r$ to 16, the scaling factor $\alpha$ to 32, and the dropout rate to 0.05, with the bias term disabled. LoRA is applied to all attention and feed-forward projection layers, including \verb|q_proj|, \verb|k_proj|, \verb|v_proj|, \verb|o_proj|, \verb|gate_proj|, \verb|up_proj|, and \verb|down_proj|. We use the same LoRA configuration across all experiments and fix the number of training epochs at 5.

\subsection{Evaluation Metrics}

We evaluate deprecated API unlearning from five complementary aspects:
\textit{deprecated API forgetting},
\textit{up-to-date API generation},
\textit{other API behavior preservation},
\textit{unrelated API behavior preservation}, and
\textit{general code-generation preservation}.

For Forget Data and Generalization Data, we generate five completions per input using sampling-based decoding with temperature $=0.8$.
Let $\hat{y}_u$ denote a completion generated by the unlearned model $f_u$, $y_d$ the deprecated reference completion, and $y$ the corresponding up-to-date reference completion.
We use API Exact Match (AEM), Edit Similarity (ES), BLEU, and ROUGE-L to evaluate deprecated API forgetting and up-to-date API generation, where suffixes ``-D'' and ``-U'' denote evaluation against the deprecated and up-to-date references, respectively.
We further introduce Other API Match Recall (OAMR) to evaluate whether other API behaviors originally exhibited by the model within the same completion contexts are preserved after unlearning.
Specificity evaluates the preservation of unrelated API behavior on separate non-target inputs, while HumanEval Pass@1 assesses the preservation of general functional code-generation capability.

\textbf{(1) API Exact Match (AEM).}
AEM measures whether a generated completion contains a specified API.
\textbf{AEM-D} is the proportion of completions containing the deprecated API, with lower values indicating stronger forgetting.
\textbf{AEM-U} is the proportion of completions containing the corresponding up-to-date API, with higher values indicating more frequent generation of the recommended replacement.
AEM-U is treated as a desirable outcome rather than a necessary condition for successful unlearning, because completions involving other APIs may also constitute plausible continuations.

\textbf{(2) Edit Similarity (ES).}
ES measures normalized similarity between a generated completion and a reference completion using Levenshtein edit distance:

\begin{equation}
\mathrm{ES}(\hat{y}_u,y_r)
=
1-
\frac{\mathrm{Edit}(\hat{y}_u,y_r)}
{\max(|\hat{y}_u|,|y_r|)},
\end{equation}

where $y_r$ denotes the reference completion and $\mathrm{Edit}(\hat{y}_u,y_r)$ is the Levenshtein edit distance between $\hat{y}_u$ and $y_r$.
For Forget Data and Generalization Data, we compute \textbf{ES-D} using $y_r=y_d$ and \textbf{ES-U} using $y_r=y$.
Lower ES-D indicates lower overall similarity to the deprecated reference, whereas higher ES-U indicates greater similarity to the up-to-date reference.
ES complements AEM by characterizing completion-level changes rather than directly measuring whether the specified API is generated.

\textbf{(3) BLEU and ROUGE-L.}
BLEU~\cite{papineni2002bleu} measures $n$-gram overlap between generated and reference completions, while ROUGE-L~\cite{lin2004rouge} measures sequence-level similarity based on the longest common subsequence.
Similar to ES, both metrics are computed separately against the deprecated and up-to-date references, yielding \textbf{BLEU-D/U} and \textbf{ROUGE-L-D/U}.
Lower deprecated-reference similarity and higher up-to-date-reference similarity indicate a more desirable adaptation outcome.
These metrics complement AEM by capturing lexical and sequence-level similarity between generated and reference completions.

\textbf{(4) Other API Match Recall (OAMR).}
OAMR measures whether \textit{other API behaviors} originally exhibited by the model within the same completion context remain observable after unlearning.
For the same input $x$, let
\[
Y_0=\{\hat{y}_0^{1},\ldots,\hat{y}_0^{5}\}, \qquad
Y_u=\{\hat{y}_u^{1},\ldots,\hat{y}_u^{5}\}
\]
denote the five completions generated by the original model $f_0$ and the unlearned model $f_u$, respectively.
After excluding the deprecated API and its corresponding up-to-date API, we extract the sets of other APIs from $Y_0$ and $Y_u$, denoted by $O(Y_0)$ and $O(Y_u)$.
OAMR is defined as

\begin{equation}
\mathrm{OAMR}(Y_0,Y_u)
=
\frac{|O(Y_0)\cap O(Y_u)|}
{|O(Y_0)|}.
\end{equation}

A higher OAMR indicates that a larger proportion of the other API behaviors originally exhibited by $f_0$ remain observable after unlearning.
Because OAMR is recall-oriented, newly generated other APIs in $O(Y_u)$ do not reduce the score.

\textbf{(5) Specificity.}
Specificity evaluates whether unlearning the deprecated API causes unintended changes to unrelated API behavior on separate non-target inputs.
For each Specificity Data input $x_s$, which involves neither the deprecated API nor its corresponding up-to-date API, let $\hat{y}_0$ and $\hat{y}_u$ denote the deterministic completions generated by $f_0$ and $f_u$, respectively.
We compute ES, BLEU, and ROUGE-L between $\hat{y}_u$ and $\hat{y}_0$, with higher scores indicating stronger preservation of the original model's generation behavior on these inputs.

To make the Specificity evaluation more challenging, the Specificity Data are constructed by selecting non-target inputs that are semantically similar to the corresponding Forget Data inputs.
Unlike OAMR, which evaluates the preservation of other API behaviors within the same target completion contexts, Specificity evaluates the preservation of unrelated API behavior on separate non-target inputs.

\textbf{(6) HumanEval Pass@1.}
We use HumanEval Pass@1~\cite{chen2021evaluating} to assess whether deprecated API unlearning affects the model's general functional code-generation capability.
Higher Pass@1 indicates better preservation of general code-generation capability after unlearning.

For Forget Data and Generalization Data, AEM, ES, BLEU, and ROUGE-L are averaged over the five sampled completions for each input and then across all instances.
OAMR is computed once per input from the two sets of five completions generated by $f_0$ and $f_u$ and is then averaged across instances.
For Specificity Data, ES, BLEU, and ROUGE-L are computed between the deterministic completions generated by $f_u$ and the corresponding pre-unlearning completions generated by $f_0$, and are then averaged across all instances.

\section{Results}
\label{sec:results}

\subsection{RQ1: How do existing machine unlearning methods perform in deprecated API unlearning while preserving desirable code-generation behaviors?} 
\label{sec:rq1}

\input{table/RQ1-forget}

\input{table/RQ1-gen}

\input{table/RQ1-S}

Tables~\ref{tab:unlearning-effectiveness-all}, \ref{tab:unlearning-generalization-all}, and \ref{tab:unlearning-specificity-all} present the performance of different machine unlearning methods on Effectiveness, Generalization, and Specificity, respectively. Overall, the results reveal a clear trade-off between suppressing deprecated API generation (AEM-D) and maintaining desirable generation behavior, including up-to-date API generation (AEM-U), other API behaviors originally exhibited by the model (OAMR), generation behavior on unrelated inputs (Specificity), and general code-generation capability (HumanEval Pass@1).

(1) GA, RLFT, and SimNPO are effective in reducing deprecated API generation on both the Effectiveness and Generalization data. Across the three models, their average AEM-D scores are all below 5.92, indicating that these methods substantially suppress the models' tendency to generate deprecated APIs. However, this strong suppression is accompanied by clear side effects. Their AEM-U and OAMR scores are also consistently low, indicating less frequent generation of up-to-date APIs and weaker preservation of other API behaviors originally exhibited by the model. The completion-level similarity metrics exhibit a consistent pattern: ES-D, BLEU-D, and ROUGE-L-D decrease substantially, confirming that the generated completions move away from the deprecated references, while ES-U, BLEU-U, and ROUGE-L-U also tend to decrease, indicating that the resulting completions do not necessarily move toward the corresponding up-to-date references. Together, these results suggest that these methods may alter API-generation behavior more broadly rather than selectively suppressing deprecated APIs. Consistently, their weaker performance on the Specificity data and HumanEval further indicates greater disruption to generation behavior on unrelated inputs and general code-generation capability.

(2) KL, DPO, NPO, and PROD exhibit a different trade-off. Compared with GA, RLFT, and SimNPO, these methods generally obtain higher AEM-D scores on the Effectiveness and Generalization data, indicating weaker suppression of deprecated APIs. Meanwhile, their higher AEM-U and OAMR scores indicate more frequent generation of up-to-date APIs and better preservation of other API behaviors originally exhibited by the model. Their ES, BLEU, and ROUGE-L results show a similar tendency: deprecated-reference similarity changes more moderately, while up-to-date-reference similarity is generally better preserved than under the more aggressive methods above. These methods therefore induce less aggressive behavioral changes, but at the cost of weaker forgetting. On the Specificity data and HumanEval, DPO and NPO also show relatively strong preservation of generation behavior on unrelated inputs and general code-generation capability, whereas KL and PROD cause more noticeable changes to the model's original behavior.

(3) GD achieves the most favorable overall trade-off. On the Effectiveness and Generalization data, GD obtains a low average AEM-D of 3.74 while maintaining relatively high average AEM-U and OAMR scores of 23.38 and 34.77, respectively. These results indicate that GD effectively suppresses deprecated API generation while maintaining relatively strong up-to-date API generation and preserving other API behaviors originally exhibited by the model. The completion-level similarity metrics further support this observation: GD substantially reduces similarity to deprecated references while retaining comparatively high similarity to up-to-date references. GD also performs strongly on the Specificity data and HumanEval, indicating fewer unintended changes to generation behavior on unrelated inputs and general code-generation capability. Overall, GD provides the most favorable balance between effective forgetting and capability preservation among the evaluated machine unlearning methods.

The observed differences are consistent with the optimization mechanisms of the evaluated methods. GA, RLFT, and SimNPO place relatively direct optimization pressure on reducing the likelihood of deprecated completions. Such strong forgetting signals can rapidly suppress deprecated API generation, but may also induce broader changes to the model's output distribution, resulting in reduced generation of up-to-date APIs, loss of other API behaviors, and degradation of general code-generation capability. In contrast, KL, DPO, NPO, and PROD incorporate distributional constraints, preference-based objectives, or probability redistribution mechanisms that moderate changes to the original model behavior. These mechanisms generally improve capability preservation, but also constrain the strength of forgetting. GD differs in that it explicitly optimizes both sides of this trade-off: gradient ascent suppresses deprecated completions on Forget Data, while gradient descent reinforces the behaviors represented in Retain Data. Because our Retain Data are selected from non-deprecated completions originally generated by the model and may contain either the corresponding up-to-date API or other APIs, this joint optimization enables GD to suppress deprecated API behavior while simultaneously preserving up-to-date API generation and other API behaviors. This is consistent with GD's combination of low AEM-D and relatively high AEM-U, OAMR, Specificity, and general code-generation performance.

As the best-performing model editing method for deprecated API adaptation among existing approaches, \textsc{AdaLoRA-L} is included as an additional baseline alongside the eight machine unlearning methods to further investigate the performance differences between model editing and machine unlearning. It exhibits substantially different behavior from the machine unlearning methods. \textsc{AdaLoRA-L} identifies API-specific layers based on gradient-derived layer importance and restricts model editing to these layers to update the deprecated API toward its predefined up-to-date replacement. On the Effectiveness data, it achieves an average AEM-D of only 0.56 across the three models while increasing the average AEM-U to 68.04, demonstrating strong suppression of deprecated APIs together with a pronounced shift toward the predefined up-to-date replacements. However, its average OAMR is only 4.47, substantially lower than that of the machine unlearning methods, indicating that other API behaviors originally exhibited by the model are rarely preserved after editing. This result is consistent with the directed replacement objective of \textsc{AdaLoRA-L}: explicitly optimizing the model toward a predefined up-to-date completion substantially increases the generation of the designated replacement, but may simultaneously narrow the original completion space. 
Moreover, \textsc{AdaLoRA-L} performs poorly on Specificity, with an average score below 60 across the three models, indicating substantial unintended changes to unrelated API behavior. This degradation extends to general code generation: its HumanEval Pass@1 decreases by an average of 89.57\% relative to the original models. Although \textsc{AdaLoRA-L} restricts parameter updates to API-specific layers to improve editing locality, our results suggest that such layer-level restriction does not sufficiently isolate the behavioral effects of large-scale deprecated-to-up-to-date API editing. Repeatedly directing the model toward predefined replacements can substantially alter generation behavior beyond the edited API mappings, which is consistent with the pronounced degradation observed on both Specificity and HumanEval. 
Therefore, despite its strong deprecated-to-up-to-date replacement performance, \textsc{AdaLoRA-L} exhibits substantial collateral changes to other API behaviors, unrelated API behavior, and general code-generation capability, making it less suitable for deprecated API adaptation in multi-solution code completion.

\begin{mybox}{Answer to RQ1}
Existing machine unlearning methods exhibit clear trade-offs between deprecated API forgetting and capability preservation. Among them, GD achieves the most favorable overall balance by effectively suppressing deprecated API generation while better maintaining up-to-date API generation, other API behaviors originally exhibited by the model, generation behavior on unrelated inputs, and general code-generation capability. In contrast, the model-editing baseline \textsc{AdaLoRA-L} strongly promotes predefined up-to-date replacements but substantially reduces the preservation of other API behaviors.
\end{mybox}

\subsection{RQ2: Does the performance of machine unlearning methods vary across different libraries? }

\input{table/RQ2}

This section investigates whether the performance of machine unlearning methods varies across different software libraries. To ensure a representative evaluation, we focus on four libraries with relatively large numbers of unlearning instances, namely PyTorch, SciPy, scikit-learn, and TensorFlow, and evaluate the eight methods on three LLMs. Tables~\ref{tab:unlearning-forget-library} --~\ref{tab:unlearning-specificity-aem} further compare the performance of different machine unlearning methods across PyTorch, SciPy, scikit-learn, and TensorFlow. Overall, the library-level results confirm the trade-off observed in RQ1 and show that GD provides the most favorable overall balance across different libraries and models.

First, aggressive forgetting methods such as RLFT, GA, and SimNPO generally achieve low AEM-D across the four libraries, indicating strong generalization of deprecated API suppression. RLFT is particularly aggressive, reducing AEM-D to near zero in most settings. However, this strong forgetting is accompanied by substantial degradation in AEM-U and OAMR. RLFT frequently reduces AEM-U to nearly zero and achieves very low OAMR across libraries, indicating that up-to-date API generation and other API behaviors originally exhibited by the model are also substantially disrupted. Its relatively low Specificity scores further show that such aggressive suppression extends to unrelated API behavior.

In contrast, DPO and NPO generally maintain higher up-to-date API generation and better preserve other and unrelated API behaviors, but their AEM-D values are also substantially higher in many settings, indicating weaker forgetting. This pattern is particularly evident for PyTorch, SciPy, and scikit-learn, where DPO and NPO frequently achieve relatively high AEM-U and OAMR while retaining considerably more deprecated API generation than the more aggressive methods. Their high Specificity scores further demonstrate strong preservation of unrelated API behavior, particularly for NPO on StarCoder2, where the scores exceed 98 across all four libraries. Such preservation, however, comes at the cost of weaker deprecated API suppression.

GD provides a more balanced behavior across these dimensions. Across most library--model combinations, GD substantially reduces AEM-D while maintaining comparatively high AEM-U and OAMR. For example, on Qwen2.5-Coder, GD increases AEM-U over the original model for all four libraries and achieves OAMR values ranging from 33.01 to 47.15. On StarCoder2 and DeepSeek-Coder, although GD does not achieve the highest AEM-U or OAMR in every library, it consistently combines relatively strong deprecated API suppression with moderate-to-high up-to-date API generation and preservation of other API behaviors. Its Specificity performance is also substantially stronger than that of aggressive methods such as GA, RLFT, and SimNPO in most settings; notably, on Qwen2.5-Coder, GD achieves Specificity scores above 94 across all four libraries. Therefore, while DPO and NPO provide stronger preservation in some settings and RLFT achieves stronger forgetting, GD maintains the most favorable overall trade-off across libraries.

The results also reveal substantial differences in unlearning difficulty across libraries. In particular, SciPy generally exhibits higher AEM-D and lower AEM-U after unlearning, indicating greater difficulty in both suppressing deprecated APIs and generating their corresponding up-to-date APIs. This pattern is especially evident for GD on Generalization data: its AEM-D on SciPy reaches 21.16, 13.97, and 16.19 for Qwen2.5-Coder, StarCoder2, and DeepSeek-Coder, respectively, compared with only 2.40, 1.82, and 0.54 on TensorFlow. Conversely, TensorFlow generally exhibits substantially lower residual deprecated API generation, suggesting that the effectiveness of deprecated API unlearning is strongly affected by the target library.

\begin{mybox}{Answer to RQ2} 
Deprecated API unlearning performance varies substantially across libraries, with SciPy generally being more difficult to unlearn and TensorFlow exhibiting stronger forgetting effectiveness. Despite these library-specific differences, GD consistently provides the most favorable overall trade-off between deprecated API forgetting and capability preservation across the evaluated libraries and models.
\end{mybox}

\subsection{RQ3: How computationally efficient are different machine unlearning methods for deprecated API unlearning?}

\input{Figure/RQ3/RQ3}

Figures~\ref{fig:time_cost} and~\ref{fig:memory_cost} present the efficiency of different machine unlearning methods in the deprecated API forgetting task, measured by average time cost and peak memory cost, respectively. Overall, computational efficiency varies substantially across methods. As shown in Figure~\ref{fig:time_cost}, RLFT incurs the lowest time cost, requiring only 3.97 hours on average across the three models. GA and SimNPO are also relatively efficient, with average time costs of 3.98 and 4.54 hours, respectively, while GD requires 8.26 hours. In contrast, DPO is the most time-consuming method, with an average time cost of 11.66 hours. As shown in Figure~\ref{fig:memory_cost}, PROD incurs the highest average peak memory cost of 28.65 GB, followed by DPO at 25.93 GB. In contrast, GA, GD, RLFT, and SimNPO are more memory-efficient, with average peak memory costs below 14 GB.

The differences in time and memory costs are closely related to the optimization procedures of the evaluated methods. GA and RLFT employ relatively direct optimization on a single model, resulting in low computational and memory overhead. GD additionally incorporates Retain Data during optimization, which increases its training time compared with GA and RLFT, while its single-model optimization keeps memory consumption relatively moderate. SimNPO is more efficient than NPO because it removes the dependence on a reference model, thereby reducing additional computation and memory usage. In contrast, DPO requires reference-model computation during preference optimization, resulting in relatively high time and memory costs. PROD operates directly on token-level output distributions and involves additional distribution-level computation, which is consistent with its particularly high peak memory consumption.

Despite these differences, the overall computational cost remains manageable under our experimental setting. For the evaluated code LLMs with LoRA-based unlearning, all methods can be executed on a single NVIDIA RTX A6000 GPU, with the highest observed peak memory usage being 47.09 GB. Notably, GD also demonstrates favorable computational efficiency. Although its average time cost of 8.26 hours is higher than those of GA, RLFT, and SimNPO, its peak memory consumption remains relatively low, below 14 GB on average, and substantially lower than those of more resource-intensive methods such as DPO and PROD. Therefore, in addition to the favorable effectiveness--preservation trade-off observed in RQ1, GD also maintains moderate time cost and relatively low memory overhead.

\begin{mybox}{Answer to RQ3}
    Machine unlearning methods exhibit substantial differences in computational efficiency. RLFT and GA incur the lowest time costs, while GA, GD, RLFT, and SimNPO exhibit relatively low memory consumption. Notably, GD, which achieves the most favorable effectiveness--preservation trade-off in RQ1, also maintains moderate time cost and relatively low memory overhead, demonstrating favorable efficiency among the evaluated methods. 
\end{mybox}

\subsection{RQ4: How does API deprecation timing relative to the model's training-data cutoff affect deprecated API unlearning performance?}

\input{Figure/RQ4/RQ4}

To answer RQ4, we investigate whether the difficulty of deprecated API unlearning depends on when an API was deprecated relative to the model's training-data cutoff. For each deprecated API in \texttt{MUDAPIBench}, we collect its official deprecation date and compare it with the training-data cutoff of each model. We divide the APIs into two groups: \textit{Pre-cutoff APIs}, which had already been deprecated before the model's training-data cutoff, and \textit{Post-cutoff APIs}, which were deprecated only after the cutoff. Importantly, APIs in the latter group were still considered valid according to the library versions available at the model's cutoff, making them particularly interesting targets for post-training knowledge removal. 
All deprecated APIs identified for Qwen2.5-Coder-3B belong to the Pre-cutoff group, so we exclude this model from RQ4 because no within-model comparison between the two groups is possible. For StarCoder2-3B, the Post-cutoff group contains 3 APIs with 115 forgetting instances, while for DeepSeek-Coder-1.3B it contains 7 APIs with 246 forgetting instances. Based on the results of RQ1, we use GD as the representative unlearning method. Figure~\ref{fig:aem_Before_After} compares the performance of GD on Pre-cutoff and Post-cutoff APIs. 

The results show that Post-cutoff APIs are substantially more difficult to forget. On StarCoder2, the average AEM-D across Forget Data and Generalization Data increases from 3.47 for Pre-cutoff APIs to 16.01 for Post-cutoff APIs, corresponding to a relative increase of approximately 362.2\%. A similar pattern is observed on DeepSeek-Coder, where the average AEM-D increases from 1.67 to 12.22, corresponding to an increase of approximately 633.6\%. These results indicate substantially more residual deprecated API generation for APIs that were deprecated only after the models' training-data cutoffs. 
Post-cutoff APIs also show a weaker tendency to generate their corresponding up-to-date replacements. On StarCoder2, the average AEM-U across Forget Data and Generalization Data decreases sharply from 15.42 for Pre-cutoff APIs to 2.26 for Post-cutoff APIs, a reduction of approximately 85.4\%. On DeepSeek-Coder, the decrease is considerably smaller, from 19.68 to 17.39, corresponding to approximately 11.6\%. Thus, the effect of deprecation timing on up-to-date API generation is particularly pronounced for StarCoder2, whereas the difference is more limited for DeepSeek-Coder. 
In contrast, Specificity does not deteriorate for Post-cutoff APIs. The Specificity score increases from 85.03 to 89.85 on StarCoder2 and from 76.17 to 81.86 on DeepSeek-Coder. This suggests that the greater difficulty of unlearning Post-cutoff APIs is primarily reflected in suppressing the target deprecated APIs and generating their up-to-date replacements, rather than in increased disruption to unrelated API behavior.

A plausible explanation lies in the different knowledge states of the two API groups at the models' training-data cutoffs. For Pre-cutoff APIs, deprecation had already occurred before the cutoff, so the pre-training corpus could contain evidence of both their declining usage and their up-to-date replacements. The model may therefore already possess some knowledge consistent with the subsequent unlearning objective. In contrast, Post-cutoff APIs were still officially valid at the cutoff. Their usages in the pre-training corpus therefore represent valid programming practices rather than outdated behavior from the model's temporal perspective. Unlearning such APIs requires the model to suppress knowledge that may have been reinforced as valid during pre-training, which may explain their substantially higher residual AEM-D after unlearning.

\begin{mybox}{Answer to RQ4}
    API deprecation timing relative to the model's training-data cutoff substantially affects unlearning difficulty. APIs deprecated after the cutoff are considerably more difficult to forget than those already deprecated before the cutoff and, particularly for StarCoder2, are less likely to shift toward their up-to-date replacements. In contrast, their Specificity does not deteriorate, suggesting that deprecation timing primarily affects target API forgetting rather than the preservation of unrelated API behavior. 
\end{mybox}

\subsection{RQ5: How are the internal model changes induced by different machine unlearning methods related to their forgetting and preservation performance?} 

 \input{Figure/RQ5/RQ5}

RQ1 reveals substantial differences among machine unlearning methods in balancing deprecated API forgetting and capability preservation. However, output-level metrics such as AEM-D, AEM-U, OAMR, Specificity, and HumanEval characterize only the behavioral consequences of unlearning and provide limited insight into how these differences arise internally. To better understand the observed trade-offs, we further examine how different unlearning methods alter the model's internal representations and parameter importance. 
Following prior work on analyzing model internals in machine unlearning~\cite{xu2025unlearning}, we conduct the analysis at two complementary levels: representation and parameter. 

At the representation level, we employ PCA Shift, PCA Similarity, and Centered Kernel Alignment (CKA) to characterize complementary aspects of hidden-state changes before and after unlearning. PCA Shift measures the magnitude of representation displacement, PCA Similarity measures whether the principal representation directions are preserved, and CKA measures the overall structural similarity of the representation space. At the parameter level, we use the Fisher Information Matrix (FIM) to characterize changes in parameter importance by measuring how parameter sensitivity differs before and after unlearning. Together, these metrics allow us to examine whether different forgetting--preservation trade-offs observed in RQ1 are associated with relatively localized and controlled internal changes or broader perturbations to the model.

For each unlearning instance, we feed the same input to the original model and its corresponding unlearned model and extract the hidden states from each Transformer layer for layer-wise comparison. For PCA Shift, we perform PCA on the hidden states of the original and unlearned models at each layer and extract their first principal components. We then project both representations onto the first principal-component direction of the original model and measure the difference between their projections to quantify representation displacement. For PCA Similarity, we compute the cosine similarity between the first principal-component directions of the original and unlearned representations, with higher values indicating stronger preservation of the principal representation direction. For CKA, we compute linear CKA between the complete hidden representations of the original and unlearned models at each layer, providing a structural measure of representation similarity beyond the principal direction. 
For the parameter-level analysis, we use FIM to estimate the sensitivity of model predictions to the parameters in each layer. We compute the Fisher information of the corresponding parameters before and after unlearning, measure their relative changes, and aggregate these changes within each layer to quantify layer-wise changes in parameter importance. This layer-wise analysis enables us to identify where the internal effects of unlearning are concentrated and to compare whether different methods induce relatively localized changes or broader perturbations across the model.

Figures~\ref{fig:pca_shift}, ~\ref{fig:pca_similarity}, ~\ref{fig:cka}, and~\ref{fig:fim} present the layer-wise PCA Shift, PCA Similarity, CKA similarity, and relative changes in FIM, respectively. Overall, when representation changes occur, they are predominantly concentrated in the higher Transformer layers. However, the magnitude and distribution of these changes vary substantially across methods and models. GD, DPO, and NPO show small PCA Shift values and maintain PCA Similarity and CKA close to 1 across the three models, indicating that they largely preserve the original representation structure. In contrast, RLFT exhibits the most pronounced representation changes, while SimNPO also shows substantial changes, with the extent and layer-wise distribution varying across models. In particular, RLFT shows a sustained decline in PCA Similarity and CKA from the middle layers to the higher layers in StarCoder2, whereas the changes in Qwen2.5-Coder and DeepSeek-Coder are more concentrated in the final layers. Other methods, including KL and PROD, generally exhibit more limited representation changes. Unlike the representation metrics, FIM changes appear across both lower and higher layers, indicating that changes in parameter importance are not restricted to the higher layers. RLFT exhibits the most pronounced FIM changes, while GA and SimNPO also show noticeable changes in some models, with KL and PROD exhibiting relatively smaller changes. These internal patterns help explain the performance trends observed in RQ1. The pronounced representation and parameter-importance changes of RLFT, together with the substantial changes of SimNPO in some models, are consistent with their strong suppression of deprecated APIs, but such broad perturbations can also affect up-to-date API knowledge and unrelated API behavior, leading to lower AEM-U and OAMR. In contrast, DPO and NPO maintain relatively stable internal representations, which aligns with their stronger preservation of up-to-date API knowledge, unrelated API behavior, and general code-generation capability, although this conservative behavior comes at the cost of weaker deprecated API forgetting. GD achieves the most favorable trade-off between forgetting and preservation, as its limited internal changes are sufficient to suppress deprecated APIs while preserving up-to-date API knowledge and unrelated API behavior. These results suggest that effective deprecated API unlearning requires controlled internal changes that are sufficient to forget deprecated API knowledge without broadly disrupting the model's original API knowledge.

\begin{mybox}{Answer to RQ5}
Different unlearning methods induce substantially different internal changes that are consistent with their forgetting--preservation trade-offs. RLFT and SimNPO exhibit relatively broad internal perturbations together with strong forgetting but weaker preservation, whereas DPO and NPO induce smaller changes but generally achieve weaker forgetting. GD combines relatively controlled representation and parameter-importance changes with strong deprecated API suppression, providing an internal perspective on its favorable forgetting--preservation trade-off observed in RQ1.
\end{mybox}

\section{Discussion}
\label{sec:Discussion}

\subsection{Case Study}

\input{Figure/Case/case}

To qualitatively illustrate the forgetting--preservation trade-off observed in RQ1, we analyze the \texttt{evaluate()} function shown in Figure~\ref{fig:case}. We focus on three types of generation behavior within the same code-completion context: deprecated API generation (AEM-D), up-to-date API generation (AEM-U), and preservation of other API behaviors originally exhibited by the model (OAMR). In this example, the context checks whether an object has reached its goal position. The deprecated API is \texttt{torch.norm}, its corresponding up-to-date API is \texttt{torch.linalg.norm}, and other plausible implementations involve APIs such as \texttt{torch.abs}, \texttt{torch.isclose}, and \texttt{torch.all}. 
The original model exhibits all three types of behavior. Among five sampled completions, one contains \texttt{torch.norm}, one uses \texttt{torch.linalg.norm}, and the remaining three involve other APIs, including \texttt{torch.abs}, \texttt{torch.isclose}, and \texttt{torch.all}. Thus, deprecated, up-to-date, and other API behaviors coexist within the original completion space.

Different adaptation methods reshape this completion space differently. \textsc{AdaLoRA-L} eliminates \texttt{torch.norm} and generates \texttt{torch.linalg.norm} in all five completions. Although this achieves strong deprecated API suppression and up-to-date API generation, none of the other API behaviors originally exhibited by the model remain observable, illustrating how directed replacement can narrow the completion space. 
GD exhibits a more balanced behavior. None of its five completions contains \texttt{torch.norm}; two generate \texttt{torch.linalg.norm}, one preserves the \texttt{torch.abs}-based behavior observed in the original model, and the remaining two use alternative position-comparison expressions. This example is consistent with GD's favorable trade-off among AEM-D, AEM-U, and OAMR observed in RQ1.

Other machine unlearning methods show different preservation patterns. KL removes \texttt{torch.norm}, but generates neither \texttt{torch.linalg.norm} nor the other APIs observed in the original samples. NPO and PROD also do not generate \texttt{torch.linalg.norm}, but preserve \texttt{torch.abs}- and \texttt{torch.all}-based behaviors, respectively. These cases illustrate that suppressing the deprecated API does not necessarily lead to generation of its up-to-date replacement, and that different methods preserve different portions of the original completion space. 
RLFT exhibits the strongest behavioral disruption. Although \texttt{torch.norm} is eliminated, four of its five completions are empty and the remaining one is not a meaningful continuation of the given context, consistent with its strong forgetting but poor preservation performance in RQ1.

Overall, this case study illustrates that deprecated API adaptation should consider not only suppression of the deprecated API, but also up-to-date API generation and preservation of other API behaviors. Among the illustrated methods, GD provides the clearest balance across these three aspects, qualitatively supporting the quantitative findings in RQ1.

\subsection{Implications}

Based on the above findings, we derive several implications for practitioners and researchers.

\textbf{(1) Prioritize the forgetting--preservation trade-off when selecting unlearning methods.}
In practical deprecated API unlearning, practitioners should consider not only the suppression of deprecated APIs, but also up-to-date API generation, preservation of other and unrelated API behaviors, general code-generation capability, and computational cost. Our results show that GD achieves the most favorable overall balance across these dimensions while maintaining moderate training time and relatively low memory overhead. We therefore recommend GD as a preferred choice for deprecated API unlearning in practical code LLM maintenance.

\textbf{(2) Consider the model's temporal knowledge state when maintaining API knowledge.}
Our results show that APIs deprecated after a model's training-data cutoff are substantially more difficult to forget than those deprecated before the cutoff. This suggests that API knowledge maintenance should consider both the API deprecation timeline and the model's training-data cutoff to determine the model's temporal knowledge state. In particular, Post-cutoff APIs may require stronger unlearning and more careful post-unlearning validation rather than applying a uniform adaptation strategy to all deprecated APIs.

\textbf{(3) Construct machine unlearning benchmarks with behavior-grounded forget targets.}
Forget data should represent not only knowledge that should be removed, but also behavior that is demonstrably present in the target model. Our results motivate incorporating pre-unlearning behavioral verification into benchmark construction, such that a candidate is included as a forget instance only when the target model actually exhibits the corresponding undesirable behavior. This can provide a more meaningful basis for evaluating whether machine unlearning truly removes existing model behavior.

\textbf{(4) Explore automated, real-world knowledge evolution as a general paradigm for constructing machine unlearning benchmarks.}
\texttt{MUDAPIBench} derives forgetting targets from real API evolution and uses a largely automated pipeline combining repository mining with model-specific behavioral verification. This construction paradigm can potentially be extended to other evolving software knowledge and provide a reference for constructing machine unlearning benchmarks in the broader AI community, reducing dependence on manually synthesized data or inaccessible pre-training corpora.

\textbf{(5) Design unlearning methods around controlled internal changes rather than forgetting strength alone.}
Our internal analyses show that stronger forgetting can be accompanied by broader representation and parameter perturbations and greater capability degradation, whereas GD achieves strong deprecated API suppression with comparatively controlled internal changes. Future unlearning methods should therefore focus on inducing sufficient but controlled changes to representations and parameters associated with the target knowledge, while minimizing unnecessary perturbations to the model's existing knowledge and capabilities.

\subsection{Threats to Validity}

(1) This study evaluates a limited number of code large language models. Although the selected models differ in architecture, parameter scale, and training data to some extent, they cannot fully represent the diversity of contemporary code LLMs. Differences in pre-training corpora, model scale, and training data cutoffs may affect the extent to which models acquire knowledge of deprecated APIs and respond to unlearning operations. Therefore, the experimental results primarily characterize the behavior of the selected models and may not generalize to other code LLMs. Future work could expand the model coverage to include code LLMs with more diverse architectures, parameter scales, and training data sources to further validate the generalizability of our findings.

(2) This study primarily evaluates machine unlearning at the API level through code completion behavior. Specifically, we examine whether the unlearned model reduces its tendency to generate deprecated APIs and whether it can instead generate their up-to-date replacements, without directly verifying the functional correctness of the generated code within complete programs. This design is motivated by the scale of our benchmark, which contains more than 15,000 evaluation instances. Constructing independent multi-line program contexts and executable test cases for each instance would incur substantial development costs, while the runtime environments and testing requirements can vary considerably across APIs. In contrast, API-level code completion provides a standardized setting for directly measuring changes in the model's API knowledge. Nevertheless, this evaluation granularity cannot fully capture the impact of unlearning on complete-program behavior. Future work could incorporate program-level functional testing to provide a more comprehensive assessment of unlearning effectiveness.

(3) The experimental data in this study are primarily drawn from mainstream Python API libraries covered by \texttt{MUDAPIBench}. Although the benchmark includes multiple representative libraries and a large number of deprecated APIs, the API evolution processes in real-world software ecosystems are considerably more diverse. The current dataset does not fully cover the broad range of APIs, programming languages, and deprecation scenarios encountered in practice. In particular, factors such as API usage frequency, invocation patterns, functional complexity, and the semantic distance between deprecated APIs and their replacements may affect the difficulty of unlearning. Consequently, the observed results may be influenced to some extent by the distribution of the benchmark. Future work could incorporate additional programming languages, API libraries, and diverse API evolution scenarios to increase dataset diversity and examine the applicability of our findings in broader settings.

(4) Our evaluation of machine unlearning methods relies on the mapping between deprecated APIs and their up-to-date replacements, as well as the official deprecation timelines of these APIs. These pieces of information provide the basis for determining which API knowledge should be forgotten and whether the corresponding knowledge should have been acquired by a model. In \texttt{MUDAPIBench}, API mappings are primarily collected and cross-validated through prior studies and official documentation. Nevertheless, the version histories and replacement relationships of some APIs may still be affected by incomplete documentation and manual curation. Moreover, the training data cutoffs of different models are not identical. The relative timing between API deprecation and a model's training data cutoff may therefore affect whether the deprecated API was actually exposed to the model during pre-training. Although we further analyze the influence of this factor on unlearning performance, it remains difficult to determine the extent to which specific API knowledge was present in the actual training corpora. Future work could integrate more comprehensive API version histories, release timelines, and model training-corpus information to construct more precise API lifecycle data and further improve the reliability of unlearning evaluation.

\section{Related Work}
\label{sec:Related Work}

\subsection{Machine Unlearning for General-Purpose AI}

Machine unlearning aims to remove specified knowledge or data influence from trained models while preserving their remaining capabilities. Existing studies in general-purpose AI and NLP have developed diverse unlearning approaches, including gradient-based optimization, distribution regularization, random supervision, preference optimization, and output-distribution modification, and have applied them to removing factual knowledge, copyrighted content, private information, and other undesirable knowledge \cite{liu2025rethinking}. Despite their different optimization mechanisms, these methods share a fundamental prerequisite: the knowledge to be forgotten must first be identified and represented as forget data. The construction of forget data therefore plays a central role in determining what knowledge an unlearning method is expected to remove and whether its effectiveness can be meaningfully evaluated.

Existing studies obtain forget data in several ways. One line of work relies on manually constructed or synthetic data, as exemplified by TOFU~\cite{maini2024tofu}, KnowUnDo~\cite{tian2024forget}, and PISTOL~\cite{qiu2024pistol}. Such datasets provide controlled forgetting targets and facilitate systematic comparison of unlearning methods. Other studies consider real-world content, such as books, news articles, or domain-specific knowledge~\cite{shi2025muse,li2024wmdp}, to evaluate unlearning under more realistic knowledge settings. 
A more direct strategy derives forget data from the model's original pre-training corpus~\cite{yao2024machine,xu2025unlearning}, establishing an explicit connection between the target data and the model's training history. However, this setting depends on access to the original pre-training corpus, which is often unavailable or only partially disclosed for modern LLMs and may therefore limit reproducibility across different models. 
Another line of work adopts a \textit{learn-then-unlearn} paradigm~\cite{tian2024forget,ren2025general,bhaila2025soft,xu2025relearn}, in which target knowledge is first injected into a model through additional training and subsequently removed using an unlearning method. This setting guarantees exposure to the target knowledge and provides a controlled environment for evaluating different unlearning methods, although the additional learning stage changes the model's original knowledge distribution. 

\subsection{Machine Unlearning for Software Engineering}

Machine unlearning in software engineering remains an emerging research area. Existing studies have investigated the removal of various forms of undesirable knowledge from code models, including buggy or vulnerable code~\cite{yang2024hotfixing}, Trojan backdoors~\cite{kazemi2024unlearning}, sensitive information and memorized code~\cite{gu2026mitigating,chu2025scrub}, copyrighted code~\cite{yao2024machine,xu2025unlearning}, and proprietary or insecure hardware designs~\cite{liang2025forgetting}. For example, Yao et al.~\cite{yao2024machine} systematically evaluated representative unlearning methods, including GA, GD, KL, and RLFT, on pre-trained code models. Yang and Lo~\cite{yang2024hotfixing} applied machine unlearning to buggy code, while Kazemi et al.~\cite{kazemi2024unlearning} combined GA with Elastic Weight Consolidation (EWC)~\cite{kirkpatrick2017overcoming} to remove Trojan backdoors. More recently, machine unlearning has been extended to evolving software knowledge. Jiang et al.~\cite{jiang2026large} evaluated several unlearning methods for deprecated APIs and proposed PROD, while Tran et al.~\cite{tran2026towards} incorporated contrastive learning into SimNPO and PROD to suppress deprecated API generation while promoting up-to-date alternatives. 
Despite this progress, the construction of forget data in existing code-unlearning studies raises two important issues: whether the forgetting targets correspond to realistic software-engineering knowledge-removal needs and whether the target behavior is actually observable in the original model before unlearning. 

Regarding the first issue, some studies construct forget data using manually designed, synthetic, or model-generated samples. For example, Liang et al.~\cite{liang2025forgetting} construct hardware-code forget data based on the GPT-generated RTLCoder corpus, while Gu et al.~\cite{gu2026mitigating} use synthetic sensitive information and corresponding code samples. Such designs provide controllable forgetting targets but represent simulated forgetting scenarios rather than knowledge-removal needs arising naturally from software evolution or development practice. Other studies derive forget data directly from a model's pre-training corpus. Yao et al.~\cite{yao2024machine} and Xu et al.~\cite{xu2025unlearning}, for example, sample GitHub code contained in the pre-training corpus of Yi-6B and treat it as the target for unlearning, simulating scenarios in which developers or organizations request the removal of particular code patterns, algorithms, or proprietary code for intellectual-property protection. This strategy establishes a direct connection between the forget data and the model's training history. However, the corresponding removal requests are simulated rather than derived from observed real-world requests or verified ownership claims. Moreover, this construction requires access to model-specific pre-training corpora, which are often unavailable for modern LLMs and therefore limits its applicability and reproducibility across models.

The second issue concerns whether a candidate forget instance corresponds to behavior that the original model actually exhibits. Existing code-unlearning studies commonly construct forget data from predefined target samples and directly apply them during unlearning without using the model's pre-unlearning generation behavior as an instance-selection criterion~\cite{yang2024hotfixing,xu2025unlearning,jiang2026large,tran2026towards}. For example, Yang and Lo~\cite{yang2024hotfixing} construct bug-related tuples based on ManySStuBs4J \cite{karampatsisHowOftenSingleStatement2020} and use them for model updating, but do not first require the original model to generate the corresponding buggy code from the given context. Similarly, Jiang et al.~\cite{jiang2026large} and Tran et al.~\cite{tran2026towards} construct forget data from predefined deprecated-API samples without requiring the target model to exhibit the corresponding deprecated-API behavior before unlearning. Consequently, a predefined forgetting target may be included even when the targeted behavior is not observable from the original model under the corresponding completion context.

To address these limitations, \texttt{MUDAPIBench} is designed around two principles for constructing realistic and model-relevant forget data. First, its forgetting targets are derived from real API deprecations caused by software-library evolution, reflecting naturally occurring software-engineering needs for knowledge removal rather than artificially defined forgetting scenarios. Second, we perform model-specific filtering before unlearning and retain a candidate instance only when the original model actually generates the corresponding deprecated API, ensuring that each forget instance targets behavior observable in the model prior to unlearning. Beyond forget-data construction, existing code-unlearning studies~\cite{yang2024hotfixing,xu2025unlearning,jiang2026large,tran2026towards,chu2025scrub} primarily assess capability preservation through general code-generation benchmarks such as HumanEval~\cite{chen2021evaluating}. While such evaluation captures changes in overall model utility, it provides limited insight into the localized behavioral effects of unlearning. In particular, existing studies do not explicitly evaluate whether unlearning a deprecated API causes unintended changes to unrelated API behavior on separate code-completion inputs. To complement general capability evaluation, \texttt{MUDAPIBench} therefore introduces Specificity Data to explicitly assess the preservation of unrelated API behavior after unlearning.  Moreover, existing evaluations do not explicitly examine whether other API behaviors originally exhibited by the model within the same target completion contexts remain observable after unlearning. Given the multi-solution nature of code completion, such other API completions may constitute plausible continuations and should not be unnecessarily disrupted. We therefore introduce OAMR to explicitly evaluate the preservation of these other API behaviors. 

Finally, for empirical comparison, we select eight representative machine unlearning methods applicable to deprecated API knowledge: GA, GD, KL, RLFT, DPO, NPO, SimNPO, and PROD. These methods cover the machine unlearning approaches adopted in the software-engineering studies reviewed above, with two exceptions. We exclude CodeEraser~\cite{chu2025scrub} because it is specifically designed for fine-grained removal of localized sensitive strings, such as credentials and secrets, which differs substantially from the shared API usage knowledge considered in our setting. We also do not separately include EWC~\cite{kirkpatrick2017overcoming}; instead, we use KL-based regularization as the representative preservation-oriented regularization strategy, as KL constraints have been more widely adopted in recent machine unlearning studies.

\subsection{LLM Adaptation to API Evolution}

API deprecation is a common consequence of software-library evolution. Prior studies have shown that APIs are deprecated for various reasons, such as improving code readability, reducing redundancy, eliminating undesirable programming practices, and fixing functional defects~\cite{guancheng2026don}. Such deprecations can affect a large number of downstream projects and require developers to continuously adapt their code to evolving libraries~\cite{guancheng2026don}. This challenge has become increasingly relevant to LLMs, whose training corpora may contain API usages from different historical library versions. Wang et al.~\cite{wang2024llms} found that 37.4\% of API calls generated by GPT-3.5 correspond to deprecated APIs, demonstrating that LLMs may retain outdated API knowledge after the underlying libraries have evolved.

Several approaches have been proposed to adapt LLMs to API evolution. Wang et al.~\cite{wang2024llms} constructed deprecated-to-up-to-date API mappings from real-world projects and proposed \texttt{REPLACEAPI}, an inference-time intervention that detects deprecated APIs in generated code, replaces them with their corresponding up-to-date APIs, and re-invokes the LLM with the reconstructed context. Although effective in reducing deprecated API generation, this approach does not update the outdated API knowledge encoded in the model and introduces additional detection and inference overhead. Lin et al.~\cite{guancheng2026don} subsequently studied API evolution from a model-editing perspective, proposed \textsc{AdaLoRA-L}, and constructed EDAPIBench for evaluation. To construct editing instances, three completions are generated for each candidate prompt under deterministic decoding, and only prompts for which the deprecated API appears in all three completions are retained. \textsc{AdaLoRA-L} then performs directed model editing by promoting the corresponding up-to-date API as the predefined editing target. While this formulation enables model-level API updating, it focuses on contexts exhibiting consistent deprecated-API behavior and formulates API adaptation as a directed deprecated-to-up-to-date replacement.

More recently, machine unlearning has been explored for deprecated API adaptation. Jiang et al.~\cite{jiang2026large} introduced deprecated API unlearning as a code-unlearning task and proposed PROD, which suppresses deprecated code at the token level by redistributing its probability mass over the remaining vocabulary. Tran et al.~\cite{tran2026towards} further incorporated contrastive learning into SimNPO and PROD to suppress deprecated API completions while promoting their corresponding up-to-date replacements. These studies demonstrate the feasibility of applying machine unlearning to evolving API knowledge, but several issues remain underexplored.

First, existing forget data do not consistently verify whether the target deprecated-API behavior is actually exhibited by each model under the corresponding code context before unlearning. Jiang et al.~\cite{jiang2026large} construct their deprecated-API data from VersiCode \cite{wu2024versicode} by using an intermediate library release as the deprecation boundary to distinguish deprecated and valid API usages. Tran et al.~\cite{tran2026towards} mainly construct their forget set from predefined deprecated-API samples derived from the dataset of Wang et al.~\cite{wang2024llms}. Neither study requires every candidate instance to be individually verified against the target model to ensure that the corresponding deprecated API is actually generated for that context before unlearning. Consequently, a candidate may be treated as a forget instance without first establishing that the targeted deprecated-API behavior is observable in the original model.

Second, existing studies cover only a limited subset of machine unlearning methods, leaving the relative effectiveness of different unlearning paradigms for deprecated API knowledge insufficiently understood. In particular, representative methods such as GD, KL, and RLFT have not been systematically investigated in this setting. Our study therefore conducts a broader empirical comparison of eight representative machine unlearning methods across three code LLMs. Interestingly, our results show that GD, which has not been evaluated in prior deprecated-API unlearning studies, achieves the most favorable overall balance among deprecated API forgetting, up-to-date API generation, preservation of other API behaviors, preservation of unrelated API behavior, general code-generation capability, and computational cost.

\section{Conclusion}
\label{sec:Conclusion}

This study systematically investigates machine unlearning as an approach to adapting code LLMs to deprecated API knowledge. Rather than treating API evolution as a directed replacement problem, we consider the multi-solution nature of code completion and examine whether deprecated API behavior can be suppressed without unnecessarily disrupting other desirable generation behaviors. To enable this investigation, we construct \texttt{MUDAPIBench}, a largely automatically constructed, behavior-grounded benchmark in which each unlearning instance is verified against the pre-unlearning behavior of the target model.

Our empirical study demonstrates that effective deprecated API unlearning requires balancing forgetting with preservation rather than maximizing forgetting strength alone. Across eight machine unlearning methods and three code LLMs, Gradient Difference (GD) provides the most favorable overall balance: it strongly suppresses deprecated API generation while better maintaining up-to-date API generation, other API behaviors, unrelated API behavior, and general code-generation capability. Its moderate training time and relatively low memory overhead further make this balance achievable without particularly high computational cost. In contrast, stronger suppression does not necessarily indicate better unlearning, as aggressive methods can remove desirable generation behaviors together with the targeted deprecated API. 
Our analyses further show that deprecated API unlearning is influenced by both the target knowledge and the way a method modifies the model. Unlearning difficulty varies across software libraries, and APIs deprecated after a model's training-data cutoff are substantially more difficult to forget than those deprecated before the cutoff, highlighting the importance of the model's temporal knowledge state. At the model-internal level, different forgetting--preservation trade-offs are associated with different degrees of representation and parameter perturbation. In particular, GD combines strong forgetting with comparatively controlled internal changes, whereas more aggressive forgetting can be accompanied by broader model perturbations and greater capability degradation.

Overall, our findings suggest that deprecated API unlearning should be viewed as a selective knowledge adaptation problem: the goal is not simply to eliminate outdated behavior, but to remove it while preserving the useful completion space surrounding it. Beyond deprecated APIs, \texttt{MUDAPIBench} also demonstrates the potential of combining real-world knowledge evolution with pre-unlearning behavioral verification to construct scalable and behavior-grounded machine unlearning benchmarks. We hope these findings can inform both the practical maintenance of evolving knowledge in code LLMs and future research on benchmark construction and more selective machine unlearning methods.

\section{Data Availability}

Our \texttt{MUDAPIBench}, the code for its construction, and machine unlearning source code are available in~\url{https://github.com/YanzhongHe/MUDAPIBench}.

\bibliographystyle{ACM-Reference-Format}
\bibliography{software}

\end{document}

%% file: table/benchmark_datanum.tex
\begin{figure}[!t]
    \centering

    \captionof{table}{The number of instances and APIs for different LLMs across evaluation dimensions in \texttt{MUDAPIBench}.}
    \vspace{-0.2cm}
    \label{tab:bench_statistics}
    \footnotesize
    \begin{tabular}{c|cc|cc|cc}
    \toprule
    Models & \multicolumn{2}{c|}{Qwen2.5-Coder} & 
    \multicolumn{2}{c|}{StarCoder2} & 
    \multicolumn{2}{c}{DeepSeek-Coder} \\
    \midrule
    Statistical value
    & \# instances & \# APIs
    & \# instances & \# APIs
    & \# instances & \# APIs \\
    \midrule
    Forget& 3,043 & 81
    & 2,117 & 78
    & 2,586 & 76 \\

    Generalization
    & 3,012 & 70
    & 2,088 & 71
    & 2,560 & 69 \\

    Specificity
    & 15,215 & 3,995
    & 10,585 & 3,059
    & 12,930 & 3,449 \\
    \bottomrule
    \end{tabular}

    \vspace{0cm}

    \includegraphics[width=\linewidth]{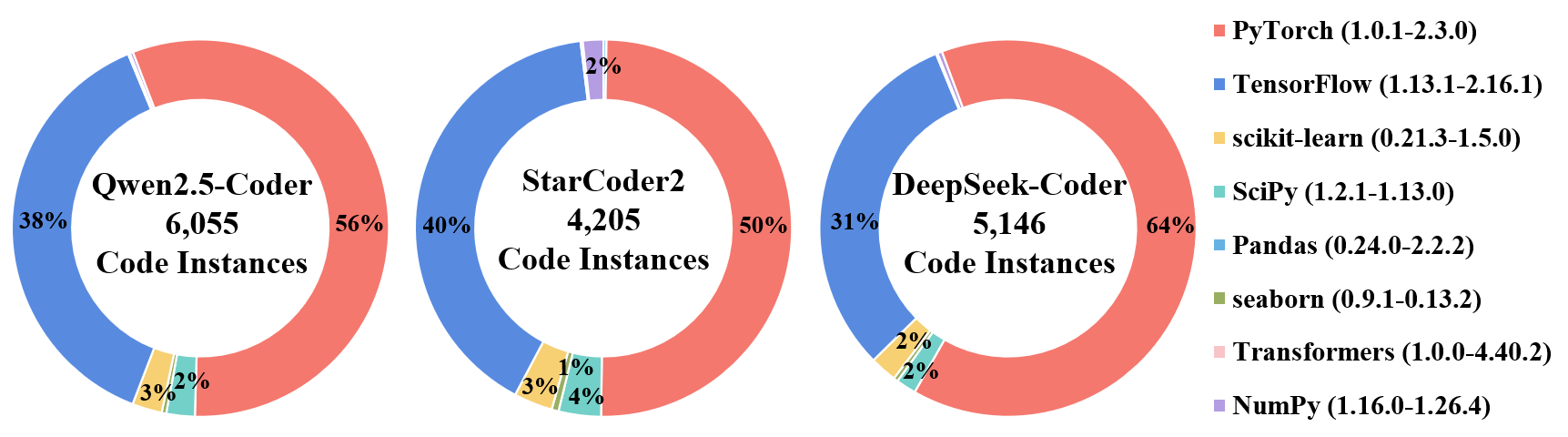}
    \captionof{figure}{The distribution of the number of target APIs from different libraries in \texttt{MUDAPIBench}.}
    \label{fig:api_distribution}

\end{figure}

\begin{figure}
    \centering
    \includegraphics[width=0.98\linewidth]{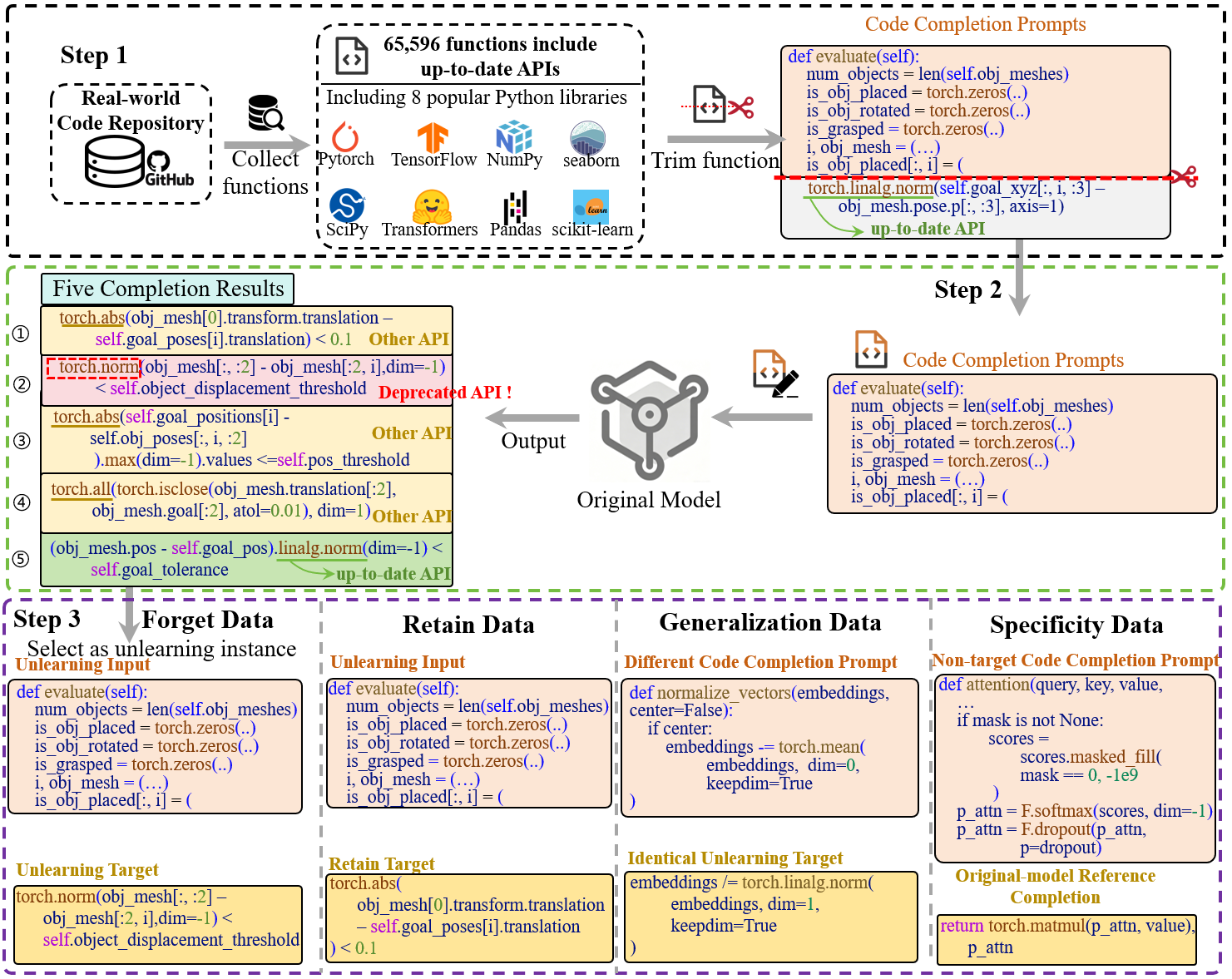}
    \captionsetup{skip=0pt}
    \caption{The construction process of \texttt{MUDAPIBench}.}
    \vspace{-0.4cm}
    \label{fig:bench_construction}
\end{figure}

%% file: table/RQ1-forget.tex
\begin{table}[t]
\centering
\setlength{\tabcolsep}{3pt}

\definecolor{darkgreen}{RGB}{0,120,0}

\newcommand{\ori}[1]{%
\makebox[38pt][l]{%
\makebox[20pt][r]{#1}%
\hspace{2pt}%
\phantom{{\fontsize{5pt}{5pt}\selectfont
\(\downarrow00.00\)}}%
}%
}

\newcommand{\metric}[1]{%
\makebox[38pt][l]{#1}%
}


\newcommand{\dec}[2]{%
\makebox[38pt][l]{%
\makebox[20pt][r]{#1}%
\hspace{2pt}%
{\fontsize{5pt}{5pt}\selectfont
\textcolor{red}{\(\downarrow#2\)}}%
}%
}

\newcommand{\inc}[2]{%
\makebox[38pt][l]{%
\makebox[20pt][r]{#1}%
\hspace{2pt}%
{\fontsize{5pt}{5pt}\selectfont
\textcolor{darkgreen}{\(\uparrow#2\)}}%
}%
}


\newcommand{\decD}[2]{%
\makebox[38pt][l]{%
\makebox[20pt][r]{#1}%
\hspace{2pt}%
{\fontsize{5pt}{5pt}\selectfont
\textcolor{darkgreen}{\(\downarrow#2\)}}%
}%
}

\newcommand{\incD}[2]{%
\makebox[38pt][l]{%
\makebox[20pt][r]{#1}%
\hspace{2pt}%
{\fontsize{5pt}{5pt}\selectfont
\textcolor{red}{\(\uparrow#2\)}}%
}%
}

\caption{The performance of eight machine unlearning methods and \textsc{AdaLoRA-L} on Forget data across three models. Here, BL and RL refer to BLEU and ROUGE-L, respectively.}
\label{tab:unlearning-effectiveness-all}

\vspace{-0.4cm}

\begin{tabular}{c|ccccccccc}
\hline
Methods
& \metric{AEM-D$_{\scriptscriptstyle\downarrow}$}
& \metric{AEM-U$_{\scriptscriptstyle\uparrow}$}
& OMAR$_{\scriptscriptstyle\uparrow}$
& \metric{ES-D$_{\scriptscriptstyle\downarrow}$}
& \metric{BL-D$_{\scriptscriptstyle\downarrow}$}
& \metric{RL-D$_{\scriptscriptstyle\downarrow}$}
& \metric{ES-U$_{\scriptscriptstyle\uparrow}$}
& \metric{BL-U$_{\scriptscriptstyle\uparrow}$}
& \metric{RL-U$_{\scriptscriptstyle\uparrow}$}
\\
\hline


\rowcolor{lightgray}
\multicolumn{10}{c}{Qwen2.5-Coder}
\\
\hline

Original model
& \ori{33.53} & \ori{19.61} & /
& \ori{47.56} & \ori{38.56} & \ori{31.53}
& \ori{20.05} & \ori{12.16} & \ori{11.24}
\\
\hline

\rowcolor{lightpink}GA
& \decD{2.08}{31.45} & \dec{3.93}{15.68} & 21.07
& \decD{31.93}{15.63} & \decD{18.23}{20.33} & \decD{18.51}{13.02}
& \dec{14.75}{5.30} & \dec{5.16}{7.00} & \dec{6.22}{5.02}
\\

\rowcolor{lightpink}GD
& \decD{5.42}{28.11} & \inc{35.30}{15.69} & \textbf{36.59}
& \decD{45.64}{1.92} & \decD{36.48}{2.08} & \decD{29.43}{2.10}
& \inc{20.13}{0.08} & \inc{12.68}{0.52} & \inc{11.45}{0.21}
\\

\rowcolor{lightblue}KL
& \decD{4.02}{29.51} & \dec{13.39}{6.22} & 30.30
& \decD{38.54}{9.02} & \decD{26.72}{11.84} & \decD{23.62}{7.91}
& \dec{17.78}{2.27} & \dec{8.54}{3.62} & \dec{9.18}{2.06}
\\

\rowcolor{lightyellow}RLFT
& \decD{\textbf{0.03}}{33.50} & \dec{0.07}{19.54} & 16.28
& \decD{\textbf{9.95}}{37.61} & \decD{\textbf{2.01}}{36.55} & \decD{\textbf{1.67}}{29.86}
& \dec{4.75}{15.30} & \dec{1.77}{10.39} & \dec{0.20}{11.04}
\\

\rowcolor{lightgreen}DPO
& \decD{10.68}{22.85} & \inc{27.60}{7.99} & 36.30
& \decD{44.68}{2.88} & \decD{35.02}{3.54} & \decD{28.30}{3.23}
& \dec{19.83}{0.22} & \dec{11.99}{0.17} & \dec{11.01}{0.23}
\\

\rowcolor{lightgreen}NPO
& \decD{13.14}{20.39} & \inc{27.39}{7.78} & 35.97
& \decD{45.74}{1.82} & \decD{36.46}{2.10} & \decD{29.24}{2.29}
& \dec{19.92}{0.13} & \inc{12.18}{0.02} & \dec{11.08}{0.16}
\\

\rowcolor{lightgreen}SimNPO
& \decD{1.58}{31.95} & \dec{0.80}{18.81} & 18.86
& \decD{28.67}{18.89} & \decD{14.65}{23.91} & \decD{15.70}{15.83}
& \dec{13.13}{6.92} & \dec{3.97}{8.19} & \dec{5.04}{6.20}
\\

\rowcolor{lightbeige}PROD
& \decD{12.85}{20.68} & \dec{14.81}{4.80} & 33.45
& \decD{37.63}{9.93} & \decD{26.70}{11.86} & \decD{21.99}{9.54}
& \dec{17.26}{2.79} & \dec{8.67}{3.49} & \dec{8.77}{2.47}
\\

\rowcolor{lightpurple}\textsc{AdaLoRA-L}
& \decD{0.80}{32.73} & \inc{\textbf{88.01}}{68.40} & 5.14
& \decD{42.13}{5.43} & \decD{30.55}{8.01} & \decD{21.85}{9.68}
& \inc{\textbf{21.98}}{1.93} & \inc{\textbf{17.64}}{5.48} & \inc{\textbf{15.93}}{4.69}
\\

\hline


\rowcolor{lightgray}
\multicolumn{10}{c}{StarCoder2}
\\
\hline

Original model
& \ori{28.34} & \ori{13.56} & /
& \ori{37.89} & \ori{26.87} & \ori{21.92}
& \ori{19.09} & \ori{11.20} & \ori{11.37}
\\
\hline

\rowcolor{lightpink}GA
& \decD{11.18}{17.16} & \dec{12.78}{0.78} & 47.45
& \decD{33.08}{4.81} & \decD{22.68}{4.19} & \decD{17.91}{4.01}
& \dec{15.36}{3.73} & \dec{8.68}{2.52} & \dec{8.65}{2.72}
\\

\rowcolor{lightpink}GD
& \decD{3.84}{24.50} & \inc{14.96}{1.40} & \textbf{48.17}
& \decD{34.13}{3.76} & \decD{22.67}{4.20} & \decD{19.38}{2.54}
& \dec{16.54}{2.55} & \dec{8.69}{2.51} & \dec{9.65}{1.72}
\\

\rowcolor{lightblue}KL
& \decD{13.50}{14.84} & \inc{14.21}{0.65} & 47.15
& \decD{36.59}{1.30} & \decD{25.73}{1.14} & \decD{20.58}{1.34}
& \dec{17.05}{2.04} & \dec{9.64}{1.56} & \dec{9.92}{1.45}
\\

\rowcolor{lightyellow}RLFT
& \decD{\textbf{0.25}}{28.09} & \dec{0.16}{13.40} & 38.69
& \decD{\textbf{11.68}}{26.21} & \decD{\textbf{3.81}}{23.06} & \decD{\textbf{1.47}}{20.45}
& \dec{4.45}{14.64} & \dec{1.74}{9.46} & \dec{0.13}{11.24}
\\

\rowcolor{lightgreen}DPO
& \decD{13.08}{15.26} & \inc{13.82}{0.26} & 47.67
& \decD{35.57}{2.32} & \decD{24.84}{2.03} & \decD{20.09}{1.83}
& \dec{16.81}{2.28} & \dec{9.48}{1.72} & \dec{9.63}{1.74}
\\

\rowcolor{lightgreen}NPO
& \decD{15.09}{13.25} & \inc{15.91}{2.35} & 47.19
& \incD{38.95}{1.06} & \incD{28.25}{1.38} & \incD{22.77}{0.85}
& \dec{\textbf{18.21}}{0.88} & \dec{\textbf{10.58}}{0.62} & \dec{\textbf{10.74}}{0.63}
\\

\rowcolor{lightgreen}SimNPO
& \decD{6.08}{22.26} & \dec{6.18}{7.38} & 44.08
& \decD{22.34}{15.55} & \decD{13.19}{13.68} & \decD{9.26}{12.66}
& \dec{9.64}{9.45} & \dec{5.04}{6.16} & \dec{4.38}{6.99}
\\

\rowcolor{lightbeige}PROD
& \decD{9.21}{19.13} & \dec{13.39}{0.17} & 46.30
& \decD{33.08}{4.81} & \decD{21.71}{5.16} & \decD{17.38}{4.54}
& \dec{16.03}{3.06} & \dec{8.48}{2.72} & \dec{9.05}{2.32}
\\
\rowcolor{lightpurple}\textsc{AdaLoRA-L}
& \decD{0.38}{27.96} & \inc{\textbf{36.12}}{22.56} & 3.78
& \decD{24.76}{13.13} & \decD{15.99}{10.88} & \decD{12.07}{9.85}
& \dec{13.24}{5.85} & \dec{9.24}{1.96} & \dec{8.02}{3.35}
\\

\hline


\rowcolor{lightgray}
\multicolumn{10}{c}{DeepSeek-Coder}
\\
\hline

Original model
& \ori{30.16} & \ori{16.15} & /
& \ori{46.36} & \ori{36.09} & \ori{28.11}
& \ori{22.62} & \ori{13.24} & \ori{11.58}
\\
\hline

\rowcolor{lightpink}GA
& \decD{4.80}{25.36} & \inc{17.96}{1.81} & 17.28
& \decD{40.86}{5.50} & \decD{28.68}{7.41} & \decD{22.94}{5.17}
& \dec{20.77}{1.85} & \dec{10.96}{2.28} & \dec{9.31}{2.27}
\\

\rowcolor{lightpink}GD
& \decD{2.03}{28.13} & \inc{19.56}{3.41} & 21.62
& \decD{42.38}{3.98} & \decD{31.18}{4.91} & \decD{25.16}{2.95}
& \dec{22.36}{0.26} & \dec{12.65}{0.59} & \dec{11.53}{0.05}
\\

\rowcolor{lightblue}KL
& \decD{9.29}{20.87} & \inc{23.74}{7.59} & 22.51
& \decD{44.95}{1.41} & \decD{34.27}{1.82} & \decD{26.26}{1.85}
& \dec{22.26}{0.36} & \dec{13.01}{0.23} & \dec{11.16}{0.42}
\\

\rowcolor{lightyellow}RLFT
& \decD{\textbf{0.38}}{29.78} & \dec{0.19}{15.96} & 17.64
& \decD{\textbf{18.16}}{28.20} & \decD{\textbf{6.04}}{30.05} & \decD{\textbf{6.50}}{21.61}
& \dec{9.16}{13.46} & \dec{2.26}{10.98} & \dec{1.40}{10.18}
\\

\rowcolor{lightgreen}DPO
& \decD{10.32}{19.84} & \inc{25.57}{9.42} & \textbf{26.50}
& \decD{46.01}{0.35} & \decD{35.46}{0.63} & \decD{26.88}{1.23}
& \inc{22.79}{0.17} & \inc{13.78}{0.54} & \inc{11.71}{0.13}
\\

\rowcolor{lightgreen}NPO
& \decD{14.11}{16.05} & \inc{25.78}{9.63} & 25.15
& \incD{47.11}{0.75} & \incD{\textbf{37.06}}{0.97} & \decD{27.90}{0.21}
& \inc{22.97}{0.35} & \inc{14.13}{0.89} & \inc{11.88}{0.30}
\\

\rowcolor{lightgreen}SimNPO
& \decD{3.83}{26.33} & \dec{14.89}{1.26} & 17.01
& \decD{39.27}{7.09} & \decD{26.65}{9.44} & \decD{21.48}{6.63}
& \dec{20.25}{2.37} & \dec{10.25}{2.99} & \dec{8.61}{2.97}
\\

\rowcolor{lightbeige}PROD
& \decD{3.98}{26.18} & \dec{4.89}{11.26} & 21.61
& \decD{31.78}{14.58} & \decD{18.29}{17.80} & \decD{16.14}{11.97}
& \dec{16.29}{6.33} & \dec{5.49}{7.75} & \dec{7.12}{4.46}
\\
\rowcolor{lightpurple}\textsc{AdaLoRA-L}
& \decD{0.50}{29.66} & \inc{\textbf{79.98}}{63.83} & 4.50
& \decD{44.36}{2.00} & \decD{33.00}{3.09} & \decD{23.76}{4.35}
& \inc{\textbf{25.09}}{2.47} & \inc{\textbf{19.06}}{5.82} & \inc{\textbf{17.47}}{5.89}
\\

\hline

\end{tabular}

\vspace{-0.5cm}
\end{table}

%% file: table/RQ1-gen.tex
\begin{table}[t]
\centering
\setlength{\tabcolsep}{3.0pt}

\definecolor{darkgreen}{RGB}{0,120,0}

\newcommand{\ori}[1]{%
\makebox[38pt][l]{%
\makebox[20pt][r]{#1}%
\hspace{2pt}%
\phantom{{\fontsize{5pt}{5pt}\selectfont
\(\downarrow00.00\)}}%
}%
}

\newcommand{\metric}[1]{%
\makebox[38pt][l]{#1}%
}


\newcommand{\dec}[2]{%
\makebox[38pt][l]{%
\makebox[20pt][r]{#1}%
\hspace{2pt}%
{\fontsize{5pt}{5pt}\selectfont
\textcolor{red}{\(\downarrow#2\)}}%
}%
}

\newcommand{\inc}[2]{%
\makebox[38pt][l]{%
\makebox[20pt][r]{#1}%
\hspace{2pt}%
{\fontsize{5pt}{5pt}\selectfont
\textcolor{darkgreen}{\(\uparrow#2\)}}%
}%
}


\newcommand{\decD}[2]{%
\makebox[38pt][l]{%
\makebox[20pt][r]{#1}%
\hspace{2pt}%
{\fontsize{5pt}{5pt}\selectfont
\textcolor{darkgreen}{\(\downarrow#2\)}}%
}%
}

\newcommand{\incD}[2]{%
\makebox[38pt][l]{%
\makebox[20pt][r]{#1}%
\hspace{2pt}%
{\fontsize{5pt}{5pt}\selectfont
\textcolor{red}{\(\uparrow#2\)}}%
}%
}

\caption{The performance of eight machine unlearning methods and \textsc{AdaLoRA-L} on Generalization data across three models. Here, BL and RL refer to BLEU and ROUGE-L, respectively.}
\label{tab:unlearning-generalization-all}

\vspace{-0.4cm}

\begin{tabular}{c|ccccccccc}
\hline
Methods
& \metric{AEM-D$_{\scriptscriptstyle\downarrow}$}
& \metric{AEM-U$_{\scriptscriptstyle\uparrow}$}
& OMAR$_{\scriptscriptstyle\uparrow}$
& \metric{ES-D$_{\scriptscriptstyle\downarrow}$}
& \metric{BL-D$_{\scriptscriptstyle\downarrow}$}
& \metric{RL-D$_{\scriptscriptstyle\downarrow}$}
& \metric{ES-U$_{\scriptscriptstyle\uparrow}$}
& \metric{BL-U$_{\scriptscriptstyle\uparrow}$}
& \metric{RL-U$_{\scriptscriptstyle\uparrow}$}
\\
\hline


\rowcolor{lightgray}
\multicolumn{10}{c}{Qwen2.5-Coder}
\\
\hline

Original model
& \ori{33.90} & \ori{19.95} & /
& \ori{47.40} & \ori{38.37} & \ori{31.25}
& \ori{20.14} & \ori{12.24} & \ori{11.39}
\\
\hline

\rowcolor{lightpink}GA
& \decD{1.95}{31.95} & \dec{4.04}{15.91} & 19.70
& \decD{31.82}{15.58} & \decD{17.88}{20.49} & \decD{18.31}{12.94}
& \dec{14.90}{5.24} & \dec{5.28}{6.96} & \dec{6.22}{5.17}
\\

\rowcolor{lightpink}GD
& \decD{5.06}{28.84} & \inc{\textbf{35.76}}{15.81} & 33.91
& \decD{44.90}{2.50} & \decD{35.79}{2.58} & \decD{28.44}{2.81}
& \inc{20.24}{0.10} & \inc{12.79}{0.55} & \inc{11.47}{0.08}
\\

\rowcolor{lightblue}KL
& \decD{3.62}{30.28} & \dec{14.40}{5.55} & 29.96
& \decD{38.50}{8.90} & \decD{26.49}{11.88} & \decD{23.44}{7.81}
& \dec{17.93}{2.21} & \dec{8.66}{3.58} & \dec{9.15}{2.24}
\\

\rowcolor{lightyellow}RLFT
& \decD{\textbf{0.07}}{33.83} & \dec{0.03}{19.92} & 15.79
& \decD{\textbf{10.04}}{37.36} & \decD{\textbf{2.01}}{36.36} & \decD{\textbf{1.67}}{29.58}
& \dec{4.81}{15.33} & \dec{1.74}{10.50} & \dec{0.17}{11.22}
\\

\rowcolor{lightgreen}DPO
& \decD{10.24}{23.66} & \inc{28.63}{8.68} & \textbf{36.14}
& \decD{44.51}{2.89} & \decD{34.91}{3.46} & \decD{28.21}{3.04}
& \dec{19.93}{0.21} & \dec{12.15}{0.09} & \dec{11.15}{0.24}
\\

\rowcolor{lightgreen}NPO
& \decD{12.86}{21.04} & \inc{28.11}{8.16} & 34.93
& \decD{45.98}{1.42} & \decD{36.66}{1.71} & \decD{29.78}{1.47}
& \dec{19.96}{0.18} & \dec{12.19}{0.05} & \dec{11.04}{0.35}
\\

\rowcolor{lightgreen}SimNPO
& \decD{1.42}{32.48} & \dec{0.59}{19.36} & 18.24
& \decD{28.61}{18.79} & \decD{14.32}{24.05} & \decD{15.47}{15.78}
& \dec{13.22}{6.92} & \dec{3.99}{8.25} & \dec{4.99}{6.40}
\\

\rowcolor{lightbeige}PROD
& \decD{12.72}{21.18} & \dec{15.48}{4.47} & 30.06
& \decD{37.77}{9.63} & \decD{26.74}{11.63} & \decD{21.88}{9.37}
& \dec{17.34}{2.80} & \dec{8.71}{3.53} & \dec{8.77}{2.62}
\\
\rowcolor{lightpurple}\textsc{AdaLoRA-L}
& \decD{0.66}{33.24} & \inc{\textbf{86.20}}{66.25} & 5.24
& \decD{42.36}{5.04} & \decD{30.72}{7.65} & \decD{22.69}{8.56}
& \inc{\textbf{21.79}}{1.65} & \inc{\textbf{17.02}}{4.78} & \inc{\textbf{14.87}}{3.48}
\\
\hline


\rowcolor{lightgray}
\multicolumn{10}{c}{StarCoder2}
\\
\hline

Original model
& \ori{28.14} & \ori{14.49} & /
& \ori{38.75} & \ori{27.62} & \ori{22.97}
& \ori{19.41} & \ori{11.46} & \ori{11.88}
\\
\hline

\rowcolor{lightpink}GA
& \decD{11.15}{16.99} & \dec{13.75}{0.74} & 45.65
& \decD{34.31}{4.44} & \decD{23.91}{3.71} & \decD{19.29}{3.68}
& \dec{15.75}{3.66} & \dec{8.92}{2.54} & \dec{9.23}{2.65}
\\

\rowcolor{lightpink}GD
& \decD{3.77}{24.37} & \inc{15.14}{0.65} & \textbf{47.39}
& \decD{34.60}{4.15} & \decD{23.28}{4.34} & \decD{20.07}{2.90}
& \dec{16.57}{2.84} & \dec{8.68}{2.78} & \dec{9.88}{2.00}
\\

\rowcolor{lightblue}KL
& \decD{13.09}{15.05} & \inc{15.03}{0.54} & 46.35
& \decD{37.12}{1.63} & \decD{26.28}{1.34} & \decD{21.38}{1.59}
& \dec{17.27}{2.14} & \dec{9.79}{1.67} & \dec{10.48}{1.40}
\\

\rowcolor{lightyellow}RLFT
& \decD{\textbf{0.34}}{27.80} & \dec{0.44}{14.05} & 37.20
& \decD{\textbf{12.04}}{26.71} & \decD{\textbf{4.08}}{23.54} & \decD{\textbf{1.69}}{21.28}
& \dec{4.56}{14.85} & \dec{1.78}{9.68} & \dec{0.20}{11.68}
\\

\rowcolor{lightgreen}DPO
& \decD{13.05}{15.09} & \inc{15.39}{0.90} & 46.04
& \decD{37.01}{1.74} & \decD{26.16}{1.46} & \decD{21.59}{1.38}
& \dec{17.28}{2.13} & \dec{9.91}{1.55} & \dec{10.32}{1.56}
\\

\rowcolor{lightgreen}NPO
& \decD{15.03}{13.11} & \inc{17.93}{3.44} & 46.32
& \incD{39.75}{1.00} & \incD{28.94}{1.32} & \incD{23.69}{0.72}
& \dec{\textbf{18.46}}{0.95} & \dec{\textbf{10.96}}{0.50} & \dec{\textbf{11.38}}{0.50}
\\

\rowcolor{lightgreen}SimNPO
& \decD{6.14}{22.00} & \dec{6.72}{7.77} & 42.90
& \decD{22.90}{15.85} & \decD{13.55}{14.07} & \decD{9.64}{13.33}
& \dec{9.93}{9.48} & \dec{5.21}{6.25} & \dec{4.70}{7.18}
\\

\rowcolor{lightbeige}PROD
& \decD{9.59}{18.55} & \dec{13.92}{0.57} & 45.35
& \decD{34.09}{4.66} & \decD{22.80}{4.82} & \decD{18.79}{4.18}
& \dec{16.23}{3.18} & \dec{8.71}{2.75} & \dec{9.41}{2.47}
\\

\rowcolor{lightpurple}\textsc{AdaLoRA-L}
& \decD{0.32}{27.82} & \inc{\textbf{35.77}}{21.28} & 4.87
& \decD{24.83}{13.92} & \decD{15.59}{12.03} & \decD{12.15}{10.82}
& \dec{13.06}{6.35} & \dec{8.91}{2.55} & \dec{7.62}{4.26}
\\
\hline


\rowcolor{lightgray}
\multicolumn{10}{c}{DeepSeek-Coder}
\\
\hline

Original model
& \ori{30.34} & \ori{16.74} & /
& \ori{46.17} & \ori{36.19} & \ori{27.95}
& \ori{22.66} & \ori{13.40} & \ori{11.74}
\\
\hline

\rowcolor{lightpink}GA
& \decD{4.36}{25.98} & \inc{19.06}{2.32} & 17.56
& \decD{40.39}{5.78} & \decD{28.61}{7.58} & \decD{22.57}{5.38}
& \dec{20.58}{2.08} & \dec{10.78}{2.62} & \dec{9.01}{2.73}
\\

\rowcolor{lightpink}GD
& \decD{2.30}{28.04} & \inc{19.59}{2.85} & 20.93
& \decD{42.54}{3.63} & \decD{31.44}{4.75} & \decD{25.60}{2.35}
& \dec{22.27}{0.39} & \dec{12.57}{0.83} & \dec{11.29}{0.45}
\\

\rowcolor{lightblue}KL
& \decD{8.89}{21.45} & \inc{24.15}{7.41} & 22.06
& \decD{44.30}{1.87} & \decD{33.99}{2.20} & \decD{25.78}{2.17}
& \dec{22.10}{0.56} & \dec{12.76}{0.64} & \dec{10.82}{0.92}
\\

\rowcolor{lightyellow}RLFT
& \decD{\textbf{0.36}}{29.98} & \dec{0.38}{16.36} & 16.86
& \decD{\textbf{18.09}}{28.08} & \decD{\textbf{6.06}}{30.13} & \decD{\textbf{6.58}}{21.37}
& \dec{9.05}{13.61} & \dec{2.28}{11.12} & \dec{1.41}{10.33}
\\

\rowcolor{lightgreen}DPO
& \decD{9.82}{20.52} & \inc{25.70}{8.96} & \textbf{27.93}
& \decD{45.64}{0.53} & \decD{35.24}{0.95} & \decD{26.92}{1.03}
& \dec{22.58}{0.08} & \inc{13.51}{0.11} & \dec{11.49}{0.25}
\\

\rowcolor{lightgreen}NPO
& \decD{13.09}{17.25} & \inc{25.40}{8.66} & 26.42
& \incD{47.15}{0.98} & \incD{37.20}{1.01} & \incD{28.15}{0.20}
& \inc{22.88}{0.22} & \inc{13.92}{0.52} & \inc{11.92}{0.18}
\\

\rowcolor{lightgreen}SimNPO
& \decD{3.55}{26.79} & \dec{15.99}{0.75} & 15.49
& \decD{38.98}{7.19} & \decD{26.77}{9.42} & \decD{21.10}{6.85}
& \dec{20.02}{2.64} & \dec{10.08}{3.32} & \dec{8.28}{3.46}
\\

\rowcolor{lightbeige}PROD
& \decD{3.59}{26.75} & \dec{4.80}{11.94} & 22.32
& \decD{31.65}{14.52} & \decD{18.26}{17.93} & \decD{16.08}{11.87}
& \dec{16.25}{6.41} & \dec{5.37}{8.03} & \dec{7.10}{4.64}
\\
\rowcolor{lightpurple}\textsc{AdaLoRA-L}
& \decD{0.52}{29.82} & \inc{\textbf{77.35}}{60.61} & 4.56
& \decD{43.93}{2.24} & \decD{32.85}{3.34} & \decD{23.69}{4.26}
& \inc{\textbf{24.51}}{1.85} & \inc{\textbf{17.91}}{4.51} & \inc{\textbf{15.73}}{3.99}
\\
\hline

\end{tabular}

\vspace{-0.1cm}
\end{table}

%% file: table/RQ1-S.tex
\begin{table}[t] \centering \normalsize \setlength{\tabcolsep}{2.2pt} \caption{The performance of eight machine unlearning methods and \textsc{AdaLoRA-L} on Specificity data across three models. Here, BL and RL refer to BLEU and ROUGE-L, respectively. } \label{tab:unlearning-specificity-all} \begin{tabular}{c|ccccc|ccccc|ccccc} \hline Methods & \multicolumn{5}{c|}{Qwen2.5-Coder} & \multicolumn{5}{c|}{StarCoder2} & \multicolumn{5}{c}{DeepSeek-Coder} \\ \cline{2-16} & AEM$_{\scriptscriptstyle\uparrow}$ & ES$_{\scriptscriptstyle\uparrow}$ & BL$_{\scriptscriptstyle\uparrow}$ & RL$_{\scriptscriptstyle\uparrow}$ & P@1$_{\scriptscriptstyle\uparrow}$ & AEM$_{\scriptscriptstyle\uparrow}$ & ES$_{\scriptscriptstyle\uparrow}$ & BL$_{\scriptscriptstyle\uparrow}$ & RL$_{\scriptscriptstyle\uparrow}$ & P@1$_{\scriptscriptstyle\uparrow}$ & AEM$_{\scriptscriptstyle\uparrow}$ & ES$_{\scriptscriptstyle\uparrow}$ & BL$_{\scriptscriptstyle\uparrow}$ & RL$_{\scriptscriptstyle\uparrow}$ & P@1$_{\scriptscriptstyle\uparrow}$ \\ \hline Original model & / & / & / & / & 51.83 & / & / & / & / & 31.71 & / & / & / & / & 57.32 \\ \hline \rowcolor{lightpink}GA & 53.44 & 54.93 & 43.88 & 44.01 & 42.68 & 84.90 & 86.59 & 82.31 & 82.49 & 29.27 & 57.01 & 60.26 & 49.74 & 49.95 & 56.71 \\ \rowcolor{lightpink}GD & \textbf{95.24} & \textbf{96.33} & \textbf{95.08} & \textbf{95.04} & 51.22 & 85.16 & 87.15 & 82.98 & 83.49 & 31.10 & 76.45 & 82.77 & 76.80 & 77.11 & 57.93 \\ \rowcolor{lightblue}KL & 68.85 & 71.81 & 63.64 & 63.73 & 50.61 & 90.09 & 91.28 & 88.32 & 88.45 & 28.66 & 68.55 & 72.04 & 63.90 & 64.22 & 58.54 \\ \rowcolor{lightyellow}RLFT & 50.04 & 15.76 & 8.13 & 8.59 & 0.61 & 45.27 & 14.63 & 10.37 & 11.62 & 3.66 & 50.23 & 43.44 & 34.14 & 34.93 & 14.63 \\ \rowcolor{lightgreen}DPO & 92.46 & 93.79 & 91.87 & 91.93 & 46.95 & 92.39 & 93.18 & 90.87 & 90.85 & 28.05 & \textbf{88.32} & \textbf{90.24} & \textbf{87.03} & \textbf{87.02} & 55.49 \\ \rowcolor{lightgreen}NPO & 91.63 & 92.96 & 90.74 & 90.75 & \textbf{53.05} & \textbf{99.66} & \textbf{99.66} & \textbf{99.54} & \textbf{99.50} & \textbf{31.71} & 87.93 & 89.96 & 86.77 & 86.66 & \textbf{59.76} \\ \rowcolor{lightgreen}SimNPO & 48.75 & 49.62 & 38.37 & 38.79 & 41.46 & 80.27 & 82.53 & 77.12 & 77.15 & 23.78 & 54.87 & 57.49 & 46.47 & 46.55 & 51.22 \\ \rowcolor{lightbeige}PROD & 76.05 & 78.13 & 71.52 & 71.28 & 45.73 & 79.65 & 82.08 & 76.64 & 77.31 & 29.88 & 58.22 & 62.91 & 52.67 & 53.93 & 49.39 \\ 
\rowcolor{lightpurple}\textsc{AdaLoRA-L}& 42.99& 43.81& 33.81& 37.90& 4.26 & 56.56& 58.65& 50.15& 53.59& 7.31 & 43.27& 45.07& 35.32& 39.72& 0.00\\
\hline \end{tabular}  \end{table}

%% file: table/RQ2.tex
\begin{table}[t]
\centering
\small
\setlength{\tabcolsep}{0pt}

\definecolor{darkgreen}{RGB}{0,120,0}

\newcommand{\ori}[1]{%
\makebox[38pt][l]{%
\makebox[20pt][r]{#1}%
\hspace{2pt}%
\phantom{{\fontsize{5pt}{5pt}\selectfont
\(\downarrow00.00\)}}%
}%
}

\newcommand{\metric}[1]{%
\makebox[38pt][l]{#1}%
}


\newcommand{\dec}[2]{%
\makebox[38pt][l]{%
\makebox[20pt][r]{#1}%
\hspace{2pt}%
{\fontsize{5pt}{5pt}\selectfont
\textcolor{red}{\(\downarrow#2\)}}%
}%
}

\newcommand{\inc}[2]{%
\makebox[38pt][l]{%
\makebox[20pt][r]{#1}%
\hspace{2pt}%
{\fontsize{5pt}{5pt}\selectfont
\textcolor{darkgreen}{\(\uparrow#2\)}}%
}%
}


\newcommand{\decD}[2]{%
\makebox[38pt][l]{%
\makebox[20pt][r]{#1}%
\hspace{2pt}%
{\fontsize{5pt}{5pt}\selectfont
\textcolor{darkgreen}{\(\downarrow#2\)}}%
}%
}

\newcommand{\incD}[2]{%
\makebox[38pt][l]{%
\makebox[20pt][r]{#1}%
\hspace{2pt}%
{\fontsize{5pt}{5pt}\selectfont
\textcolor{red}{\(\uparrow#2\)}}%
}%
}

\caption{The AEM-D and AEM-U performance of different machine unlearning methods on Forget data across four code libraries and three code language models. Here, PT, SP, SK, and TF refer to PyTorch, SciPy, scikit-learn, and TensorFlow, respectively.}
\label{tab:unlearning-forget-library}

\vspace{-0.4cm}

\begin{tabular}{c|ccc|ccc|ccc|ccc}
\hline

\multirow{2}{*}{Methods}
& \multicolumn{3}{c|}{PT}
& \multicolumn{3}{c|}{SP}
& \multicolumn{3}{c|}{SK}
& \multicolumn{3}{c}{TF}
\\
\cline{2-13}

& \metric{AEM-D$_{\scriptscriptstyle\downarrow}$}
& \metric{AEM-U$_{\scriptscriptstyle\uparrow}$}
& OMAR
& \metric{AEM-D$_{\scriptscriptstyle\downarrow}$}
& \metric{AEM-U$_{\scriptscriptstyle\uparrow}$}
& OMAR
& \metric{AEM-D$_{\scriptscriptstyle\downarrow}$}
& \metric{AEM-U$_{\scriptscriptstyle\uparrow}$}
& OMAR
& \metric{AEM-D$_{\scriptscriptstyle\downarrow}$}
& \metric{AEM-U$_{\scriptscriptstyle\uparrow}$}
& OMAR
\\

\hline


\rowcolor{lightgray}
\multicolumn{13}{c}{Qwen2.5-Coder}
\\
\hline

Original model
& \ori{34.27} & \ori{10.73} & /
& \ori{31.73} & \ori{3.73} & /
& \ori{29.87} & \ori{27.79} & /
& \ori{33.55} & \ori{33.52} & /
\\
\hline

\rowcolor{lightpink}GA
& \decD{2.44}{31.83} & \dec{1.11}{9.62} & 9.60
& \decD{7.47}{24.26} & \dec{2.93}{0.80} & 23.48
& \decD{12.21}{17.66} & \dec{19.22}{8.57} & 15.98
& \decD{0.45}{33.10} & \dec{6.98}{26.54} & 12.44
\\

\rowcolor{lightpink}GD
& \decD{6.28}{27.99} & \inc{\textbf{25.03}}{14.30} & \textbf{41.94}
& \decD{19.47}{12.26} & \inc{\textbf{4.27}}{0.54} & \textbf{47.15}
& \decD{11.69}{18.18} & \inc{37.40}{9.61} & \textbf{35.22}
& \decD{2.43}{31.12} & \inc{\textbf{53.87}}{20.35} & 41.88
\\

\rowcolor{lightblue}KL
& \decD{4.66}{29.61} & \dec{6.05}{4.68} & 27.56
& \decD{15.47}{16.26} & \dec{2.13}{1.60} & 35.19
& \decD{13.25}{16.62} & \inc{29.87}{2.08} & 24.05
& \decD{1.39}{32.16} & \dec{24.07}{9.45} & 35.3
\\

\rowcolor{lightyellow}RLFT
& \decD{\textbf{0.01}}{34.26} & \dec{0.00}{10.73} & 0.01
& \decD{\textbf{0.00}}{31.73} & \dec{0.27}{3.46} & 0.36
& \decD{\textbf{0.52}}{29.35} & \dec{0.52}{27.27} & 0.00
& \decD{\textbf{0.04}}{33.51} & \dec{0.09}{33.43} & 0.17
\\

\rowcolor{lightgreen}DPO
& \decD{14.38}{19.89} & \inc{16.07}{5.34} & 41.08
& \decD{20.80}{10.93} & \dec{2.93}{0.80} & 43.91
& \decD{10.91}{18.96} & \inc{35.58}{7.79} & 28.65
& \decD{4.57}{28.98} & \inc{46.48}{12.96} & \textbf{44.47}
\\

\rowcolor{lightgreen}NPO
& \decD{17.90}{16.37} & \inc{14.08}{3.35} & 41.77
& \decD{19.73}{12.00} & \dec{1.87}{1.86} & 45.85
& \decD{11.43}{18.44} & \inc{\textbf{38.96}}{11.17} & 34.28
& \decD{6.09}{27.46} & \inc{48.66}{15.14} & 41.64
\\

\rowcolor{lightgreen}SimNPO
& \decD{1.59}{32.68} & \dec{0.32}{10.41} & 6.00
& \decD{7.20}{24.53} & \dec{1.87}{1.86} & 15.41
& \decD{13.25}{16.62} & \dec{10.39}{17.40} & 13.07
& \decD{0.32}{33.23} & \dec{0.43}{33.09} & 6.02
\\

\rowcolor{lightbeige}PROD
& \decD{13.97}{20.30} & \dec{7.54}{3.19} & 33.19
& \decD{12.53}{19.20} & \dec{1.60}{2.13} & 32.61
& \decD{14.55}{15.32} & \dec{25.71}{2.08} & 33.23
& \decD{11.66}{21.89} & \dec{25.59}{7.93} & 38.58
\\

\hline


\rowcolor{lightgray}
\multicolumn{13}{c}{StarCoder2}
\\
\hline

Original model
& \ori{30.61} & \ori{4.05} & /
& \ori{28.50} & \ori{2.00} & /
& \ori{24.29} & \ori{18.57} & /
& \ori{26.00} & \ori{25.91} & /
\\
\hline

\rowcolor{lightpink}GA
& \decD{13.16}{17.45} & \dec{3.58}{0.47} & 20.68
& \decD{11.50}{17.00} & \dec{1.25}{0.75} & 30.61
& \decD{11.43}{12.86} & \dec{15.71}{2.86} & 15.10
& \decD{8.92}{17.08} & \dec{24.73}{1.18} & 27.63
\\

\rowcolor{lightpink}GD
& \decD{4.49}{26.12} & \inc{\textbf{4.94}}{0.89} & 21.68
& \decD{12.50}{16.00} & \inc{\textbf{3.00}}{1.00} & 32.42
& \decD{8.29}{16.00} & \inc{22.00}{3.43} & 16.63
& \decD{1.46}{24.54} & \inc{28.02}{2.11} & 33.26
\\

\rowcolor{lightblue}KL
& \decD{15.53}{15.08} & \inc{4.26}{0.21} & 23.36
& \decD{14.25}{14.25} & \dec{1.50}{0.50} & 28.53
& \decD{10.86}{13.43} & \inc{20.00}{1.43} & 19.93
& \decD{11.39}{14.61} & \inc{27.20}{1.29} & 31.64
\\

\rowcolor{lightyellow}RLFT
& \decD{\textbf{0.00}}{30.61} & \dec{0.00}{4.05} & 0.06
& \decD{\textbf{0.00}}{28.50} & \dec{0.00}{2.00} & 0
& \decD{\textbf{2.57}}{21.72} & \dec{2.57}{16.00} & 2.94
& \decD{\textbf{0.33}}{25.67} & \dec{0.19}{25.72} & 0.43
\\

\rowcolor{lightgreen}DPO
& \decD{16.34}{14.27} & \inc{4.09}{0.04} & 23.01
& \decD{14.00}{14.50} & \dec{1.75}{0.25} & 27.70
& \decD{13.14}{11.15} & \inc{19.43}{0.86} & 19.18
& \decD{8.99}{17.01} & \inc{26.38}{0.47} & 30.42
\\

\rowcolor{lightgreen}NPO
& \decD{19.41}{11.20} & \inc{4.56}{0.51} & \textbf{25.68}
& \decD{12.75}{15.75} & \dec{1.50}{0.50} & \textbf{38.81}
& \decD{12.29}{12.00} & \inc{\textbf{22.29}}{3.72} & \textbf{26.93}
& \decD{10.45}{15.55} & \inc{\textbf{30.49}}{4.58} & \textbf{37.54}
\\

\rowcolor{lightgreen}SimNPO
& \decD{5.74}{24.87} & \dec{1.65}{2.40} & 9.57
& \decD{7.75}{20.75} & \dec{1.00}{1.00} & 23.20
& \decD{8.29}{16.00} & \dec{11.43}{7.14} & 17.55
& \decD{6.28}{19.72} & \dec{11.62}{14.29} & 15.15
\\

\rowcolor{lightbeige}PROD
& \decD{11.70}{18.91} & \inc{4.41}{0.36} & 18.45
& \decD{9.25}{19.25} & \dec{1.25}{0.75} & 26.51
& \decD{12.86}{11.43} & \dec{17.43}{1.14} & 14.12
& \decD{5.93}{20.07} & \dec{25.27}{0.64} & 30.64
\\

\hline


\rowcolor{lightgray}
\multicolumn{13}{c}{DeepSeek-Coder}
\\
\hline

Original model
& \ori{31.96} & \ori{6.66} & /
& \ori{27.56} & \ori{6.67} & /
& \ori{26.78} & \ori{25.08} & /
& \ori{27.31} & \ori{35.13} & /
\\
\hline

\rowcolor{lightpink}GA
& \decD{6.97}{24.99} & \inc{7.07}{0.41} & 12.24
& \decD{4.44}{23.12} & \dec{1.78}{4.89} & 17.15
& \decD{2.71}{24.07} & \inc{28.81}{3.73} & 6.56
& \decD{0.64}{26.67} & \inc{40.10}{4.97} & 18.64
\\

\rowcolor{lightpink}GD
& \decD{2.36}{29.60} & \inc{7.06}{0.40} & 16.66
& \decD{11.56}{16.00} & \decD{4.89}{1.78} & 26.06
& \decD{5.76}{21.02} & \inc{\textbf{32.88}}{7.80} & 17.62
& \decD{0.36}{26.95} & \inc{44.18}{9.05} & 23.37
\\

\rowcolor{lightblue}KL
& \decD{13.66}{18.30} & \inc{11.01}{4.35} & 22.43
& \decD{13.78}{13.78} & \dec{2.22}{4.45} & 26.10
& \decD{5.42}{21.36} & \inc{30.51}{5.43} & 17.27
& \decD{0.92}{26.39} & \inc{50.06}{14.93} & 21.21
\\

\rowcolor{lightyellow}RLFT
& \decD{\textbf{0.17}}{31.79} & \dec{0.01}{6.65} & 0.49
& \decD{\textbf{0.44}}{27.12} & \dec{0.00}{6.67} & 7.72
& \decD{\textbf{1.02}}{25.76} & \dec{3.05}{22.03} & 2.84
& \decD{\textbf{0.76}}{26.55} & \dec{0.38}{34.75} & 2.13
\\

\rowcolor{lightgreen}DPO
& \decD{14.39}{17.57} & \inc{\textbf{13.28}}{6.62} & \textbf{27.21}
& \decD{21.33}{6.23} & \dec{4.44}{2.23} & 27.93
& \decD{9.83}{16.95} & \inc{26.44}{1.36} & 17.98
& \decD{2.09}{25.22} & \inc{51.92}{16.79} & \textbf{27.25}
\\

\rowcolor{lightgreen}NPO
& \decD{20.31}{11.65} & \inc{11.76}{5.10} & 26.94
& \decD{21.33}{6.23} & \dec{1.78}{4.89} & \textbf{29.15}
& \decD{12.88}{13.90} & \inc{28.81}{3.73} & \textbf{20.82}
& \decD{2.24}{25.07} & \inc{\textbf{55.06}}{19.93} & 25.30
\\

\rowcolor{lightgreen}SimNPO
& \decD{5.59}{26.37} & \dec{5.33}{1.33} & 10.73
& \decD{4.44}{23.12} & \dec{2.67}{4.00} & 15.53
& \decD{2.71}{24.07} & \dec{22.71}{2.37} & 6.03
& \decD{0.33}{26.98} & \dec{34.24}{0.89} & 16.57
\\

\rowcolor{lightbeige}PROD
& \decD{4.12}{27.84} & \dec{1.44}{5.22} & 10.92
& \decD{8.89}{18.67} & \dec{0.89}{5.78} & 17.56
& \decD{6.44}{20.34} & \dec{7.80}{17.28} & 10.46
& \decD{3.18}{24.13} & \dec{11.49}{23.64} & 16.61
\\

\hline

\end{tabular}

\vspace{-0.6cm}
\end{table}

\begin{table}[t]
\centering
\small
\setlength{\tabcolsep}{0pt}

\definecolor{darkgreen}{RGB}{0,120,0}

\newcommand{\ori}[1]{%
\makebox[38pt][l]{%
\makebox[20pt][r]{#1}%
\hspace{2pt}%
\phantom{{\fontsize{5pt}{5pt}\selectfont
\(\downarrow00.00\)}}%
}%
}

\newcommand{\metric}[1]{%
\makebox[38pt][l]{#1}%
}

\newcommand{\decD}[2]{%
\makebox[38pt][l]{%
\makebox[20pt][r]{#1}%
\hspace{2pt}%
{\fontsize{5pt}{5pt}\selectfont
\textcolor{darkgreen}{\(\downarrow#2\)}}%
}%
}

\newcommand{\incD}[2]{%
\makebox[38pt][l]{%
\makebox[20pt][r]{#1}%
\hspace{2pt}%
{\fontsize{5pt}{5pt}\selectfont
\textcolor{red}{\(\uparrow#2\)}}%
}%
}

\newcommand{\decU}[2]{%
\makebox[38pt][l]{%
\makebox[20pt][r]{#1}%
\hspace{2pt}%
{\fontsize{5pt}{5pt}\selectfont
\textcolor{red}{\(\downarrow#2\)}}%
}%
}

\newcommand{\incU}[2]{%
\makebox[38pt][l]{%
\makebox[20pt][r]{#1}%
\hspace{2pt}%
{\fontsize{5pt}{5pt}\selectfont
\textcolor{darkgreen}{\(\uparrow#2\)}}%
}%
}

\caption{The AEM-D, AEM-U, and OMAR performance of different machine unlearning methods on Generalization data across four code libraries and three code language models. Here, PT, SP, SK, and TF refer to PyTorch, SciPy, scikit-learn, and TensorFlow, respectively.}
\label{tab:unlearning-gen-all-lib}

\vspace{-0.4cm}

\begin{tabular}{c|ccc|ccc|ccc|ccc}
\hline

\multirow{2}{*}{Methods}
& \multicolumn{3}{c|}{PT}
& \multicolumn{3}{c|}{SP}
& \multicolumn{3}{c|}{SK}
& \multicolumn{3}{c}{TF}
\\
\cline{2-13}

& \metric{AEM-D$_{\scriptscriptstyle\downarrow}$}
& \metric{AEM-U$_{\scriptscriptstyle\uparrow}$}
& OMAR
& \metric{AEM-D$_{\scriptscriptstyle\downarrow}$}
& \metric{AEM-U$_{\scriptscriptstyle\uparrow}$}
& OMAR
& \metric{AEM-D$_{\scriptscriptstyle\downarrow}$}
& \metric{AEM-U$_{\scriptscriptstyle\uparrow}$}
& OMAR
& \metric{AEM-D$_{\scriptscriptstyle\downarrow}$}
& \metric{AEM-U$_{\scriptscriptstyle\uparrow}$}
& OMAR
\\

\hline


\rowcolor{lightgray}
\multicolumn{13}{c}{Qwen2.5-Coder}
\\
\hline

Original model
& \ori{34.11} & \ori{11.05} & /
& \ori{35.65} & \ori{4.06} & /
& \ori{31.08} & \ori{24.59} & /
& \ori{34.40} & \ori{33.78} & /
\\
\hline

\rowcolor{lightpink}GA
& \decD{1.97}{32.14} & \decU{1.55}{9.50} & 8.90
& \decD{8.70}{26.95} & \decU{2.32}{1.74} & 14.18
& \decD{16.76}{14.32} & \decU{11.08}{13.51} & 19.10
& \decD{0.47}{33.93} & \decU{7.39}{26.39} & 11.30
\\

\rowcolor{lightpink}GD
& \decD{5.46}{28.65} & \incU{\textbf{24.67}}{13.62} & 40.12
& \decD{21.16}{14.49} & \incU{\textbf{6.38}}{2.32} & \textbf{36.99}
& \decD{18.11}{12.97} & \incU{\textbf{32.97}}{8.38} & 33.01
& \decD{2.40}{32.00} & \incU{\textbf{55.07}}{21.29} & 38.92
\\

\rowcolor{lightblue}KL
& \decD{3.86}{30.25} & \decU{6.78}{4.27} & 26.68
& \decD{15.94}{19.71} & \decU{2.90}{1.16} & 26.09
& \decD{18.65}{12.43} & \decU{22.43}{2.16} & 31.84
& \decD{1.27}{33.13} & \decU{25.97}{7.81} & 32.67
\\

\rowcolor{lightyellow}RLFT
& \decD{\textbf{0.01}}{34.10} & \decU{0.01}{1.04} & 0
& \decD{\textbf{0.00}}{35.65} & \decU{0.00}{4.06} & 0
& \decD{\textbf{0.27}}{30.81} & \decU{0.27}{24.32} & 2.19
& \decD{\textbf{0.11}}{34.29} & \decU{0.04}{33.74} & 0.50
\\

\rowcolor{lightgreen}DPO
& \decD{13.20}{20.91} & \incU{16.65}{5.60} & \textbf{41.07}
& \decD{23.48}{12.17} & \incU{\textbf{6.38}}{2.32} & 32.57
& \decD{17.84}{13.24} & \incU{31.62}{7.03} & 34.62
& \decD{4.56}{29.84} & \incU{47.91}{14.13} & \textbf{41.59}
\\

\rowcolor{lightgreen}NPO
& \decD{17.11}{17.00} & \incU{14.31}{3.26} & 40.40
& \decD{23.19}{12.46} & \incU{4.93}{0.87} & 34.21
& \decD{21.62}{9.46} & \incU{32.70}{8.11} & \textbf{40.32}
& \decD{5.79}{28.61} & \incU{50.15}{16.37} & 41.03
\\

\rowcolor{lightgreen}SimNPO
& \decD{1.35}{32.76} & \decU{0.39}{10.66} & 5.41
& \decD{4.93}{30.72} & \decU{1.45}{2.61} & 11.31
& \decD{15.41}{15.67} & \decU{5.68}{18.91} & 12.72
& \decD{0.27}{34.13} & \decU{0.34}{33.44} & 6.47
\\

\rowcolor{lightbeige}PROD 
& \decD{13.60}{20.51} & \decU{8.24}{2.81} & 31.26
& \decD{12.17}{23.48} & \decU{2.90}{1.16} & 29.84
& \decD{16.22}{14.86} & \decU{14.86}{9.73} & 28.71
& \decD{11.91}{22.49} & \decU{26.91}{6.87} & 34.37
\\

\hline


\rowcolor{lightgray}
\multicolumn{13}{c}{StarCoder2}
\\
\hline

Original model
& \ori{30.06} & \ori{4.85} & /
& \ori{32.88} & \ori{3.84} & /
& \ori{27.71} & \ori{22.00} & / 
& \ori{25.88} & \ori{26.73} & /
\\
\hline

\rowcolor{lightpink}GA
& \decD{12.03}{18.03} & \decU{4.06}{0.79} & 21.21
& \decD{16.16}{16.72} & \decU{2.47}{1.37} & 28.75
& \decD{11.43}{16.28} & \decU{20.29}{1.71} & 18.41
& \decD{9.95}{15.93} & \decU{26.47}{0.26} & 27.55
\\

\rowcolor{lightpink}GD
& \decD{4.06}{26.00} & \decU{4.81}{0.04} & 22.46
& \decD{13.97}{18.91} & \decU{3.29}{0.55} & 32.79
& \decD{11.71}{16.00} & \incU{26.29}{4.29} & \textbf{34.21}
& \decD{1.82}{24.06} & \incU{28.34}{1.61} & 32.72
\\

\rowcolor{lightblue}KL
& \decD{14.33}{15.73} & \decU{4.41}{0.44} & 25.07
& \decD{18.36}{14.52} & \decU{2.74}{1.10} & \textbf{44.11}
& \decD{12.57}{15.14} & \incU{26.00}{4.00} & 31.15
& \decD{11.68}{14.20} & \incU{28.25}{1.52} & 29.15
\\

\rowcolor{lightyellow}RLFT
& \decD{\textbf{0.04}}{30.02} & \decU{0.00}{4.85} & 0.09
& \decD{\textbf{1.10}}{31.78} & \decU{0.00}{3.84} & 0
& \decD{\textbf{0.86}}{26.85} & \decU{2.86}{19.14} & 2.26
& \decD{\textbf{0.64}}{25.24} & \decU{0.85}{25.88} & 0.53
\\

\rowcolor{lightgreen}DPO
& \decD{15.46}{14.60} & \decU{4.54}{0.31} & 24.09
& \decD{17.53}{15.35} & \decU{2.74}{1.10} & 29.40
& \decD{11.43}{16.28} & \incU{25.71}{3.71} & 30.24
& \decD{10.21}{15.67} & \incU{28.93}{2.20} & 29.71
\\

\rowcolor{lightgreen}NPO
& \decD{18.77}{11.29} & \incU{\textbf{5.38}}{0.53} & \textbf{26.46}
& \decD{19.18}{13.70} & \incU{\textbf{4.11}}{0.27} & 35.83
& \decD{12.57}{15.14} & \incU{\textbf{32.29}}{10.29} & 25.79
& \decD{10.83}{15.05} & \incU{\textbf{33.48}}{6.75} & \textbf{35.59}
\\

\rowcolor{lightgreen}SimNPO
& \decD{5.10}{24.96} & \decU{1.49}{3.36} & 10.00
& \decD{15.07}{17.81} & \decU{2.19}{1.65} & 18.78
& \decD{10.86}{16.85} & \decU{15.71}{6.29} & 24.84
& \decD{6.35}{19.53} & \decU{13.01}{13.72} & 16.28
\\

\rowcolor{lightbeige}PROD
& \decD{12.11}{17.95} & \decU{4.25}{0.60} & 19.21
& \decD{17.53}{15.35} & \decU{2.74}{1.10} & 25.63
& \decD{12.00}{15.71} & \incU{23.14}{1.14} & 30.60
& \decD{5.97}{19.91} & \decU{26.23}{0.50} & 30.49
\\

\hline


\rowcolor{lightgray}
\multicolumn{13}{c}{DeepSeek-Coder}
\\
\hline

Original model
& \ori{31.65} & \ori{7.16} & /
& \ori{30.00} & \ori{1.43} & / 
& \ori{30.53} & \ori{32.63} & /
& \ori{27.98} & \ori{35.58} & /
\\
\hline

\rowcolor{lightpink}GA
& \decD{5.98}{25.67} & \incU{7.77}{0.61} & 12.70
& \decD{10.95}{19.05} & \incU{\textbf{2.38}}{0.95} & 19.12
& \decD{7.37}{23.16} & \incU{36.14}{3.51} & 9.87
& \decD{0.72}{27.26} & \incU{42.16}{6.58} & 16.15
\\

\rowcolor{lightpink}GD
& \decD{2.38}{29.27} & \incU{7.41}{0.25} & 17.27
& \decD{16.19}{13.81} & \decU{0.95}{0.48} & \textbf{27.59}
& \decD{8.77}{21.76} & \incU{40.35}{7.72} & 17.54
& \decD{0.54}{27.44} & \incU{42.98}{7.40} & 25.53
\\

\rowcolor{lightblue}KL
& \decD{12.55}{19.10} & \incU{11.49}{4.33} & 22.42
& \decD{19.52}{10.48} & \incU{1.90}{0.47} & 24.82
& \decD{7.72}{22.81} & \incU{34.74}{2.11} & \textbf{17.98}
& \decD{1.39}{26.59} & \incU{50.35}{14.77} & 21.30
\\

\rowcolor{lightyellow}RLFT
& \decD{\textbf{0.21}}{31.44} & \decU{0.02}{7.14} & 0.63
& \decD{\textbf{0.48}}{29.52} & \decU{0.48}{0.95} & 3.51
& \decD{\textbf{1.05}}{29.48} & \decU{7.72}{24.91} & 2.63
& \decD{\textbf{0.64}}{27.34} & \decU{0.51}{35.07} & 1.76
\\

\rowcolor{lightgreen}DPO
& \decD{13.16}{18.49} & \incU{\textbf{13.96}}{6.80} & \textbf{29.09}
& \decD{20.95}{9.05} & \incU{1.90}{0.47} & 26.75
& \decD{12.28}{18.25} & \incU{34.74}{2.11} & \textbf{17.98}
& \decD{2.36}{25.62} & \incU{50.12}{14.54} & \textbf{28.72}
\\

\rowcolor{lightgreen}NPO
& \decD{18.17}{13.48} & \incU{12.25}{5.09} & 28.72
& \decD{23.81}{6.19} & \decU{0.00}{1.43} & 27.02
& \decD{13.68}{16.85} & \incU{\textbf{41.75}}{9.12} & 16.01
& \decD{2.75}{25.23} & \incU{\textbf{52.22}}{16.64} & 26.88
\\

\rowcolor{lightgreen}SimNPO
& \decD{4.82}{26.83} & \decU{5.66}{1.50} & 10.79
& \decD{11.43}{18.57} & \decU{0.00}{1.43} & 17.19
& \decD{7.02}{23.51} & \incU{36.84}{4.21} & 7.46
& \decD{0.46}{27.52} & \incU{36.74}{1.16} & 12.26
\\

\rowcolor{lightbeige}PROD
& \decD{3.73}{27.92} & \decU{1.33}{5.83} & 13.10
& \decD{8.57}{21.43} & \decU{0.48}{0.95} & 15.48
& \decD{6.67}{23.86} & \decU{10.88}{21.75} & 12.50
& \decD{2.88}{25.10} & \decU{10.86}{24.72} & 16.77
\\

\hline

\end{tabular}

\end{table}

\begin{table*}[t]
\centering
\small
\setlength{\tabcolsep}{4pt}

\caption{The AEM performance of different machine unlearning methods on Specificity data across four code libraries and three code language models. Here, PT, SP, SK, and TF refer to PyTorch, SciPy, scikit-learn, and TensorFlow, respectively. }
\label{tab:unlearning-specificity-aem}

\begin{tabular}{c|cccc|cccc|cccc}
\hline

\multirow{2}{*}{Methods}
& \multicolumn{4}{c|}{Qwen2.5-Coder}
& \multicolumn{4}{c|}{StarCoder2}
& \multicolumn{4}{c}{DeepSeek-Coder}
\\

\cline{2-13}

& PT & SP & SK & TF
& PT & SP & SK & TF
& PT & SP & SK & TF
\\

\hline

Original model& 100.00 & 100.00 & 100.00 & 100.00
& 100.00 & 100.00 & 100.00 & 100.00
& 100.00 & 100.00 & 100.00 & 100.00
\\

\hline

\rowcolor{lightpink}
GA
& 52.48 & 61.06 & 59.77 & 53.39
& 83.94 & 86.45 & 82.76 & 85.98
& 56.37 & 66.21 & 61.72 & 57.22
\\

\rowcolor{lightpink}
GD
& \textbf{95.40} & \textbf{97.00} & \textbf{95.33} & \textbf{94.72}
& 83.66 & 89.49 & 85.36 & 86.48
& 75.28 & 83.26 & 84.19 & 77.08
\\

\rowcolor{lightblue}
KL
& 68.51 & 72.96 & 69.52 & 68.42
& 89.46 & 89.81 & 88.09 & 90.97
& 67.87 & 76.26 & 73.36 & 68.75
\\

\rowcolor{lightyellow}
RLFT
& 50.49 & 51.43 & 46.79 & 49.62
& 44.14 & 57.85 & 44.74 & 45.04
& 51.75 & 61.14 & 52.03 & 46.88
\\

\rowcolor{lightgreen}
DPO
& 92.51 & 93.11 & 93.91 & 92.03
& 91.83 & 92.99 & 91.59 & 92.94
& \textbf{88.16} & 88.64 & 88.01 & 88.56
\\

\rowcolor{lightgreen}
NPO
& 91.74 & 93.99 & 92.82 & 91.07
& \textbf{99.71} & \textbf{99.73} & \textbf{98.46} & \textbf{99.67}
& 87.29 & \textbf{89.59} & \textbf{91.10} & \textbf{88.70}
\\

\rowcolor{lightgreen}
SimNPO
& 47.95 & 55.65 & 53.67 & 48.79
& 79.37 & 82.30 & 77.56 & 81.14
& 54.41 & 64.84 & 59.45 & 54.57
\\

\rowcolor{lightbeige}
PROD
& 76.12 & 81.64 & 75.21 & 75.27
& 78.90 & 82.21 & 78.06 & 80.21
& 57.29 & 70.37 & 63.83 & 58.14
\\

\hline

\end{tabular}

\vspace{-0.5cm}
\end{table*}

%% file: Figure/RQ3/RQ3.tex
\begin{figure}[!h]
    \centering
    \begin{minipage}[t]{0.49\textwidth}
        \centering
        \includegraphics[width=0.99\linewidth]{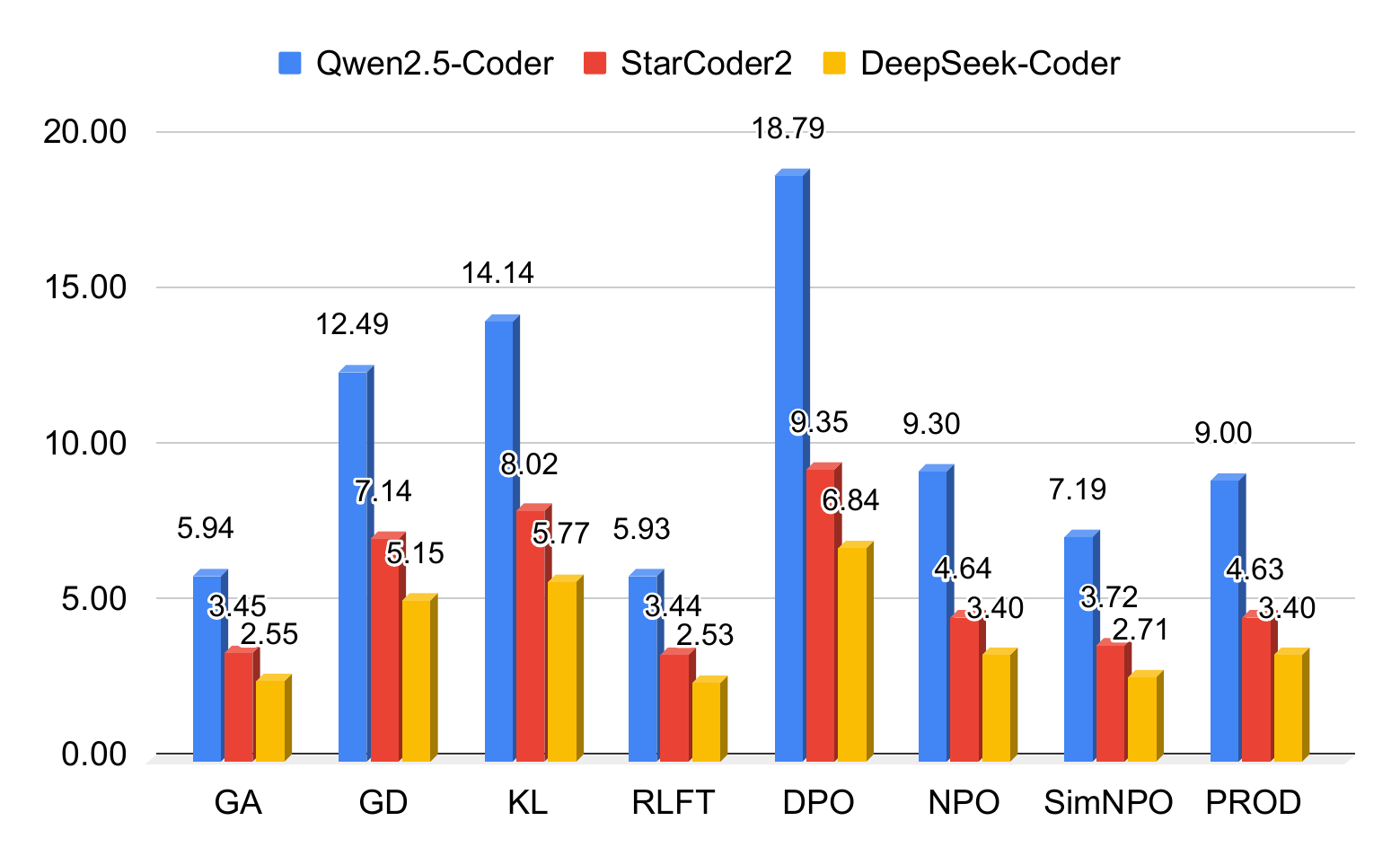}
        \captionsetup{skip=0pt}
        \caption{The average time cost (hours) of the machine unlearning methods.}
        \label{fig:time_cost}
    \end{minipage}
    \begin{minipage}[t]{0.49\textwidth}
        \centering
        \includegraphics[width=0.99\linewidth]{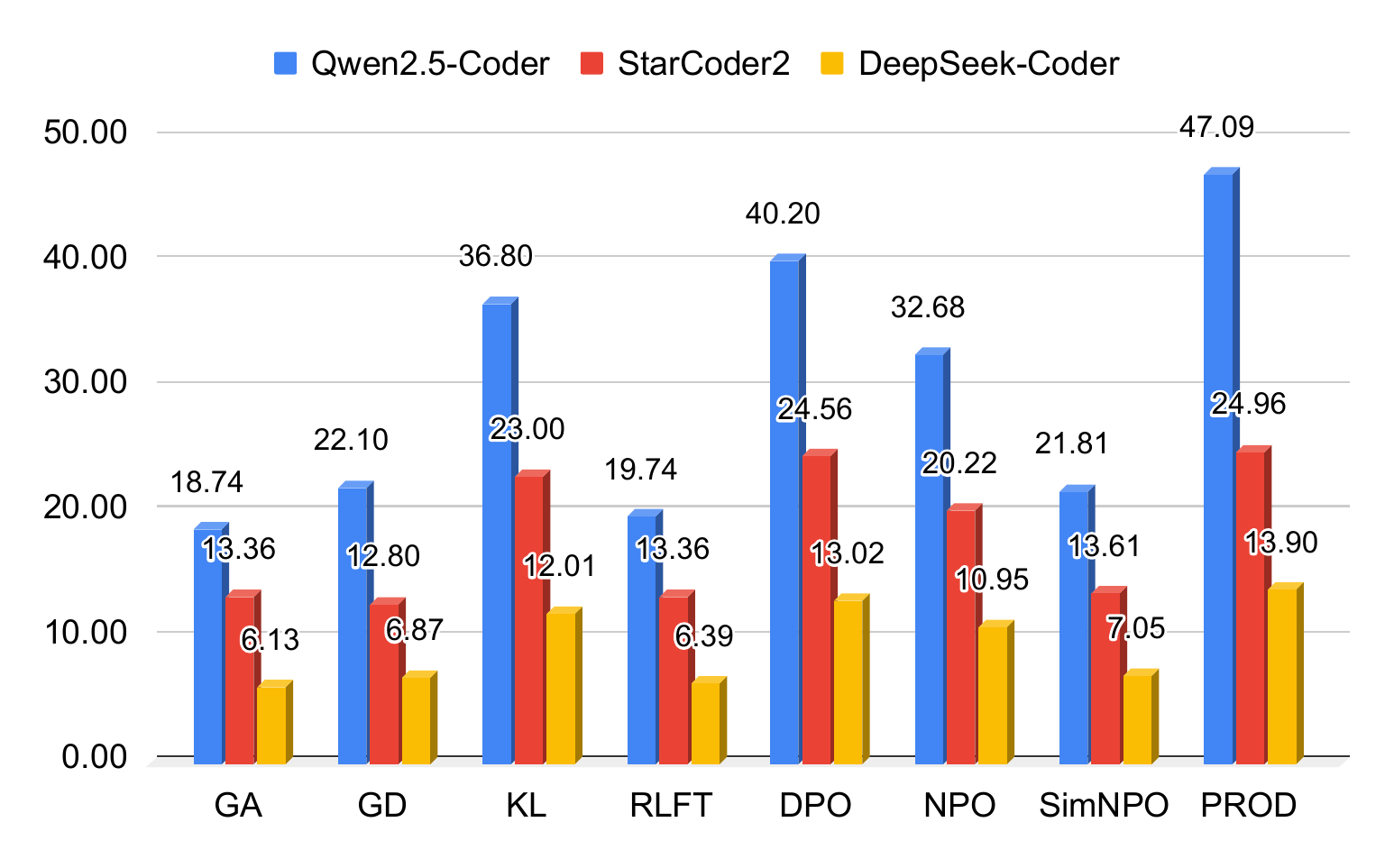}
        \captionsetup{skip=0pt}
        \caption{The average peak memory cost (GB) of the machine unlearning methods.}
        \label{fig:memory_cost}
    \end{minipage}
\vspace{-0.4cm}
\end{figure}

%% file: Figure/RQ4/RQ4.tex
\begin{figure*}[t]
    \centering
    \label{Before-After}

    \begin{subfigure}[t]{0.33\textwidth}
        \centering
        \includegraphics[width=\linewidth]{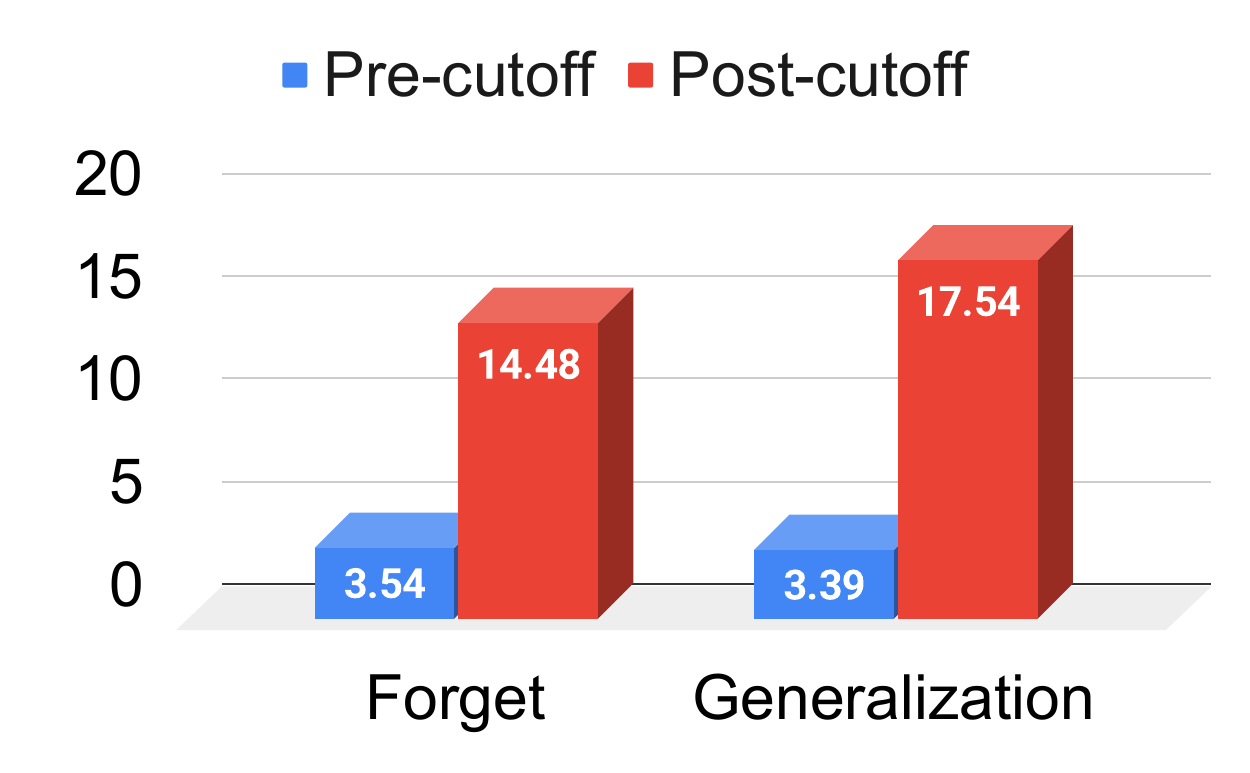}
        \caption{AEM-D on StarCoder2}
    \end{subfigure}
    \hfill
    \begin{subfigure}[t]{0.33\textwidth}
        \centering
        \includegraphics[width=\linewidth]{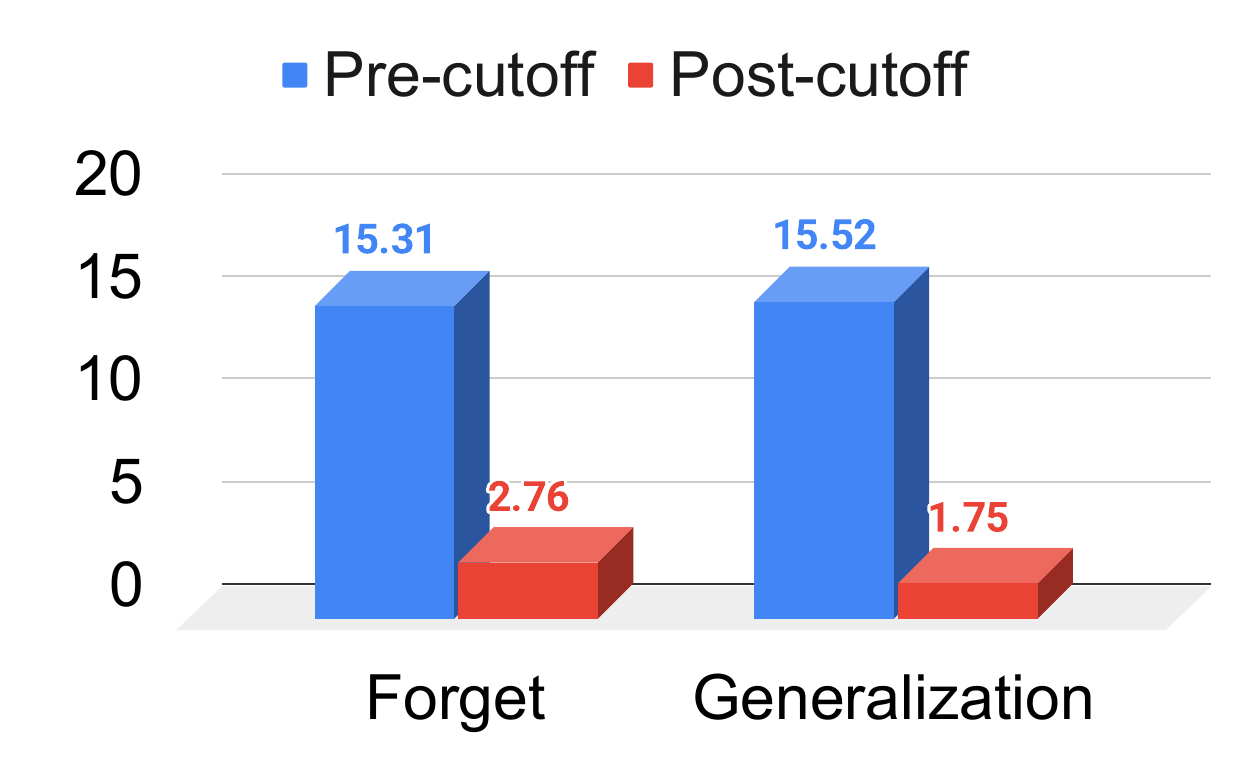}
        \caption{AEM-U on StarCoder2}
    \end{subfigure}
    \hfill
    \begin{subfigure}[t]{0.33\textwidth}
        \centering
        \includegraphics[width=\linewidth]{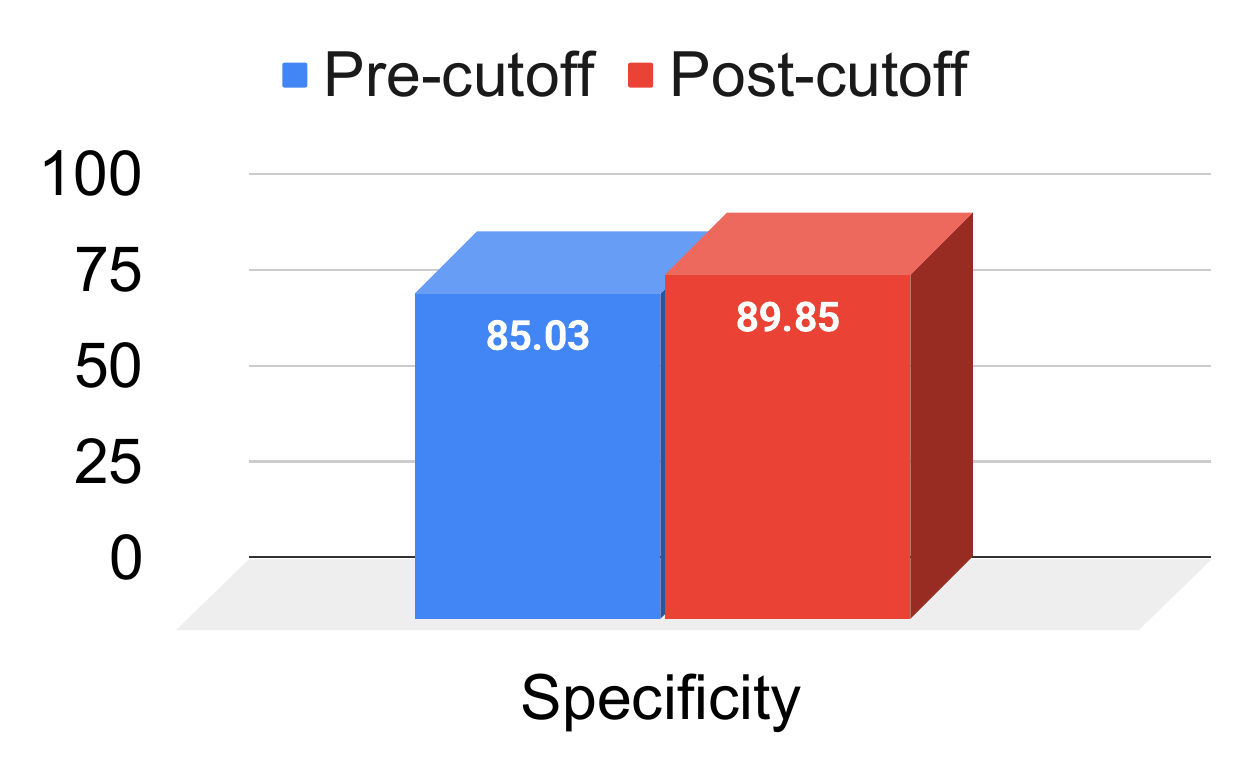}
        \caption{AEM on StarCoder2}
    \end{subfigure}

    \vspace{0.15cm}

    \begin{subfigure}[t]{0.33\textwidth}
        \centering
        \includegraphics[width=\linewidth]{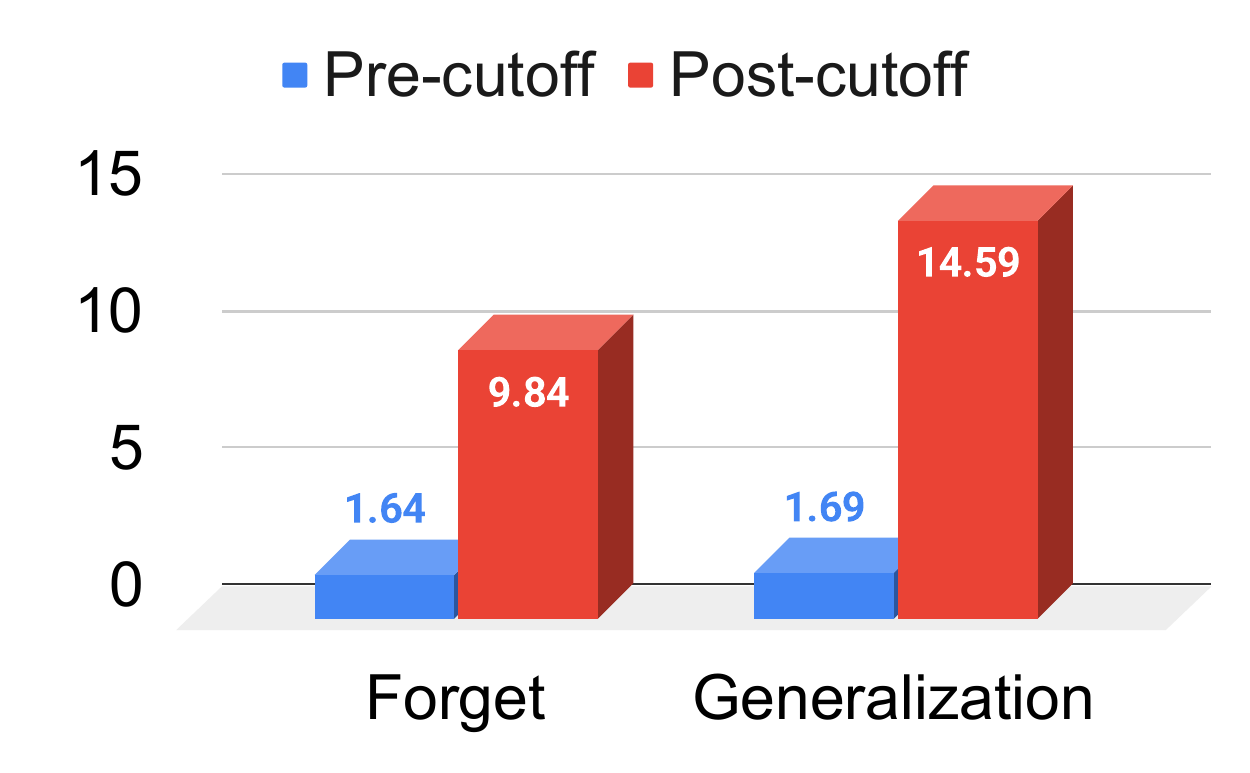}
        \caption{AEM-D on DeepSeek-Coder}
    \end{subfigure}
    \hfill
    \begin{subfigure}[t]{0.33\textwidth}
        \centering
        \includegraphics[width=\linewidth]{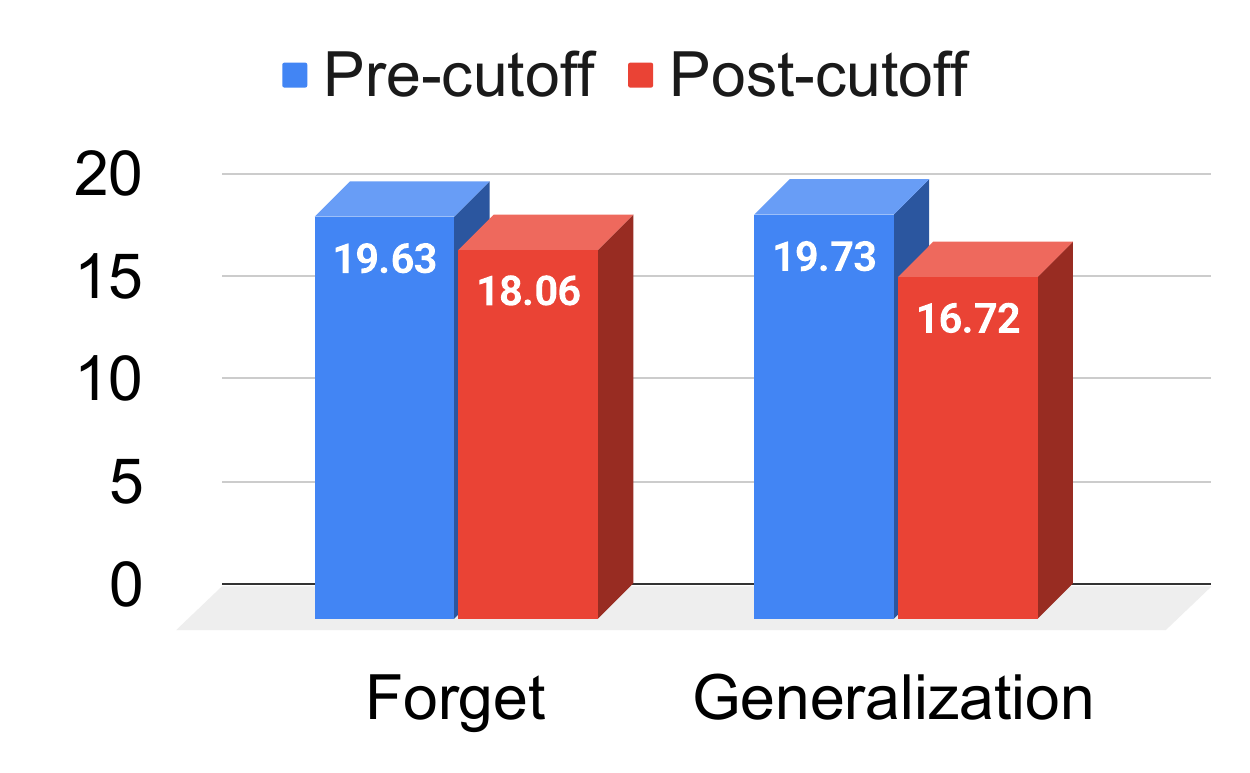}
        \caption{AEM-U on DeepSeek-Coder}
    \end{subfigure}
    \hfill
    \begin{subfigure}[t]{0.33\textwidth}
        \centering
        \includegraphics[width=\linewidth]{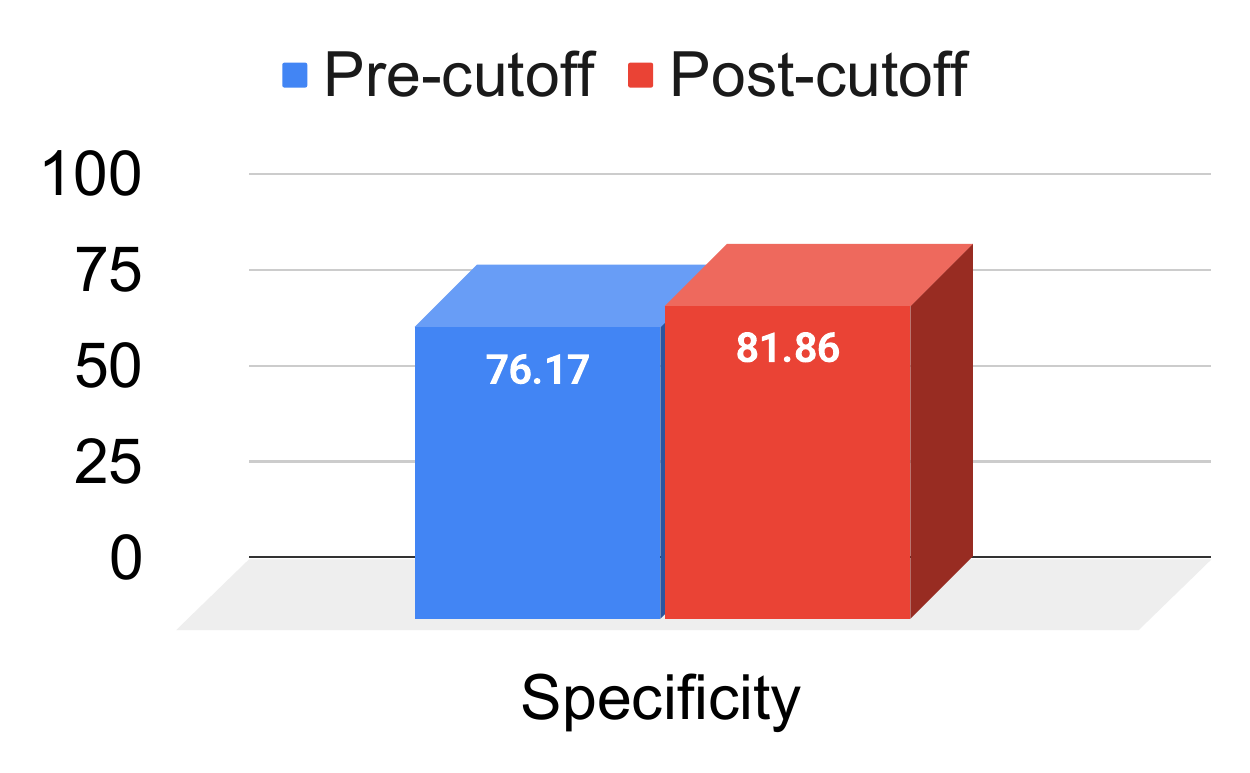}
        \caption{AEM on DeepSeek-Coder}
    \end{subfigure}

    \caption{The comparison of GD performance on APIs deprecated before (\textit{Pre-cutoff}) and after (\textit{Post-cutoff}) each model's training-data cutoff across Forget Data, Generalization Data, and Specificity Data.}
    \label{fig:aem_Before_After}
\end{figure*}

%% file: Figure/RQ5/RQ5.tex
%

\begin{figure*}[t]
    \centering
    \includegraphics[width=0.33\textwidth]{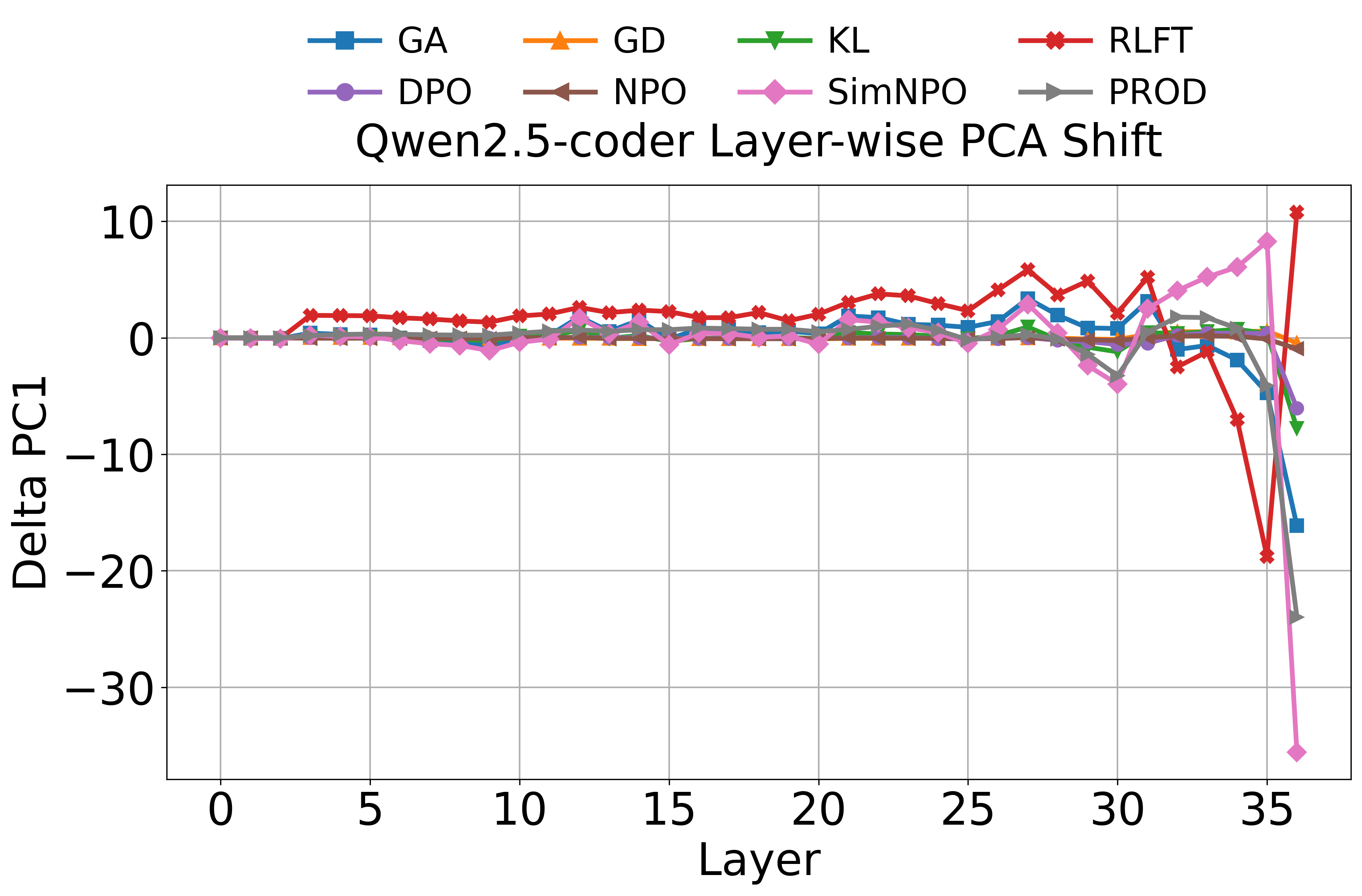}
    \hfill
    \includegraphics[width=0.33\textwidth]{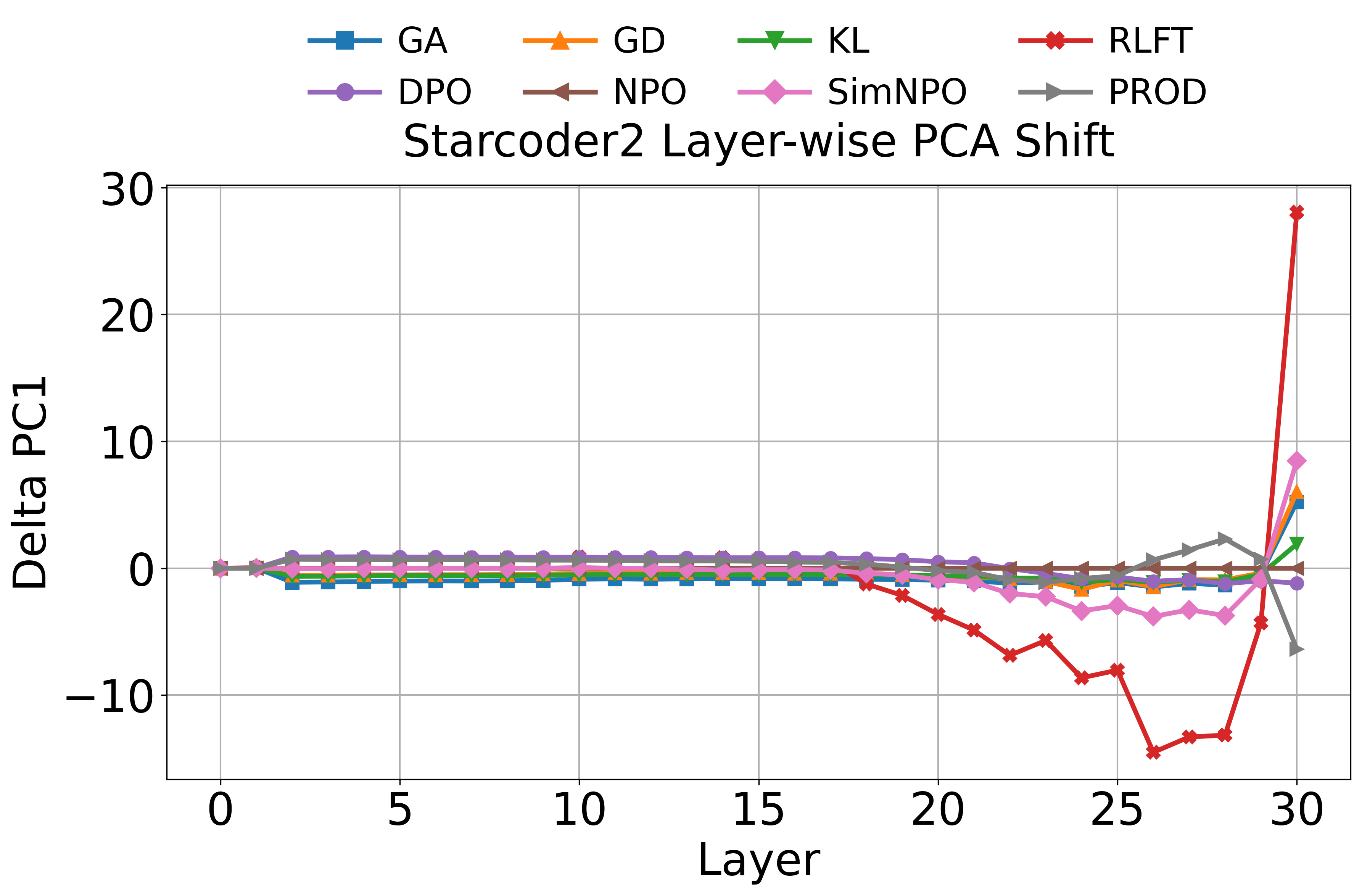}
    \hfill
    \includegraphics[width=0.33\textwidth]{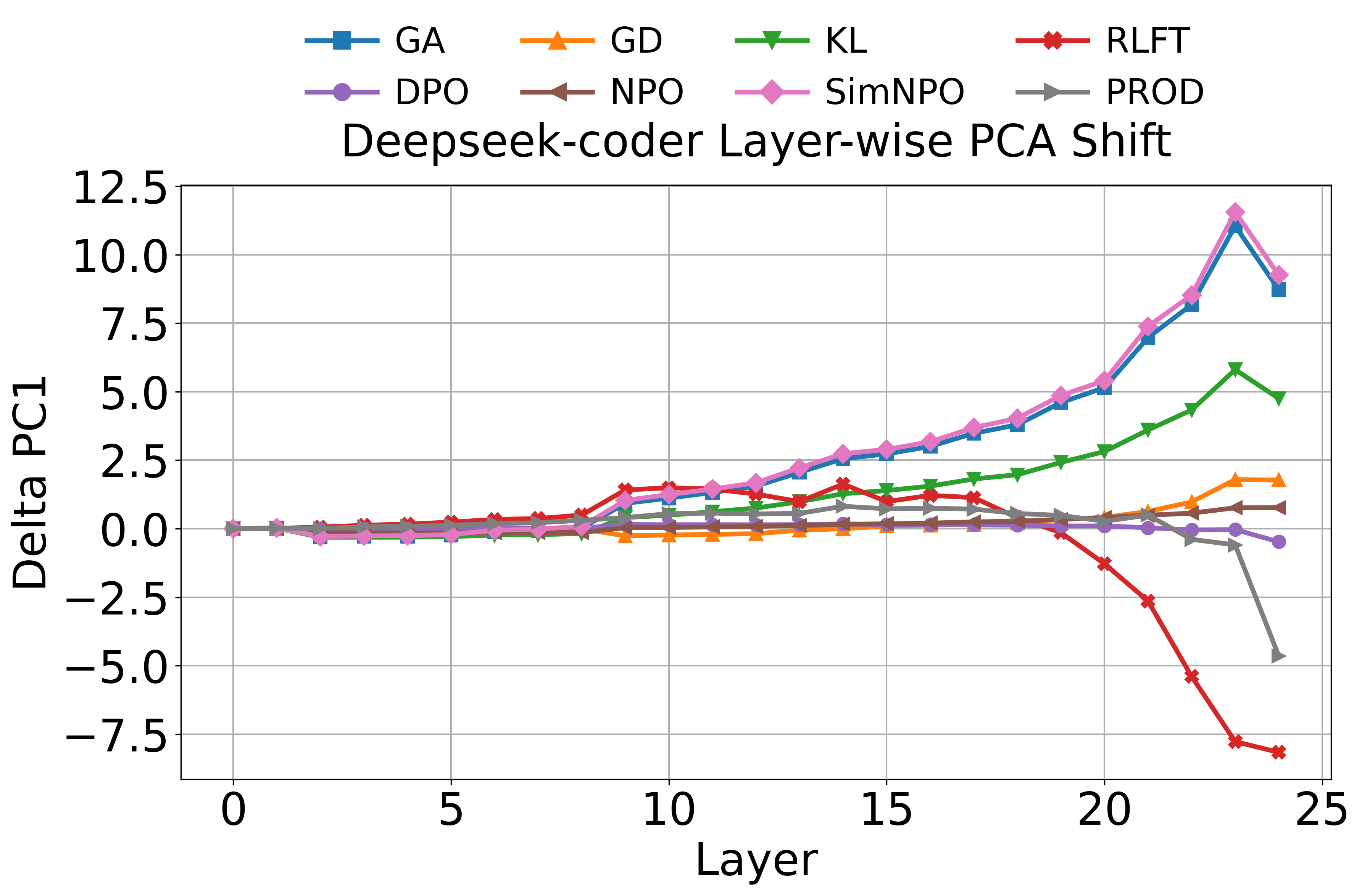}

    \caption{The PCA shift of different machine unlearning methods across model layers.}
    \label{fig:pca_shift}
\end{figure*}

\begin{figure*}[t]
    \centering
    \includegraphics[width=0.33\textwidth]{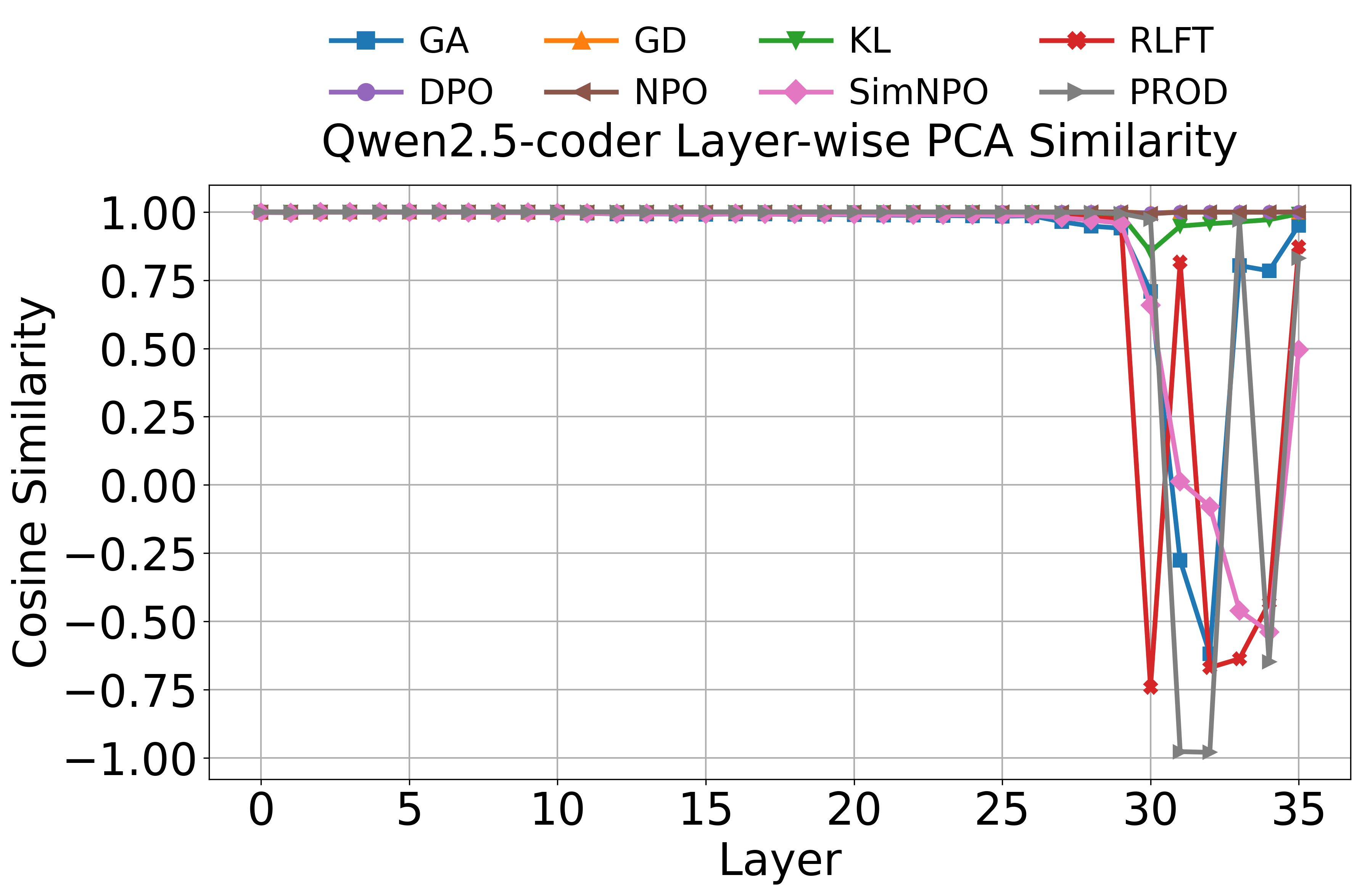}
    \hfill
    \includegraphics[width=0.33\textwidth]{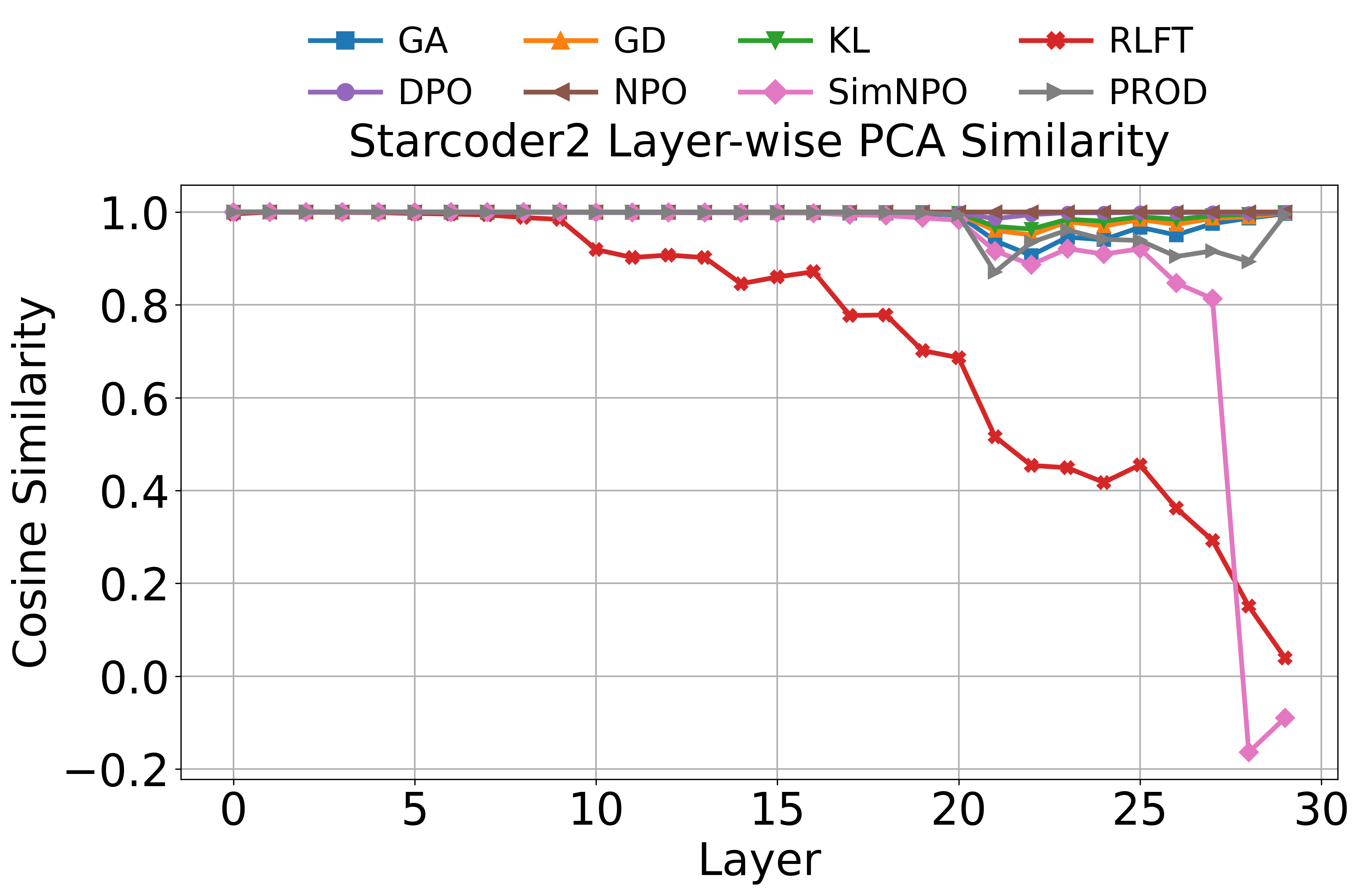}
    \hfill
    \includegraphics[width=0.33\textwidth]{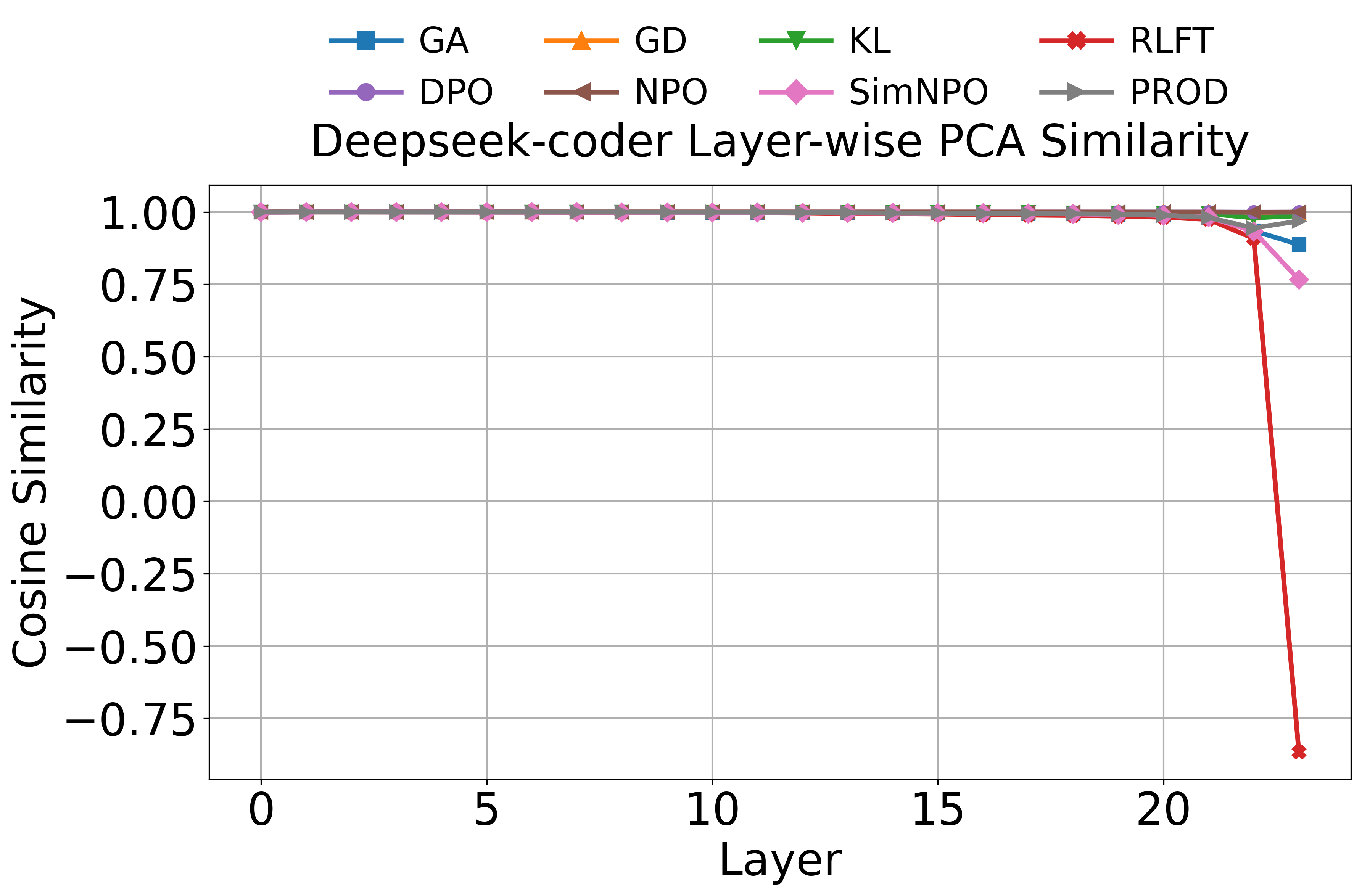}

    \caption{The PCA similarity of different machine unlearning methods across model layers.}
    \label{fig:pca_similarity}
\end{figure*}

\begin{figure*}[t]
    \centering
    \includegraphics[width=0.33\textwidth]{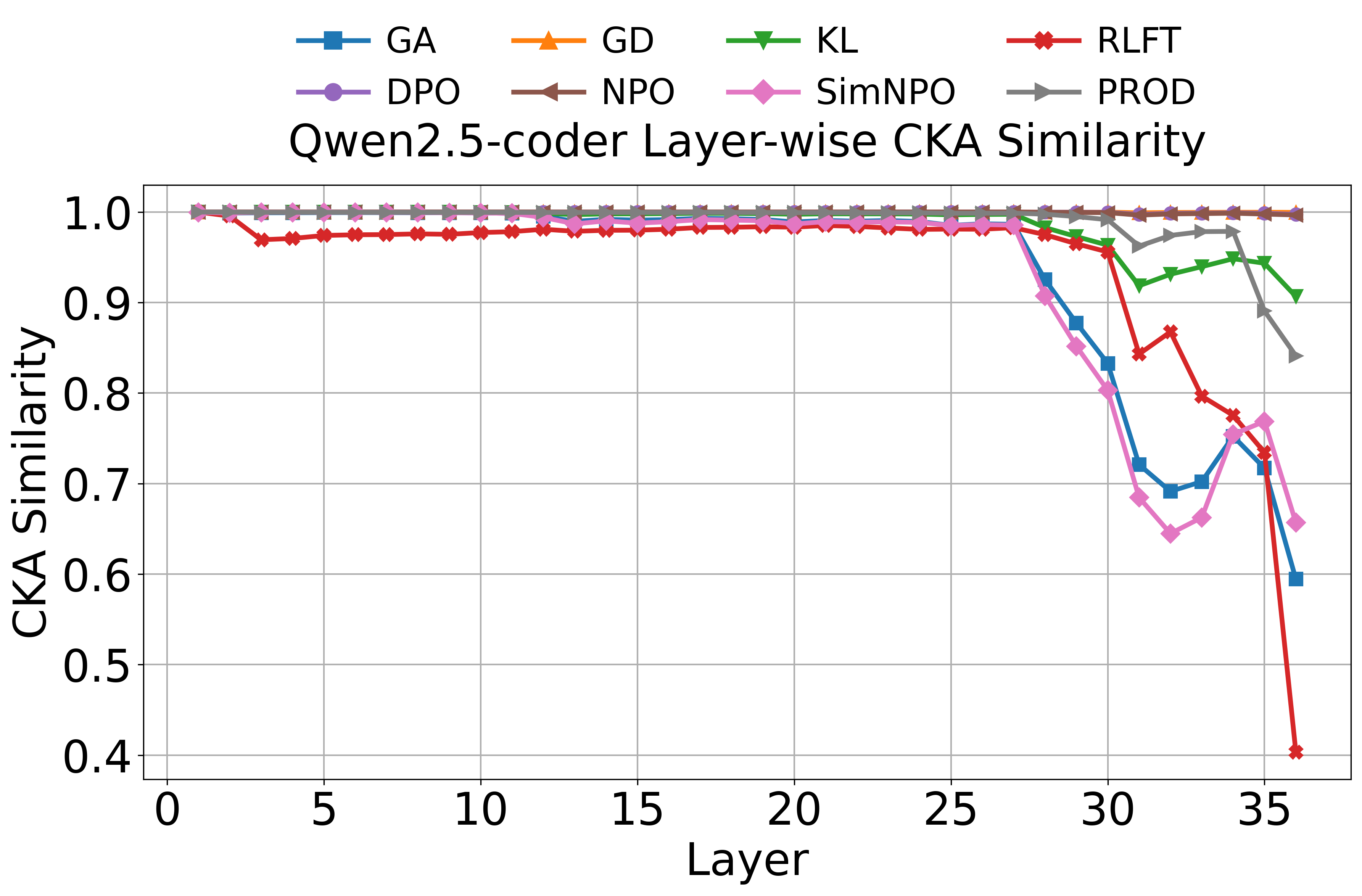}
    \hfill
    \includegraphics[width=0.33\textwidth]{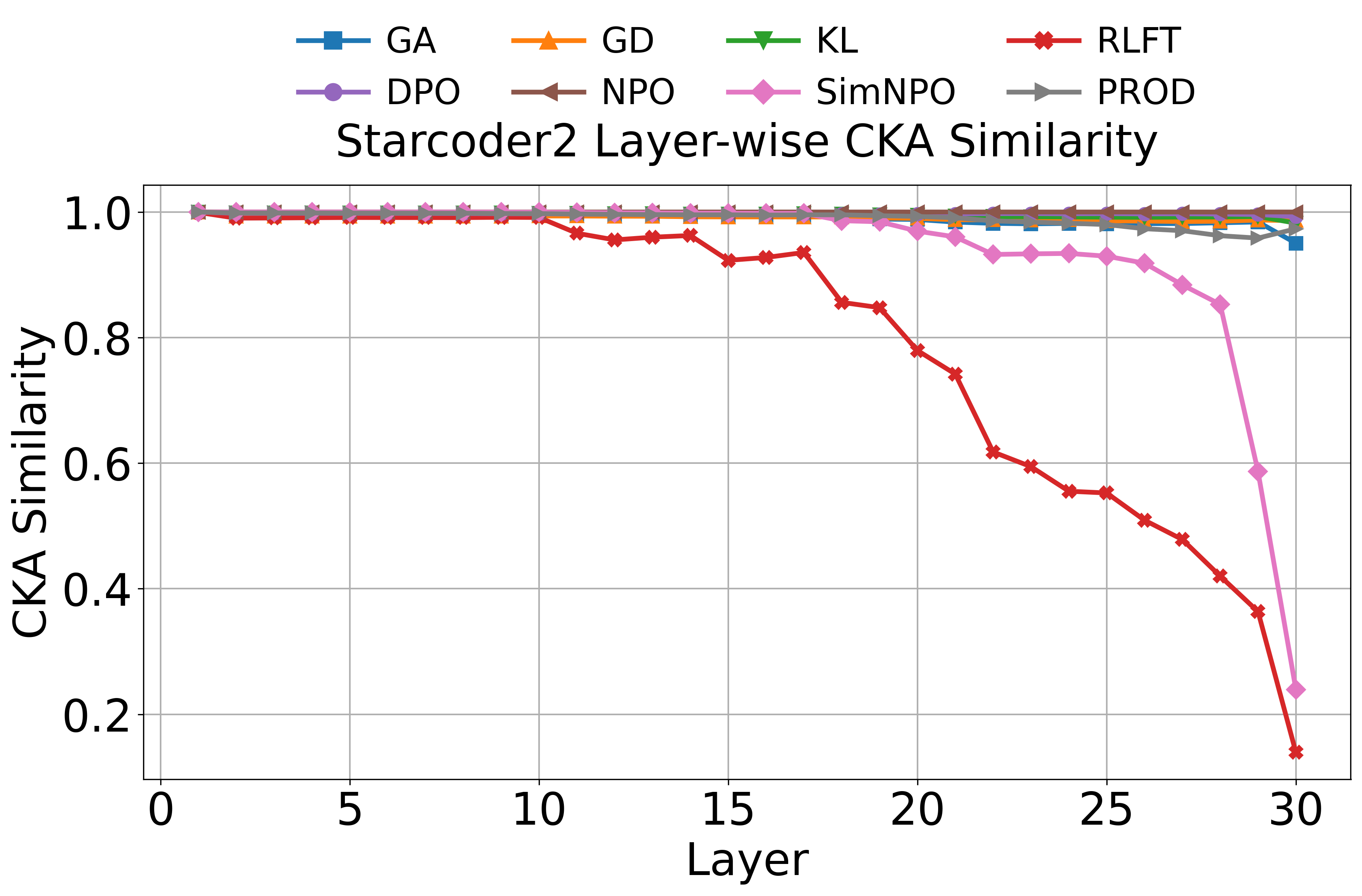}
    \hfill
    \includegraphics[width=0.33\textwidth]{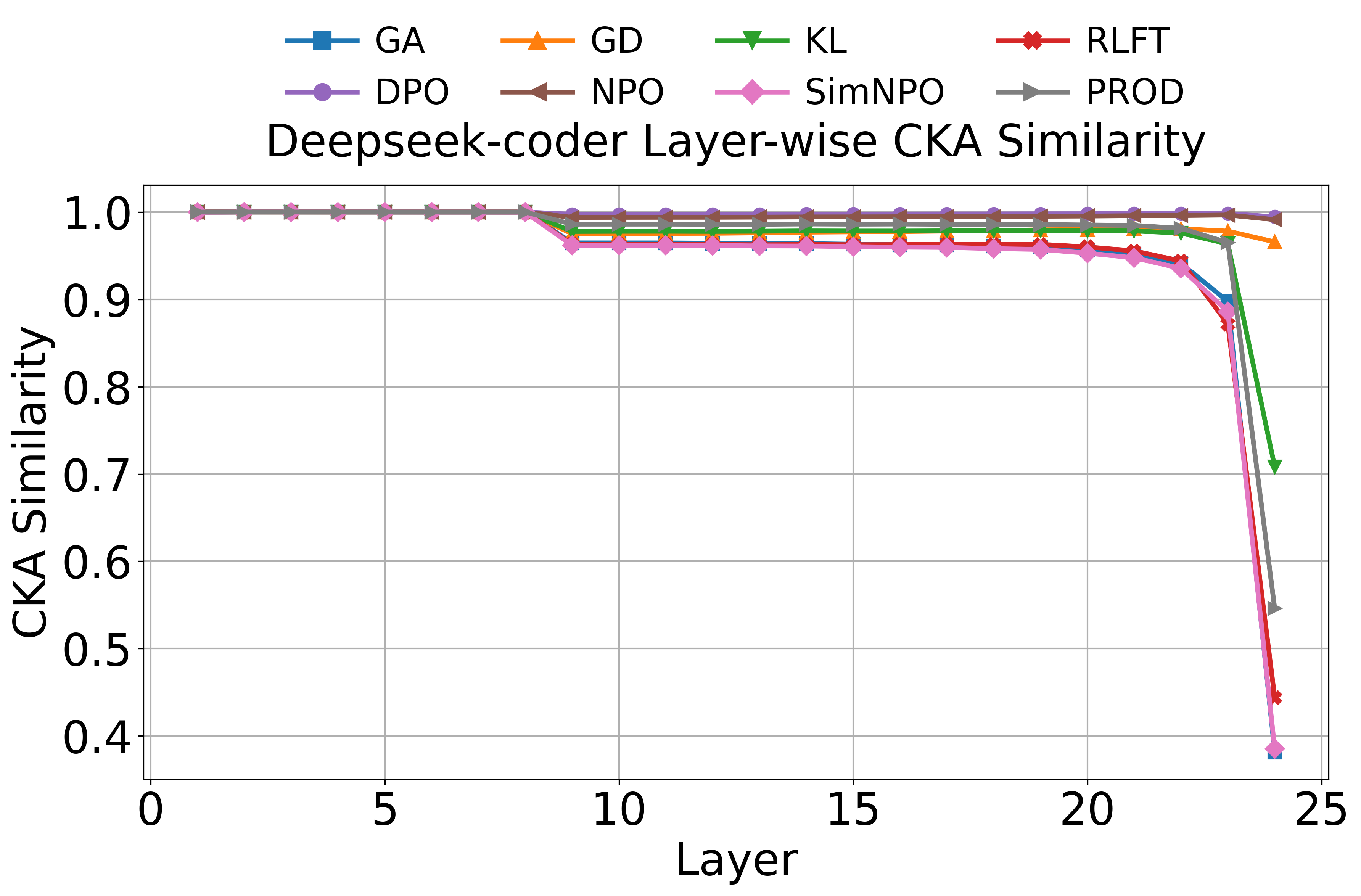}

    \caption{The CKA similarity of different machine unlearning methods across model layers.}
    \label{fig:cka}
\end{figure*}

\begin{figure*}[t]
    \centering
    \includegraphics[width=0.33\textwidth]{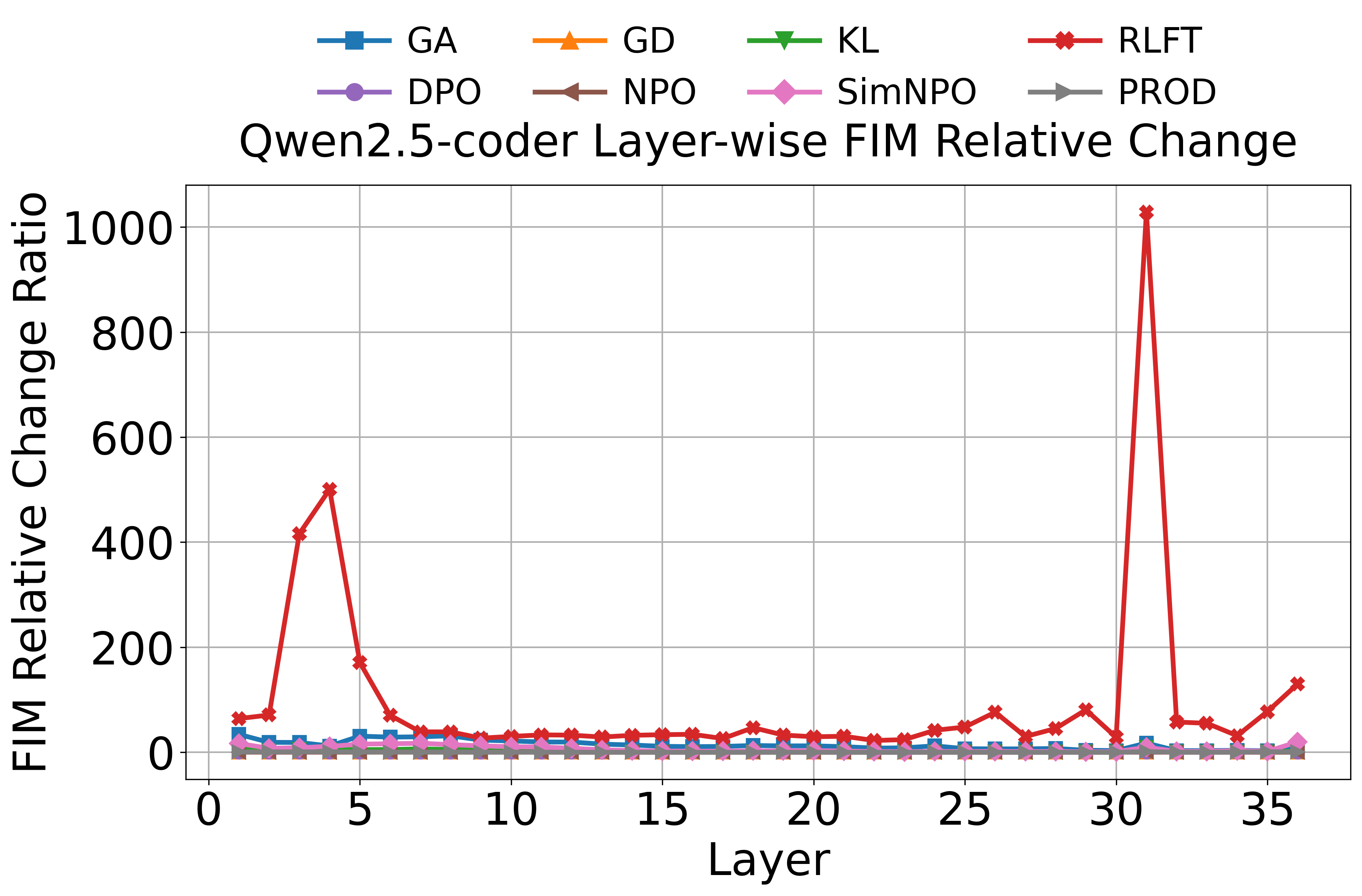}
    \hfill
    \includegraphics[width=0.33\textwidth]{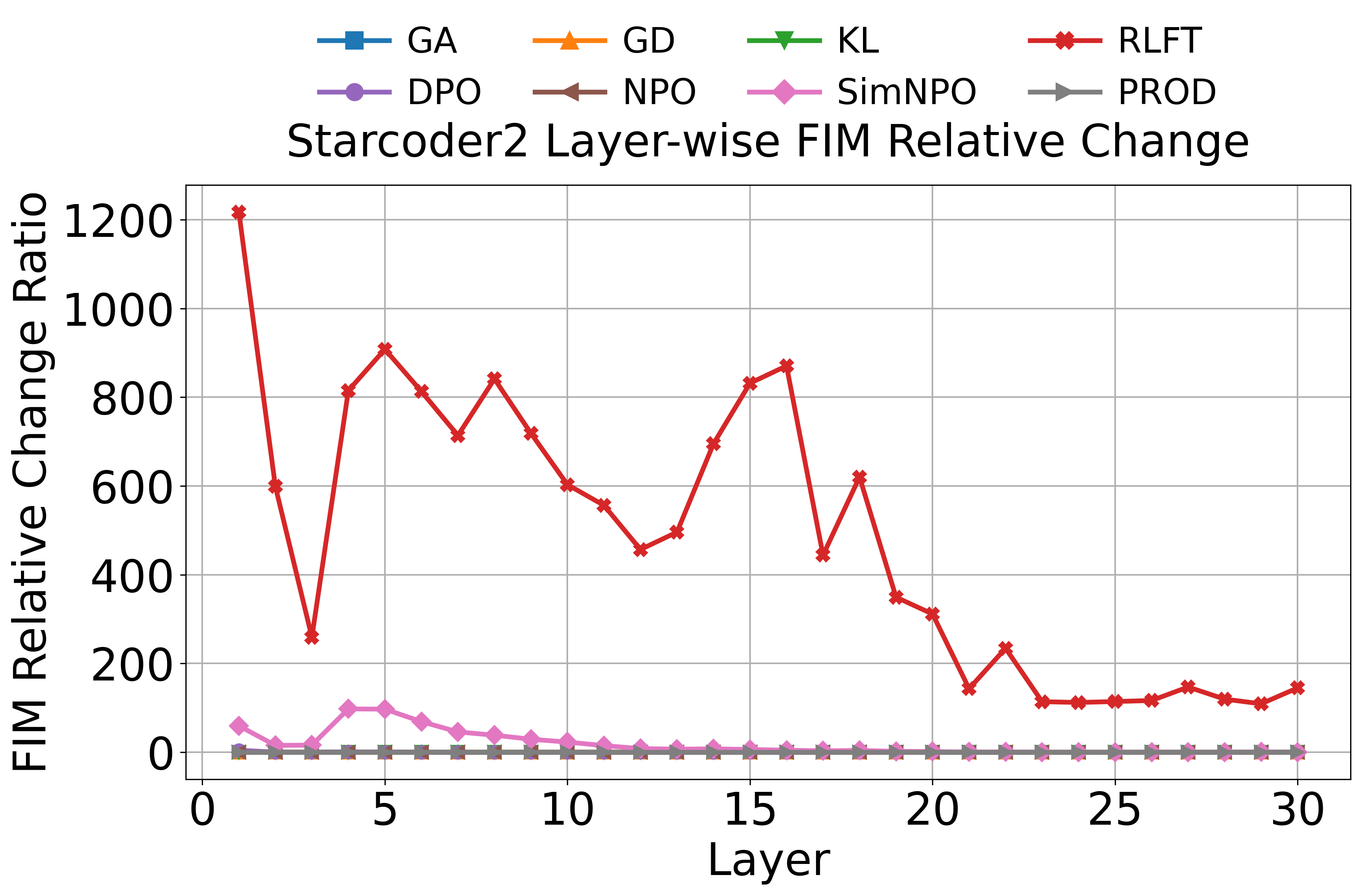}
    \hfill
    \includegraphics[width=0.33\textwidth]{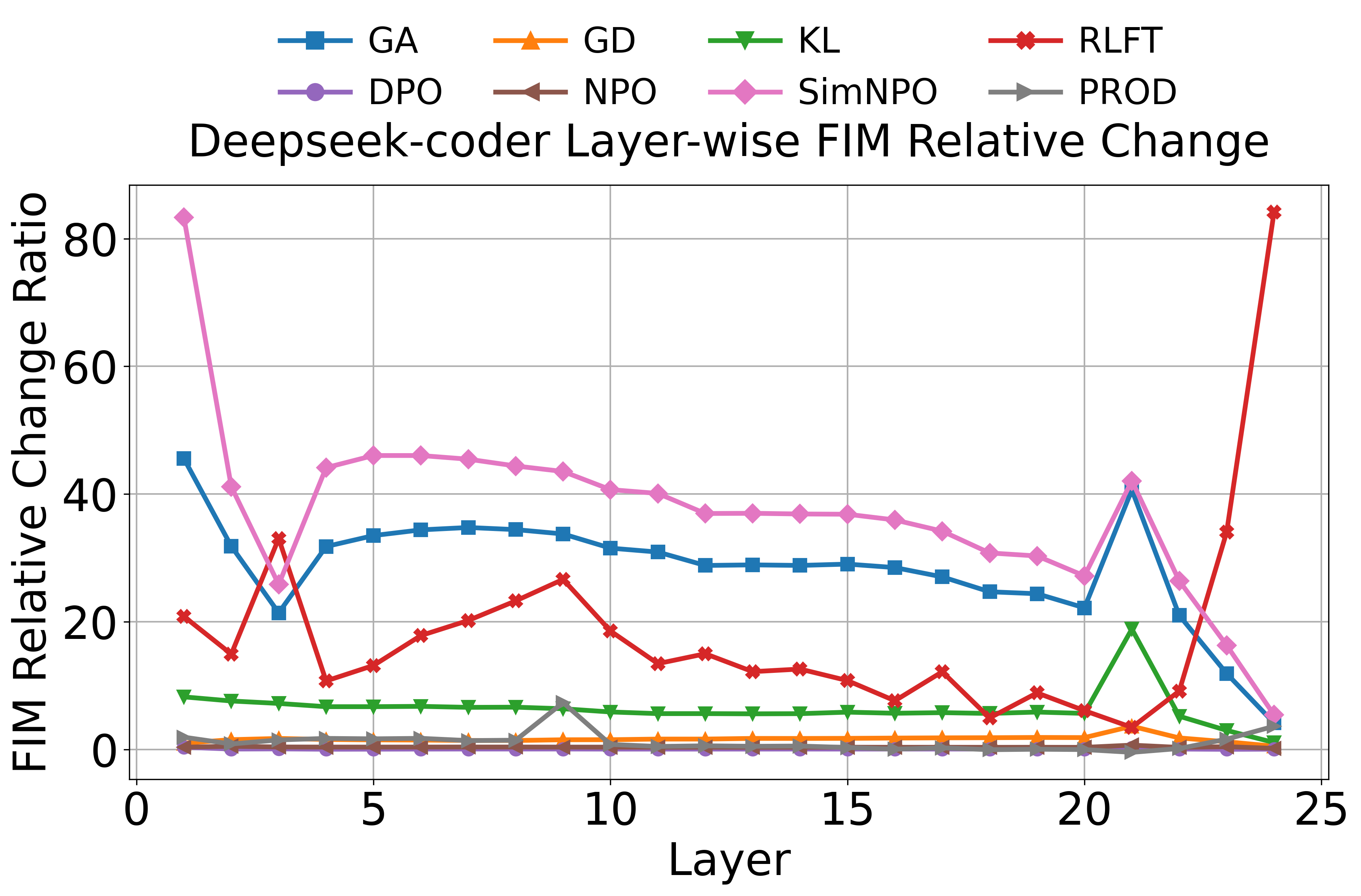}

    \caption{The relative changes in FIM across model layers.}
    \label{fig:fim}
\end{figure*}

%% file: Figure/Case/case.tex
\begin{figure}[t]
    \centering
    \includegraphics[width=1\linewidth]{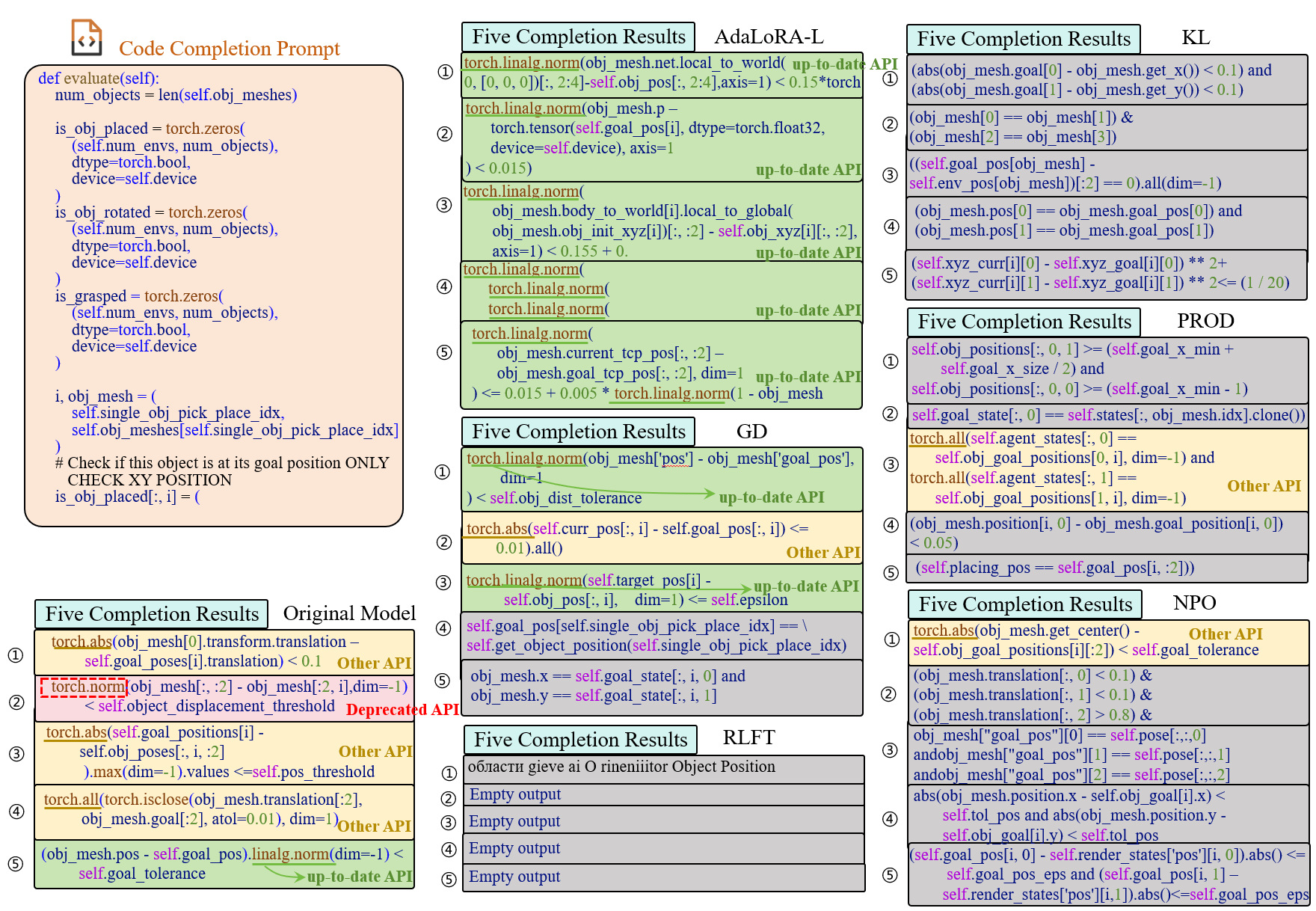}
    \caption{The five code completions generated by the model after applying machine unlearning methods and \textsc{AdaLoRA-L}.}
    \label{fig:case}
\end{figure}

%% file: software.bib
@article{sun2026fly,
  title={On-the-Fly Generation-Quality Enhancement of Deep Code Models via Model Collaboration},
  author={Sun, Weifeng and Huang, Naiqi and Yan, Meng and Liu, Zhongxin and Li, Hongyan and Lei, Yan and Lo, David},
  journal={ACM Transactions on Software Engineering and Methodology},
  volume={35},
  number={6},
  pages={1--40},
  year={2026},
  publisher={ACM New York, NY}
}

@article{jiang2026survey,
  title={A survey on large language models for code generation},
  author={Jiang, Juyong and Wang, Fan and Shen, Jiasi and Kim, Sungju and Kim, Sunghun},
  journal={ACM Transactions on Software Engineering and Methodology},
  volume={35},
  number={2},
  pages={1--72},
  year={2026},
  publisher={ACM New York, NY}
}

@inproceedings{yu2025realisticcodebench,
  title={RealisticCodeBench: Towards More Realistic Evaluation of Large Language Models for Code Generation},
  author={Yu, Xiao and Chen, Haoxuan and Liu, Lei and Hu, Xing and Keung, Jacky Wai and Xia, Xin},
  booktitle={2025 40th IEEE/ACM International Conference on Automated Software Engineering (ASE)},
  pages={3021--3033},
  year={2025},
  organization={IEEE}
}

@inproceedings{karampatsisHowOftenSingleStatement2020,
  title = {How Often Do Single-Statement Bugs Occur? The {ManySStuBs4J} Dataset},
  author = {Karampatsis, Rafael-Michael and Sutton, Charles},
  year = {2020},
  booktitle = {Proceedings of the 17th International Conference on Mining Software Repositories},
  pages = {573--577},
  publisher = {ACM},
  doi = {10.1145/3379597.3387491}
}

@inproceedings{yao2024machine,
  title={Machine unlearning of pre-trained large language models},
  author={Yao, Jin and Chien, Eli and Du, Minxin and Niu, Xinyao and Wang, Tianhao and Cheng, Zezhou and Yue, Xiang},
  booktitle={Proceedings of the 62nd annual meeting of the association for computational linguistics (volume 1: Long papers)},
  pages={8403--8419},
  year={2024}
}

@article{yang2024hotfixing,
  title={Hotfixing large language models for code},
  author={Yang, Zhou and Lo, David},
  journal={arXiv preprint arXiv:2408.05727},
  year={2024}
}

@article{kazemi2024unlearning,
  title={Unlearning Trojans in Large Language Models: A Comparison Between Natural Language and Source Code},
  author={Kazemi, Mahdi and Hussain, Aftab and Rabin, Md Rafiqul Islam and Alipour, Mohammad Amin and Lin, Sen},
  journal={arXiv preprint arXiv:2408.12416},
  year={2024}
}

@article{gu2026mitigating,
  title={Mitigating sensitive information leakage in LLMs4Code through machine unlearning},
  author={Gu, Shanzhi and Qu, Zhaoyang and Geng, Ruotong and Geng, Mingyang and Wang, Shangwen and Xu, Chuanfu and Wang, Haotian and Lin, Zhipeng and Dong, Dezun},
  journal={Neural Networks},
  pages={108606},
  year={2026},
  publisher={Elsevier}
}

@article{chu2025scrub,
  title={Scrub it out! erasing sensitive memorization in code language models via machine unlearning},
  author={Chu, Zhaoyang and Wan, Yao and Zhang, Zhikun and Wang, Di and Yang, Zhou and Zhang, Hongyu and Zhou, Pan and Shi, Xuanhua and Jin, Hai and Lo, David},
  journal={arXiv preprint arXiv:2509.13755},
  year={2025}
}

@article{liang2025forgetting,
  title={When Forgetting Builds Reliability: LLM Unlearning for Reliable Hardware Code Generation},
  author={Liang, Yiwen and Li, Qiufeng and Wang, Shikai and Cao, Weidong},
  journal={arXiv preprint arXiv:2512.05341},
  year={2025}
}

@inproceedings{jiang2026large,
  title={Large language model unlearning for source code},
  author={Jiang, Xue and Dong, Yihong and Zhang, Huangzhao and Wang, Tangxinyu and Fang, Zheng and Ma, Yingwei and Cao, Rongyu and Li, Binhua and Jin, Zhi and Jiao, Wenpin and others},
  booktitle={Proceedings of the AAAI Conference on Artificial Intelligence},
  volume={40},
  number={37},
  pages={31346--31355},
  year={2026}
}

@article{tran2026towards,
  title={Towards Knowledge Alignment in Code LLMs: Contrastive Unlearning for Evolving APIs},
  author={Tran, Huy Q and Vu, Dang H and Dinh, Tuyen N and Nguyen, Anh HD and Vu, Anh NH and Bui, Anh MT and Nguyen, Phuong T},
  journal={arXiv preprint arXiv:2606.30810},
  year={2026}
}

@article{xu2025unlearning,
  title={Unlearning isn't deletion: Investigating reversibility of machine unlearning in llms},
  author={Xu, Xiaoyu and Yue, Xiang and Liu, Yang and Ye, Qingqing and Zheng, Huadi and Hu, Peizhao and Du, Minxin and Hu, Haibo},
  journal={arXiv preprint arXiv:2505.16831},
  year={2025}
}

@article{dorna2026openunlearning,
  title={Openunlearning: Accelerating llm unlearning via unified benchmarking of methods and metrics},
  author={Dorna, Vineeth and Mekala, Anmol and Zhao, Wenlong and McCallum, Andrew and Kolter, Zico and Lipton, Zachary and Maini, Pratyush},
  journal={Advances in Neural Information Processing Systems},
  volume={38},
  year={2026}
}

@article{maini2024tofu,
  title={Tofu: A task of fictitious unlearning for llms},
  author={Maini, Pratyush and Feng, Zhili and Schwarzschild, Avi and Lipton, Zachary C and Kolter, J Zico},
  journal={arXiv preprint arXiv:2401.06121},
  year={2024}
}

@inproceedings{tian2024forget,
  title={To forget or not? towards practical knowledge unlearning for large language models},
  author={Tian, Bozhong and Liang, Xiaozhuan and Cheng, Siyuan and Liu, Qingbin and Wang, Mengru and Sui, Dianbo and Chen, Xi and Chen, Huajun and Zhang, Ningyu},
  booktitle={Findings of the Association for Computational Linguistics: EMNLP 2024},
  pages={1524--1537},
  year={2024}
}

@inproceedings{ren2025general,
  title={A general framework to enhance fine-tuning-based llm unlearning},
  author={Ren, Jie and Dai, Zhenwei and Tang, Xianfeng and Liu, Hui and Zeng, Jingying and Li, Zhen and Goutam, Rahul and Wang, Suhang and Xing, Yue and He, Qi},
  booktitle={Findings of the Association for Computational Linguistics: ACL 2025},
  pages={18464--18476},
  year={2025}
}

@inproceedings{bhaila2025soft,
  title={Soft prompting for unlearning in large language models},
  author={Bhaila, Karuna and Van, Minh-Hao and Wu, Xintao},
  booktitle={Proceedings of the 2025 Conference of the Nations of the Americas Chapter of the Association for Computational Linguistics: Human Language Technologies (Volume 1: Long Papers)},
  pages={4046--4056},
  year={2025}
}

@inproceedings{xu2025relearn,
  title={Relearn: Unlearning via learning for large language models},
  author={Xu, Haoming and Zhao, Ningyuan and Yang, Liming and Zhao, Sendong and Deng, Shumin and Wang, Mengru and Hooi, Bryan and Oo, Nay and Chen, Huajun and Zhang, Ningyu},
  booktitle={Proceedings of the 63rd Annual Meeting of the Association for Computational Linguistics (Volume 1: Long Papers)},
  pages={5967--5987},
  year={2025}
}

@article{zhang2024negative,
  title={Negative preference optimization: From catastrophic collapse to effective unlearning},
  author={Zhang, Ruiqi and Lin, Licong and Bai, Yu and Mei, Song},
  journal={arXiv preprint arXiv:2404.05868},
  year={2024}
}

@article{lozhkov2024starcoder,
  title={Starcoder 2 and the stack v2: The next generation},
  author={Lozhkov, Anton and Li, Raymond and Allal, Loubna Ben and Cassano, Federico and Lamy-Poirier, Joel and Tazi, Nouamane and Tang, Ao and Pykhtar, Dmytro and Liu, Jiawei and Wei, Yuxiang and others},
  journal={arXiv preprint arXiv:2402.19173},
  year={2024}
}

@article{hui2024qwen2,
  title={Qwen2. 5-coder technical report},
  author={Hui, Binyuan and Yang, Jian and Cui, Zeyu and Yang, Jiaxi and Liu, Dayiheng and Zhang, Lei and Liu, Tianyu and Zhang, Jiajun and Yu, Bowen and Lu, Keming and others},
  journal={arXiv preprint arXiv:2409.12186},
  year={2024},
  doi={10.48550/arXiv.2409.12186}
}

@article{guo2024deepseek,
  title={DeepSeek-Coder: when the large language model meets programming--the rise of code intelligence},
  author={Guo, Daya and Zhu, Qihao and Yang, Dejian and Xie, Zhenda and Dong, Kai and Zhang, Wentao and Chen, Guanting and Bi, Xiao and Wu, Yifan and Li, YK and others},
  journal={arXiv preprint arXiv:2401.14196},
  year={2024},
  doi={10.48550/arXiv.2401.14196}
}

@article{pan2024codev,
  title={Codev-Bench: How Do LLMs Understand Developer-Centric Code Completion?},
  author={Pan, Zhenyu and Cao, Rongyu and Cao, Yongchang and Ma, Yingwei and Li, Binhua and Huang, Fei and Liu, Han and Li, Yongbin},
  journal={arXiv preprint arXiv:2410.01353},
  year={2024},
  doi={10.48550/arXiv.2410.01353}
}

@article{sultana2024code,
  title={Code vulnerability detection: A comparative analysis of emerging large language models},
  author={Sultana, Shaznin and Afreen, Sadia and Eisty, Nasir U},
  journal={arXiv preprint arXiv:2409.10490},
  year={2024},
  doi={10.48550/arXiv.2409.10490}
}

@article{ji2025causality,
  title={Causality-Aided Evaluation and Explanation of Large Language Model-Based Code Generation},
  author={Ji, Zhenlan and Ma, Pingchuan and Li, Zongjie and Wang, Zhaoyu and Wang, Shuai},
  journal={Proceedings of the ACM on Software Engineering},
  volume={2},
  number={ISSTA},
  pages={1374--1397},
  year={2025},
  publisher={ACM New York, NY, USA},
  doi={10.1145/3728938}
}

@inproceedings{sun2024ai,
  title={Ai coders are among us: Rethinking programming language grammar towards efficient code generation},
  author={Sun, Zhensu and Du, Xiaoning and Yang, Zhou and Li, Li and Lo, David},
  booktitle={Proceedings of the 33rd ACM SIGSOFT International Symposium on Software Testing and Analysis},
  pages={1124--1136},
  year={2024},
  doi={10.1145/3650212.3680347}
}

@inproceedings{zhang2023repocoder,
  title={Repocoder: Repository-level code completion through iterative retrieval and generation},
  author={Zhang, Fengji and Chen, Bei and Zhang, Yue and Keung, Jacky and Liu, Jin and Zan, Daoguang and Mao, Yi and Lou, Jian-Guang and Chen, Weizhu},
  booktitle={Proceedings of the 2023 Conference on Empirical Methods in Natural Language Processing},
  pages={2471--2484},
  year={2023},
  doi={10.18653/v1/2023.emnlp-main.151}
}

@article{yu2024fight,
  title={Fight fire with fire: How much can we trust chatgpt on source code-related tasks?},
  author={Yu, Xiao and Liu, Lei and Hu, Xing and Keung, Jacky Wai and Liu, Jin and Xia, Xin},
  journal={IEEE Transactions on Software Engineering},
  volume={50},
  number={12},
  pages={3435--3453},
  year={2024},
  publisher={IEEE},
  doi={10.1109/TSE.2024.3492204}
}

@inproceedings{wang2024llms,
  title={Llms meet library evolution: Evaluating deprecated api usage in llm-based code completion},
  author={Wang, Chong and Huang, Kaifeng and Zhang, Jian and Feng, Yebo and Zhang, Lyuye and Liu, Yang and Peng, Xin},
  booktitle={2025 ieee/acm 47th international conference on software engineering (icse)},
  pages={885--897},
  year={2025},
  organization={IEEE},
  doi={10.1109/ICSE55347.2025.00245}
}

@online{source-graph,
    title = {Sourcegraph},
    url="https://sourcegraph.com/search",
}

@inproceedings{wang2020empirical,
  title={An empirical study of usages, updates and risks of third-party libraries in java projects},
  author={Wang, Ying and Chen, Bihuan and Huang, Kaifeng and Shi, Bowen and Xu, Congying and Peng, Xin and Wu, Yijian and Liu, Yang},
  booktitle={2020 IEEE International conference on software maintenance and evolution (ICSME)},
  pages={35--45},
  year={2020},
  organization={IEEE},
  doi={10.1109/ICSME46990.2020.00014}
}

@article{zhan2021research,
  title={Research on third-party libraries in android apps: A taxonomy and systematic literature review},
  author={Zhan, Xian and Liu, Tianming and Fan, Lingling and Li, Li and Chen, Sen and Luo, Xiapu and Liu, Yang},
  journal={IEEE Transactions on Software Engineering},
  volume={48},
  number={10},
  pages={4181--4213},
  year={2021},
  publisher={IEEE},
  doi={10.1109/TSE.2021.3114381}
}

@article{kula2018empirical,
  title={An empirical study on the impact of refactoring activities on evolving client-used apis},
  author={Kula, Raula Gaikovina and Ouni, Ali and German, Daniel M and Inoue, Katsuro},
  journal={Information and Software Technology},
  volume={93},
  pages={186--199},
  year={2018},
  publisher={Elsevier},
  doi={10.1016/j.infsof.2017.09.007}
}

@article{hu2023empirical,
  title={An empirical study of the Python/C API on evolution and bug patterns},
  author={Hu, Mingzhe and Zhang, Yu},
  journal={Journal of Software: Evolution and Process},
  volume={35},
  number={2},
  pages={e2507},
  year={2023},
  publisher={Wiley Online Library},
  doi={10.1002/smr.2507}
}

@inproceedings{wang2020exploring,
  title={Exploring how deprecated python library apis are (not) handled},
  author={Wang, Jiawei and Li, Li and Liu, Kui and Cai, Haipeng},
  booktitle={Proceedings of the 28th acm joint meeting on european software engineering conference and symposium on the foundations of software engineering},
  pages={233--244},
  year={2020},
  doi={10.1145/3368089.3409735}
}

@inproceedings{feng2020codebert,
  title={Codebert: A pre-trained model for programming and natural languages},
  author={Feng, Zhangyin and Guo, Daya and Tang, Duyu and Duan, Nan and Feng, Xiaocheng and Gong, Ming and Shou, Linjun and Qin, Bing and Liu, Ting and Jiang, Daxin and others},
  booktitle={Findings of the association for computational linguistics: EMNLP 2020},
  pages={1536--1547},
  year={2020},
  doi={10.18653/v1/2020.findings-emnlp.139}
}

@inproceedings{papineni2002bleu,
  title={Bleu: a method for automatic evaluation of machine translation},
  author={Papineni, Kishore and Roukos, Salim and Ward, Todd and Zhu, Wei-Jing},
  booktitle={Proceedings of the 40th annual meeting of the Association for Computational Linguistics},
  pages={311--318},
  year={2002},
  doi={10.3115/1073083.1073135}
}

@inproceedings{lin2004rouge,
  title={Rouge: A package for automatic evaluation of summaries},
  author={Lin, Chin-Yew},
  booktitle={Text summarization branches out},
  pages={74--81},
  year={2004}
}

@article{chen2021evaluating,
  title={Evaluating large language models trained on code},
  author={Chen, Mark and Tworek, Jerry and Jun, Heewoo and Yuan, Qiming and Pinto, Henrique Ponde De Oliveira and Kaplan, Jared and Edwards, Harri and Burda, Yuri and Joseph, Nicholas and Brockman, Greg and others},
  journal={arXiv preprint arXiv:2107.03374},
  year={2021}
}

@article{fan2026simplicity,
  title={Simplicity prevails: Rethinking negative preference optimization for llm unlearning},
  author={Fan, Chongyu and Liu, Jiancheng and Lin, Licong and Jia, Jinghan and Zhang, Ruiqi and Mei, Song and Liu, Sijia},
  journal={Advances in Neural Information Processing Systems},
  volume={38},
  pages={1540--1567},
  year={2026}
}

@article{hu2021lora,
  title={Lora: Low-rank adaptation of large language models},
  author={Hu, Edward J and Shen, Yelong and Wallis, Phillip and Allen-Zhu, Zeyuan and Li, Yuanzhi and Wang, Shean and Wang, Lu and Chen, Weizhu},
  journal={arXiv preprint arXiv:2106.09685},
  year={2021}
}

@inproceedings{guancheng2026don,
  title={Don’t Use a Cannon to Kill a Fly: Lightweight Model Editing for LLMs to Correct Deprecated API Recommendations},
  author={Guancheng, LIN and Yu, Xiao and Keung, Jacky and Hu, Xing and Xia, Xin and Liu, Alex X},
  booktitle={35th ACM SIGSOFT International Symposium on Software Testing and Analysis (ISSTA 2026)},
  year={2026},
  organization={Association for Computing Machinery}
}

@inproceedings{golatkar2020eternal,
  title={Eternal sunshine of the spotless net: Selective forgetting in deep networks},
  author={Golatkar, Aditya and Achille, Alessandro and Soatto, Stefano},
  booktitle={2020 IEEE/CVF Conference on Computer Vision and Pattern Recognition (CVPR)},
  pages={9301--9309},
  year={2020},
  organization={IEEE}
}

@inproceedings{liu2022continual,
  title={Continual learning and private unlearning},
  author={Liu, Bo and Liu, Qiang and Stone, Peter},
  booktitle={Conference on Lifelong Learning Agents},
  pages={243--254},
  year={2022},
  organization={PMLR}
}

@article{qiu2024pistol,
  title={Pistol: Dataset compilation pipeline for structural unlearning of llms},
  author={Qiu, Xinchi and Shen, William F and Chen, Yihong and Cancedda, Nicola and Stenetorp, Pontus and Lane, Nicholas D},
  journal={arXiv preprint arXiv:2406.16810},
  volume={2406},
  year={2024}
}

@inproceedings{shi2025muse,
  title={Muse: Machine unlearning six-way evaluation for language models},
  author={Shi, Weijia and Lee, Jaechan and Huang, Yangsibo and Malladi, Sadhika and Zhao, Jieyu and Holtzman, Ari and Liu, Daogao and Zettlemoyer, Luke and Smith, Noah and Zhang, Chiyuan},
  booktitle={International Conference on Learning Representations},
  volume={2025},
  pages={27797--27818},
  year={2025}
}

@article{wu2024versicode,
  title={Versicode: Towards version-controllable code generation},
  author={Wu, Tongtong and Wu, Weigang and Wang, Xingyu and Xu, Kang and Ma, Suyu and Jiang, Bo and Yang, Ping and Xing, Zhenchang and Li, Yuan-Fang and Haffari, Gholamreza},
  journal={arXiv preprint arXiv:2406.07411},
  year={2024}
}

@article{li2024wmdp,
  title={The wmdp benchmark: Measuring and reducing malicious use with unlearning},
  author={Li, Nathaniel and Pan, Alexander and Gopal, Anjali and Yue, Summer and Berrios, Daniel and Gatti, Alice and Li, Justin D and Dombrowski, Ann-Kathrin and Goel, Shashwat and Phan, Long and others},
  journal={arXiv preprint arXiv:2403.03218},
  year={2024}
}

@article{kullback1951information,
  title={On information and sufficiency},
  author={Kullback, Solomon and Leibler, Richard A},
  journal={The annals of mathematical statistics},
  volume={22},
  number={1},
  pages={79--86},
  year={1951},
  publisher={JSTOR}
}

@article{le2025survey,
  title={A survey on large language models unlearning: taxonomy, evaluations, and future directions},
  author={Le-Khac, Uyen N and Truong, Vinh NX},
  journal={Artificial Intelligence Review},
  volume={58},
  number={12},
  pages={399},
  year={2025},
  publisher={Springer}
}

@article{kirkpatrick2017overcoming,
  title={Overcoming catastrophic forgetting in neural networks},
  author={Kirkpatrick, James and Pascanu, Razvan and Rabinowitz, Neil and Veness, Joel and Desjardins, Guillaume and Rusu, Andrei A and Milan, Kieran and Quan, John and Ramalho, Tiago and Grabska-Barwinska, Agnieszka and others},
  journal={Proceedings of the national academy of sciences},
  volume={114},
  number={13},
  pages={3521--3526},
  year={2017},
  publisher={National Academy of Sciences}
}

@article{liu2025rethinking,
  title={Rethinking machine unlearning for large language models},
  author={Liu, Sijia and Yao, Yuanshun and Jia, Jinghan and Casper, Stephen and Baracaldo, Nathalie and Hase, Peter and Yao, Yuguang and Liu, Chris Yuhao and Xu, Xiaojun and Li, Hang and others},
  journal={Nature Machine Intelligence},
  volume={7},
  number={2},
  pages={181--194},
  year={2025},
  publisher={Nature Publishing Group UK London}
}
